\documentclass[10pt]{article}
\def\p{\partial}
\def\no{\nonumber}
\def\la{\langle}
\def\ra{\rangle}
\def\p{\prime}

\DeclareMathAlphabet{\mathpzc}{OT1}{pzc}{s}{it}
\usepackage{mathrsfs}
\usepackage{tabularx}
\usepackage{subfigure}
\usepackage{graphicx}
\usepackage{amsfonts}
\usepackage{amssymb}
\usepackage{amsbsy}
\usepackage{amsmath}
\usepackage{mathrsfs}
\usepackage{latexsym}
\usepackage{bm}
\usepackage{subfigure}
\usepackage{color}
\usepackage{wasysym}
\usepackage{mathbbol}
\usepackage{bigints}
\usepackage{dsfont}

\begin{document}
\title{{\bf{\Large Constructing an entangled Unruh Otto engine and its efficiency}}}
\author{
 {\bf {\normalsize Dipankar Barman}$
$\thanks{E-mail: dipankar1998@iitg.ac.in}},\, 
 {\bf {\normalsize Bibhas Ranjan Majhi}$
$\thanks{E-mail: bibhas.majhi@iitg.ac.in}}\\
 {\normalsize Department of Physics, Indian Institute of Technology Guwahati,}
\\{\normalsize Guwahati 781039, Assam, India}
\\[0.3cm]
}

\maketitle

\begin{abstract}
Uniformly accelerated frame mimics a thermal bath whose temperature is proportional to the proper acceleration. Using this phenomenon we give a detailed construction of an Otto cycle between two energy eigenstates of a system, consists of two entangled qubits. In the isochoric stages the thermal bath is being provided via the vacuum fluctuations of the background field for a monopole interaction by accelerating them. We find that making of Otto cycle is possible when one qubit is accelerating in the right Rindler wedge and other one is moving in the left Rindler wedge; i.e. in anti-parallel motion, with the initial composite  state is a non-maximally entangled one. However, the efficiency greater than that of the usual single qubit quantum Otto engine is not possible. We provide values of the available parameters which make Otto cycle possible. On the other hand, Otto cycle is not possible if one considers the non-maximally entangled state for parallel motion. Moreover, for both initial symmetric and anti-symmetric Bell states we do not find any possibility of  the cycle for qubits' parallel and anti-parallel motion. 

\end{abstract}

\section{Introduction}
Thermodynamics of a quantum system has been a very active area and considerable amount of effort has been devoted to develop a fruitful theoretical formalism in order to explore the quantum thermodynamics. 
The definition of work, energy and heat of a quantum system is properly addressed in this paradigm.
Consider a quantum system, described by a density operator $\rho(t)$, is evolving under a time dependent Hamiltonian $H_0(t)$. The variation of the expectation value of the energy, $\Delta\langle{E}\rangle$  from time $t_i$ to $t_f$ satisfies \cite{book:quanthermo}
\begin{eqnarray}
\Delta\langle{E}\rangle=\langle{Q}\rangle+\langle{W}\rangle~,
\label{B1}
\end{eqnarray}
where we identify
\begin{eqnarray}
&&\Delta\langle{E}\rangle=\int_{t_i}^{t_f}dt\frac{d}{dt}\langle{E}\rangle~;
\label{B2n2}
\\ &&\langle{Q}\rangle=\int_{t_i}^{t_f}dt~\text{Tr}\Big(\frac{d\rho(t)}{dt}H_0\Big)~;
\label{B3}
\end{eqnarray}
and
\begin{eqnarray}
\langle{W}\rangle=\int_{t_i}^{t_f}dt~\text{Tr}\Bigg(\rho(t)\frac{dH_0}{dt}\Big)~,
\label{B4}
\end{eqnarray}
as total energy change, heat transfer and work done on the system, respectively.
The above concepts of different thermodynamic quantities has been successfully implemented to construct quantum version of different classical engines; e.g. Carnot engine, Otto engine, {\it etc}. \cite{Bender_2000,PhysRevLett.93.140403,Kieu:2006ux, Quan_2007, Cakmak:2020ts}. A Quantum Otto Engine (QOE) can be constructed with a single qubit \cite{PhysRevLett.93.140403,Kieu:2006ux}, which has four steps. The system undergoes two adiabatic processes and two isochoric processes. In adiabatic process, there is work done on or by the system and in isochoric process, system has heat exchange with the environment. The efficiency turns out to be 
\begin{equation}
\eta_0 = 1 - \frac{\omega_1}{\omega_2}~,
\label{B5n5}
\end{equation}
with $\eta_0<1$.
Here $\omega_1$ ($\omega_2$) is the energy gap before (after) the adiabatic expansion of the levels of the qubit. Note that, contrary to the classical Otto engine in which efficiency depends on the temperatures of the thermal baths during the isochoric processes, $\eta_0$ for QOE depends on the energy gaps between the quantum levels of the qubit system.  

On the other hand the combination of relativity and quantum mechanics, best described by quantum field theory, brings several interesting and important phenomenon, like Hawking \cite{HAWKING:1974us, hawking1975} and Unruh \cite{Davies:1974th,Unruh:1976db, PhysRevD.29.1047} effects, in front of us. These two phenomenon naturally create a thermal environment for a specific class of observers.
Recently, using the concept of Unruh phenomenon \cite{Davies:1974th,Unruh:1976db, PhysRevD.29.1047}, a Unruh quantum Otto engine (UQOE) has been proposed in \cite{Arias:2018wc,Gray:2018uw,XU2020135201}. This, contrary to QOE, is a completely relativistic set up and therefore the measured time in each stages is denoted by qubit's proper time (so the role of time parameters $t$ in Eqs. (\ref{B2n2}) -- (\ref{B4}) is played by proper time $\tau$ of qubit's frame). According to the Unruh effect, a uniformly accelerating observer can see particles in the Minkowski vacuum and this acts as the thermal bath for the accelerating frame. The temperature of the bath which is proportional to the proper acceleration can be utilised as heat source for an Otto cycle. So the hot (cold) thermal baths during the isochoric processes can be mimicked by giving uniformly acceleration (deceleration) to the qubit. This is known as Unruh Quantum Otto Engine (UQOE). The efficiency of this cycle came out to be that given in (\ref{B5n5}) (see \cite{Arias:2018wc,Gray:2018uw}, for details). 

Using the idea of UQOE, one of the authors of this paper with Kane very recently proposed a quantum
Otto cycle consists of two entangled qubits  \cite{PhysRevD.104.L041701}. The underlying inspiration comes from the fact that Unruh effect is greatly influenced by entanglement between the accelerated observer with another accelerated frame \cite{Arias:2016tj, RODRIGUEZCAMARGO2018266, Picanco:2020uu, Barman:2021vj} . Moreover the entanglement phenomenon itself is observer dependent quantity \cite{Alsing_2004, PhysRevLett.95.120404, PhysRevLett.106.210502,Chowdhury:2021ieg}. In this regard it may also be noted that the quantum entanglement harvesting between two causally disconnected accelerated detectors is possible within a relativistic setup. All these events imply a possibility of considerable influence on the efficiency of engines when one considers the cycle between entangled states of two accelerated qubits.  The analysis in \cite{PhysRevD.104.L041701} revealed that the efficiency for the cycle between an
entangled state and their collective excited state, depending on the relative accelerations of the qubits
during the isochoric process, may vary from a standard quantum Otto cycle. This is named as Entangled Unruh Quantum Otto Engine (EUQOE). 

In this paper we will readdress this EUQOE and its nature of efficiency in a much more broader perspective. Moreover the analysis in \cite{PhysRevD.104.L041701} is found to have few limitations and is not complete. Let us now mention them.
\begin{itemize}
\item To make a cycle efficient for work output, the amount of work done has to be positive. In addition during the isochoric processes the heat absorption (rejection) must happen in order to make a fruitful engine. We will see later that these are supplemented by condition given by Eq. (\ref{Tr+}) (see also below Eq. (25) of \cite{PhysRevD.104.L041701}).  In order to fulfilment of these one must choose the values of the available parameters (e.g. accelerations of the qubits, the spacing of energy levels of the qubits, the interaction time during isochoric phases, {\it  etc}.), appearing in the system. This identification of the parameter space of all these quantities has not been done earlier. Therefore it is not clear whether any such parameter space is available in all practical purposes in order to construct a EUQOE. 
\item The time evolution of the state of system has been done by assuming the satisfaction of commutation between different time Hamiltonians. But we know that in general this may not be always true and in that situation time-ordering is necessary.
\item In the existing analysis considered that the two qubits are accelerating in the same side of the Rindler wedge (right Rindler wedge (RRW)) during the isochoric stages. But we know that accelerated frame can be constructed in the left Rindler wedge (LRW) as well. Also since the frames, accelerating in these two opposite wedges, can harvest  quantum entanglement \cite{Koga:2019fqh,Barman:2021wk}, it would be interesting to incorporate this situation in constructing EUQOE. Such generalization was absent in \cite{PhysRevD.104.L041701}. Therefore, the question of the possibility of making a EUQOE when one detector is accelerating in the right and another in the left Rindler wedge remains an option to investigate.
\end{itemize}

Here we aim to present a complete picture of the EUQOE. All the gaps, mentioned above, will be addressed here.
We scan the parameter space for which making of a EUQOE is possible. Here for simplicity, we always fix the ratio of the acceleration of the detectors. We observed that the necessary conditions are not getting satisfied when both detectors are in the right Rindler wedge with any initially entangled state of the detectors. 
 We also observe that either maximally symmetric or anti-symmetric both do not lead to any fruitful cycle when the qubits are accelerating in two opposite Rindler wedges. The construction of Otto engine is prohibited by the dissatisfaction of the required conditions. However, making of EUQOE is possible with initially non-maximally entangled states for detectors are in anti-parallel motion. However, the efficiency get suppressed for all the parameter values. The efficiency of the cycle depends on the acceleration of the first detector, ratio of the detectors $A$'s acceleration to detector $B$'s acceleration and the energy gap of the composite system. Therefore we can regulate the efficiency of the cycle by tuning these parameters.

The organization of the paper is as follows.
In Section \ref{sec:2} we begin with a brief description on our model. This section also describes the stages of our EUQOE and  provides the expression of the efficiency of the cycle in terms of elements of time evolved density matrix of the two detectors' system. In the next Section \ref{sec:calc} we provide the expressions of elements of the density matrix for the detectors are accelerating in same and opposite Rindler wedges. Thereafter, we analyzed for the possibility of making an EUQOE in various possible parameter spaces. Finally, we conclude this article in Section \ref{sec:conclu} with discussion of our results. Five appendices are also provided at the end in order to show the explicit steps to evaluate several important relations and results.

\section{Set Up}\label{sec:2}

We construct a similar system as considered in \cite{PhysRevD.104.L041701}. It consists of two identical qubits (chosen as two level Unruh De-Witt detectors \cite{book:Birrell}),  $A$ and $B$, having energy levels with energy $g_{j}=-\omega/2$ for the ground state and $e_{j}=\omega/2$ for the exited state. The corresponding eigenkets are $|g_j\rangle$ and $|e_j\rangle$ (where $j=A, B$), respectively. The free Hamiltonian of the detectors can be chosen as \cite{Arias:2016tj, PhysRev.93.99} 
\begin{eqnarray}\label{freeH}
H_0=\frac{\omega(\tau)}{2}\Big(\frac{d\tau_{A}}{d\tau} ~S_{A}^{z}\otimes\mathds{1}_{B}+\frac{d\tau_B}{d\tau}~\mathds{1}_{A}\otimes {S}_{B}^{z}\Big)~,
\end{eqnarray}
where $S_{j}^{z}=(|{e_j}\rangle\langle{e_j}|-|{g_j}\rangle\langle{g_j}|), ~~\omega$ is the energy gap of the detectors and $\tau_{A}, ~\tau_{B}$ are the proper times of the respective detectors.  In the above $\tau$ is the proper time of our observer which we choose as the detector $A$ and hence $\tau=\tau_A$. Same Hamiltonian has been chosen earlier in various investigations \cite{Arias:2016tj, PhysRev.93.99}. The energy eigenstates and the corresponding eigenvalues can be obtained for the composite system as 
\begin{eqnarray}
&&E_e=\omega,~~~|e\rangle=|e_{A}\rangle|e_{B}\rangle;\nonumber\\
&&E_s=0,~~~|s\rangle=\frac{1}{\sqrt{2}}(|e_{A}\rangle|g_{B}\rangle+|g_{A}\rangle|e_{B}\rangle);\nonumber\\
&&E_a=0,~~~|a\rangle=\frac{1}{\sqrt{2}}(|e_{A}\rangle|g_{B}\rangle-|g_{A}\rangle|e_{B}\rangle);\nonumber\\
&&E_g=-\omega,~~~|g\rangle=|g_{A}\rangle|g_{B}\rangle;
\end{eqnarray}
Here, as discussed in \cite{ PhysRevD.104.L041701, Barman:2021vj} transition from $|g\ra\to|e\ra$ or $|e\ra\to|g\ra$ is not possible for the following type of interaction between the background real scalar field and monopole detectors  \cite{PhysRevD.97.125011, PhysRevA.97.062338, Koga:2019fqh, Barman:2021vj, Barman:2021wk}:
\begin{eqnarray}
H_{int}=\sum_{j=A,B}c_{j} \chi_{j}(\tau)m_{j}(\tau_j)\phi(x_j(\tau_j))\frac{d\tau_j}{d\tau}~.
\end{eqnarray}
In the above $c_j$ is coupling constant of interaction. We choose $c_{A}=c_{B}=c$, as the detectors in our case are identical. For our present purpose we will use this type of interaction.
The monopole operator of $j$-th detector, $m_j$ at the initial time is defined by
\begin{eqnarray}
m_j (0)=|{e_j}\rangle\langle{g_j}|+|{g_j}\rangle\langle{e_j}|~.
\end{eqnarray}
It has been shown in \cite{Barman:2021vj} that the transition probability form  $|a\ra\to|e\ra$ and  $|a\ra\to|g\ra$ are same, as they have same energy spacing. Similarly the probabilities for the transitions  $|s\ra\to|e\ra$ and  $|s\ra\to|g\ra$ are also same. Notice that $|s\ra$ and $|a\ra$ are two Bell states appearing in the composite system which are maximally entangled ones. Moreover, for a general initial state, $|D\ra=b_{1}|e_{A}g_{B}\ra+b_{2}|g_{A}e_{B}\ra$ with $E_{D}=0$,  transition probabilities 
$|D\ra\to|e\ra$ and $|D\ra\to|g\ra$ are also same. With the above information here we can try to construct the Otto cycle between $|s\ra$ and $|e\ra$ or $|g\ra$, $|a\ra$ and $|e\ra$ or $|g\ra$ (also between $|D\ra$ and $|e\ra$ or $|g\ra$), for maximally and non-maximally (depending on the values of $b_1$ and $b_2$ with $b_1^2+b_2^2=1$) entangled states. We will check all of these possibilities here.

We choose the detector $A$ as our primary observer. Therefore we set $\tau=\tau_{A}$ and define $d{\tau_B}/d\tau=\alpha$ which is constant during detectors' motions with uniform velocities as well as uniform accelerations (see \cite{RODRIGUEZCAMARGO2018266} for relation between $\tau_A$ band $\tau_B$; an additional discussion has been given in Appendix \ref{A} as well). Later for notational convenience, we will add a subscript with $\alpha$: $v$ for constant velocity and $a_{k}~(k=H, ~C)$ for uniform acceleration in heating  process ($H$) and cooling process ($C$). We have calculated $\alpha$'s for different stages of the cycle in Appendix \ref{A}. The free Hamiltonian (\ref{freeH}) can be represented as
\begin{equation}\label{H0-hamiltonian}
H_0=\frac{\omega}{2}\begin{pmatrix}
(1+\alpha)&0&0&0\\
0&(1-\alpha)&0&0\\
0&0&(-1+\alpha)&0\\
0&0&0&(-1-\alpha)
\end{pmatrix}=\omega{h_\alpha}~,
\end{equation}
in the basis $\{|e_{A}e_{B}\ra$, $|e_{A}g_{B}\ra$, $|g_{A}e_{B}\ra$ $|g_{A}g_{B}\ra\}$.

\subsection{Stages of cycle}\label{sec:2.1}
The stages of our EUQOE are already being discussed in \cite{PhysRevD.104.L041701} which are along the line of those given in  \cite{Arias:2018wc,Gray:2018uw} for UQOE. For convenience, here again let us briefly describe the stages of the cycle. 
\begin{itemize}
\item
\textit{Adiabatic expansion:} In step I, the detectors $A$ and $B$ travel at velocities $-v_A$ and $-v_B~(v_{B}$ for anti-parallel motion) for time intervals $\Delta\tau_A^1$ and   $\Delta\tau_B^1$, respectively (here superscripts for the stage number and subscripts for detectors). In this process work is done on the composite system, so that the energy gap increased from the initial gap $\omega_1$ to $\omega_2$. However, the population density of the states remain unchanged.

\item
\textit{Heat absorption through isochoric process:} After step I, the detectors started accelerating with proper accelerations $a_{A_{H}}$ and $a_{B_{H}}$, respectively. Due to acceleration, the detectors perceives the Minkowski vacuum as thermal bath as a result of interaction between the detectors and the background quantum fields. The system then absorbs the heat. The detectors accelerate for the time intervals $\Delta\tau_A^2$ and $\Delta\tau_B^2$, when their velocities changes from $-v_A$ to $v_A$ and $-v_B$ to $v_B$ (or $v_{B}$ to $-v_{B}$), respectively.
 As the density matrix evolve under the interaction, it goes from initial state $\rho_0$ to $\rho_0+\delta{\rho^H}$. The expressions of $\delta\rho^{H}$ for general $\rho_0 = |D\ra\la D|$ is explicitly calculated in Appendix \ref{B} (see final expression in (\ref{deltarho})). 

\item
\textit{Adiabatic Compression:}
The interaction with the background field in this stage is turned off and the population density of the states remain unchanged. The detectors started moving with constant velocity, $v_A$ and $v_B$ (or $-v_{B}$) for the time intervals of $\Delta\tau_{A}^{3}$ and $\Delta\tau_{B}^{3}$, respectively. The energy spacing of the system is allowed to decrease from $\omega_2$ to $\omega_1$.
\item
\textit{Heat ejection through isochoric process:} In the final stage, the detectors started decelerating with proper uniform decelerations $a_{A_{C}}$ and $a_{B_{C}}$, respectively. As a result, their velocities changes from $v_A$ to $-v_A$ and $v_B$ to $-v_B$  (or $-v_{B}$ to $v_{B}$), respectively. The temperature of background quantum vacuum is taken to be lower by considering $a_{j_{C}}<a_{j_{H}}$. This can be done by taking longer interaction time intervals $\Delta\tau_A^4$ and $\Delta\tau_B^4$ compared to those in stage II. Then the system transfers heat to the environment.   The density of state goes from $\rho_0+\delta\rho^H$ to $\rho_0+\delta\rho^H+\delta\rho^C$. To maintain the cyclicity of the engine, we impose a constraint, $\delta\rho^H+\delta\rho^C=0$.
\end{itemize}

\subsection{Efficiency}\label{sec:2.2}
For the cycle to be efficient to work, one has to insure the positivity of work done and heat absorption by the engine. This is being confirmed by the following impositions \cite{PhysRevD.104.L041701}:
\begin{equation}\label{Tr+}
 \text{Tr}(\delta\rho^H h_{\alpha_k})>0~;~~~~(k=v, a_H, a_C)~.
 \end{equation}
 This condition must be satisfied by the EUQOE. The conservation of energy requires \cite{PhysRevD.104.L041701}
\begin{eqnarray}\label{E-conserv}
\omega_2\text{Tr}(\delta\rho^Hh_{\alpha_{a_H}})-\omega_1\text{Tr}(\delta\rho^Hh_{\alpha_{a_C}})
=(\omega_2-\omega_1)\text{Tr}(\delta\rho^H h_{\alpha_v})~.
\end{eqnarray}
The efficiency of our EUQOE can be obtained as \cite{PhysRevD.104.L041701}
 \begin{eqnarray}\label{eta-E}
 \eta_E=\Big(1-\frac{\omega_1}{\omega_2}\Big)\frac{\text{Tr}(\delta\rho^Hh_{\alpha_v})}{\text{Tr}(\delta\rho^Hh_{\alpha_{a_{H}}})}=\eta_0\frac{\text{Tr}(\delta\rho^Hh_{\alpha_v})}{\text{Tr}(\delta\rho^Hh_{\alpha_{a_{H}}})}~.
 \end{eqnarray}
 Simultaneous satisfaction of the conditions (\ref{Tr+}), (\ref{E-conserv}) and (\ref{eta-E}) guarantees that $\eta_E<1$ (see \cite{PhysRevD.104.L041701} for all these details). Note that the factor, appearing on the right hand side with $\eta_0$, contains the variation of the density matrix ($\delta\rho^{H}$) only for the second stage, i.e. in the isochoric heating process. This factor is always greater than zero due to the positivity of the work done, heat absorption and heat rejection by the engine, and so $(\eta_E/\eta_0)>0$. Now in order to make EUQOE more efficient than the UQOE we must have $(\eta_E/\eta_0)>1$; i.e. $\frac{\text{Tr}(\delta\rho^Hh_{\alpha_v})}{\text{Tr}(\delta\rho^Hh_{\alpha_{a_{H}}})}>1$, under the constraint $\eta_0<1$ and $\eta_E<1$. In this case, the efficiency ($\eta_E$) of the engine will be increased due to entanglement phenomenon. However, the efficiency ($\eta_{E}$) will always be less than one. We mention that the expression for efficiency (\ref{eta-E}) reduced to a very simple form in \cite{PhysRevD.104.L041701} for the initial state as $|a\ra$ and $|s\ra$. We found that such is due to the calculation of traces in (\ref{eta-E}) without proper time ordering in the perturbative expansion. Here we will improve this analysis by incorporating the time ordering prescription.

The explicit expressions of the trace quantities in (\ref{eta-E}), incorporating the time ordering, have been calculated in Appendix \ref{C} through perturbation technique, described in Appendix \ref{B}. The first relevant contribution is coming in the second order.  Considering till this order we obtain $\text{Tr}(\delta\rho^H h_{\alpha})$ as (see Eq. (\ref{EffAppenC}))
\begin{eqnarray}\label{traceFinal}
\text{Tr}(\delta\rho^H h_{\alpha})=\{b_{2}^{2}\mathcal{P}_A(\omega)-b_{1}^{2}\mathcal{P}_A(-\omega))+b_{1}b_{2}\Delta\mathcal{P}_{AB}\}+\alpha\{(b_{1}^{2}\mathcal{P}_{B}(\omega)-b_{2}^{2}\mathcal{P}_{B}(-\omega))+b_{1}b_{2}\Delta\mathcal{P}_{AB}\}~.
\end{eqnarray}
 The expressions of the relevant quantities are given by
 \begin{equation}\label{Pi-Def}
 \mathcal{P}_j(\omega_{2})=c^2\int \int d\tau_{j}d\tau_{j}^{\p} \chi_{j}(\tau_{j})  \chi_{j}(\tau_{j}^{\p}) G^{+}(\bar{x}_j^{\p},\bar{x}_j)e^{ i\omega_{2}(\tau_j-\tau_j^{\p})};~~(j=A,B),
 \end{equation}
 and
 \begin{eqnarray}\label{deltaPij}
\Delta\mathcal{P}_{AB}= \mathcal{P}_{AB}(\omega_{2},-\omega_{2})-\mathcal{P}_{AB}(-\omega_{2},\omega_{2})~,
 \end{eqnarray}
 where
  \begin{equation}\label{PAB-Def}
\mathcal{P}_{AB}( \omega_{2},- \omega_{2})=c^2\int\int{d\tau_A}{d\tau_B^{\p}} \chi_{A}(\tau_{A})  \chi_{B}(\tau_{B}^{\p})e^{i\omega_{2}(\tau_A- \tau_B^{\p})}G^{+}(\bar{x}_B^{\p},\bar{x}_A),~
\end{equation}  
with $ \mathcal{P}_j(-\omega_{2})$ and $ \mathcal{P}_{AB}(-\omega_{2},\omega_{2})$ are obtained by replacing $\omega_{2}\to-\omega_{2}$ in the {\it integrands} of (\ref{Pi-Def}) and (\ref{PAB-Def}). For maximally entangled initial states, $b_{1}=\pm b_{2}=1/\sqrt{2}$, (\ref{traceFinal}) get further simplified by introducing $\Delta\mathcal{P}_{j}=\mathcal{P}_{j}(\omega)-\mathcal{P}_{j}(-\omega)$. In the above quantities, $G^{+}(\bar{x}_j^{\p},\bar{x}_j)$ is the positive frequency Wightman function. These integrations are calculated during the stage II of the cycle, where $\omega_{2}$ is energy gap of the engine. We do not need to evaluate the same for stage IV of the cycle due to the cyclicity condition.

\section{Calculation of efficiency}\label{sec:calc}
In order to find the explicit value of $\eta_E$ in Eq. (\ref{eta-E}) one needs to evaluate the integrations (\ref{Pi-Def}) and (\ref{PAB-Def}).
Note that in the second stage, the interaction time is $\Delta\tau_{j}^{2}$ for the $j$-th detector ($j=A,B$). This time is the time required for changing the velocity from $-v_{j}$ to $v_{j}$ and can be determined in the following way. 

For RRW, the relations between the Rindler proper time and the Minkowski coordinates are
\begin{equation}
t_j = \frac{1}{a_{j_{k}}}\sinh(a_{j_{k}}\tau_j); \,\,\,\,\ x_j = \frac{1}{a_{j_{k}}}\cosh(a_{j_{k}}\tau_j)~, 
\label{B2}
\end{equation}
where index $k$ can be $H$ ($C$) denoting heating process (cooling process). Now velocity is given by $v_j = (dx_j/dt_j) = (dx_j/d\tau_j)/(dt_j/d\tau_j)=\tanh(a_{j_{k}}\tau_{j})$ and this determines the time interval as $(2/a_{j_{k}})\tanh^{-1}(v_{j})$ when the detector is accelerating on the RRW. Here we define $\mathcal{T}_{j_{k}}/2=(1/a_{j_{k}})\tanh^{-1}(v_{j})$ for that, and then the integration limits will be from $-\mathcal{T}_{j_{k}}/2$ to $\mathcal{T}_{j_{k}}/2$. 

 On the other hand, if the detector is accelerating in the LRW, then the coordinate relations are
\begin{equation}
t_j = \frac{1}{a_{j_{k}}}\sinh(a_{j_{k}}\tau_j); \,\,\,\,\ x_j =- \frac{1}{a_{j_{k}}}\cosh(a_{j_{k}}\tau_j)~.
\label{B3}
\end{equation}
Now the velocities are given by $v_{j}=-\tanh(a_{j_{k}}\tau_{j})$ and hence the interaction time interval is given by $(2/a_{j_{k}})\tanh^{-1}(v_{j})$. So the integration limits are from $-\mathcal{T}_{j_{k}}/2$ to $\mathcal{T}_{j_{k}}/2$. 

However, evaluation of the integrations, appearing in $ \Delta\mathcal{P}_j$ and $\Delta\mathcal{P}_{AB}$, for finite integration limits can not be done analytically. But a suitable choice of a switching function $ \chi_j(\tau_j)$, which vanishes rapidly beyond the finite integration limits, can allow to extend the interaction time limits from \( -\infty \) to \( +\infty \). In that case there is a possibility of obtaining an analytical expression. Following the argument of \cite{Arias:2018wc} and as suggested there (later used in \cite{Gray:2018uw} as well), we choose a Lorentzian like compact symmetric switching function:
\begin{equation} \label{switch}
 \chi_j(\tau_j)=\frac{(\mathcal{T}_{j_{k}}/2)^2}{\tau_j^2+(\mathcal{T}_{j_{k}}/2)^2}~.
\end{equation}
This function is non-vanishing for $-\mathcal{T}_{j_{k}}/2<\tau_j<\mathcal{T}_{j_{k}}/2$, and approximately zero beyond this domain. We could choose a Gaussian like switching function \cite{Arias:2018wc}, however that  will not help us to extend the integration limit from  \( \pm\mathcal{T}_{j_{k}}/2 \) to \( \pm\infty \). Because at \( \tau_{j}\to i \times\infty \), the gaussian exponent ($\exp({-\tau_{j}^{2}/\mathcal{T}_{j_{k}}^{2}})$) diverges and that fails Jordon's lemma for contour integration.

\subsection{Both detectors are moving parallely}\label{sec:calc:parallel}
The two qubits (the detectors here) are accelerating with constant accelerations ($a_{H_{A}}$ and $a_{H_{B}}$) in the RRW during the second stage of the cycle. In this case their trajectories are given by (\ref{B2}) and the relation between the proper times between them turns out to be (from (\ref{tAtB}), see also \cite{RODRIGUEZCAMARGO2018266})
\begin{equation}
\tau_{B}=\alpha_{a_{H}}\tau_{A}~,
\label{B5}
\end{equation}
where $\alpha_{a_{H}}=a_{A_{H}}/a_{B_{H}} $ (since this is the heating stage). Using the switching function (\ref{switch}) and extending the limits of integration from $-\infty$ to $+\infty$ in (\ref{Pi-Def}) and (\ref{PAB-Def}), one case perform the integrations analytically. In the evaluation of these, since here our designated observer's frame is qubit $A$, all the integration variables will be expressed in terms of $\tau_A$ by using (\ref{B5}).

The explicit calculations for $\mathcal{P}_{A}$ and $\mathcal{P}_{B}$ are presented in Appendix \ref{D}. The values of these quantities are given by 
 \begin{eqnarray}\label{PAexpression}
&&\mathcal{P}_{A}(\omega_{2})=\frac{(a_{A_{H}}\mathcal{T}_{A_{H}}/2)^2~e^{-\mathcal{T}_{A_{H}} \omega_{2} }}{16\sin^{2}(a_{A_{H}}\mathcal{T}_{A_{H}}/2)}\no\\&&~~~~~~~~~~~+
\frac{a_{A_{H}}^2 \mathcal{T}_{A_{H}}^2 ~e^{-\frac{2 \pi  \omega_{2} }{a_{A_{H}}}} }{64 \pi ^2}
\left(\Phi \left(e^{-\frac{2 \pi  \omega_{2} }{a_{A_{H}}}},2,1+\frac{a_{A_{H}} \mathcal{T}_{A_{H}}}{2 \pi }\right)-\Phi \left(e^{-\frac{2 \pi  \omega_{2} }{a_{A_{H}}}},2,1-\frac{a_{A_{H}} \mathcal{T}_{A_{H}}}{2 \pi }\right)\right)\no\\&&~~~~~~~~+
\frac{a_{A_{H}} \mathcal{T}_{A_{H}}^2 \omega_{2}~  e^{-\frac{2 \pi  \omega_{2} }{a_{A_{H}}}}}{32 \pi }
\left(\Phi \left(e^{-\frac{2 \pi  \omega_{2} }{a_{A_{H}}}},1,1+\frac{a_{A_{H}} \mathcal{T}_{A_{H}}}{2 \pi }\right)-\Phi \left(e^{-\frac{2 \pi  \omega_{2} }{a_{A_{H}}}},1,1-\frac{a_{A_{H}} \mathcal{T}_{A_{H}}}{2 \pi }\right)\right),
\end{eqnarray}
\begin{eqnarray}\label{PAnexpression}
&&\mathcal{P}_{A}(-\omega_{2})=\frac{\omega_{2}\mathcal{T}_{A_{H}}}{8}+\mathcal{P}_{A}(\omega_{2});~~~~~~\Delta\mathcal{P}_{A}(\omega_{2})=\mathcal{P}_{A}(\omega_{2})-\mathcal{P}_{A}(-\omega_{2})
\end{eqnarray}
 \begin{eqnarray}\label{PBexpression}
&&\mathcal{P}_{B}(\omega_{2})=\frac{(a_{A_{H}}\mathcal{T}_{A_{H}}/2)^2~e^{-\mathcal{T}_{A_{H}}\alpha_{a_{H}} \omega_{2} }}{16\sin^{2}(a_{A_{H}}\mathcal{T}_{A_{H}}/2)}\no\\&&~~~~~~~~~+
\frac{a_{A_{H}}^2 \mathcal{T}_{A_{H}}^2 ~e^{-\frac{2 \pi  \omega_{2} }{a_{B_{H}}}} }{64 \pi ^2}
\left(\Phi \left(e^{-\frac{2 \pi  \omega_{2} }{a_{B_{H}}}},2,1+\frac{a_{A_{H}} \mathcal{T}_{A_{H}}}{2 \pi }\right)-\Phi \left(e^{-\frac{2 \pi  \omega_{2} }{a_{B_{H}}}},2,1-\frac{a_{A_{H}} \mathcal{T}_{A_{H}}}{2 \pi }\right)\right)\no\\&&~~~~~+
\frac{a_{A_{H}} \mathcal{T}_{A_{H}}^2\alpha_{a_{H}} \omega_{2}  ~e^{-\frac{2 \pi  \omega_{2} }{a_{B_{H}}}}}{32 \pi }
\left(\Phi \left(e^{-\frac{2 \pi  \omega_{2} }{a_{B_{H}}}},1,1+\frac{a_{A_{H}} \mathcal{T}_{A_{H}}}{2 \pi }\right)-\Phi \left(e^{-\frac{2 \pi  \omega_{2} }{a_{B_{H}}}},1,1-\frac{a_{A_{H}} \mathcal{T}_{A_{H}}}{2 \pi }\right)\right),~~
\end{eqnarray}and
 \begin{eqnarray}\label{PBnexpression}
&&\mathcal{P}_{B}(-\omega_{2})=\frac{\mathcal{T}_{A_{H}}\omega_{2}\alpha_{a_{H}}}{8}+\mathcal{P}_{B}(\omega_{2});~~~~~~\Delta\mathcal{P}_{B}(\omega_{2})=\mathcal{P}_{B}(\omega_{2})-\mathcal{P}_{B}(-\omega_{2})
\end{eqnarray}
In the above, $\Phi$ is the transcendental Lerch-Hurwitz function, defined as
\begin{equation}
\Phi(z,s,a) = \sum_{k=0}^\infty \frac{z^k}{(k+a)^s}~.
\end{equation}

 Similarly $\Delta\mathcal{P}_{AB}$ has been calculated in Appendix \ref{E.1}. We obtain $\mathcal{P}_{AB}$ for $\alpha_{a_{H}}<1$ case as
 \begin{eqnarray}\label{deltaPABa}
&&\Delta\mathcal{P}_{AB}=-\frac{\alpha_{a_{H}} a_{B_{H}}\mathcal{T}_{A_{H}}^{3}\text{csch}(a_{A_{H}} \kappa )e^{-\frac{1}{2} (1-\alpha_{a_{H}} ) \mathcal{T}_{A_{H}} \omega_{2}}}{16 ~ \kappa~( \kappa^{2}+\mathcal{T}_{A_{H}}^{2}) }( \kappa  \sin (\alpha_{a_{H}}  \kappa  \omega_{2} )+\kappa  \sin (\kappa  \omega_{2} )+\mathcal{T}_{A_{H}} \cos (\alpha_{a_{H}}  \kappa  \omega_{2} )\no\\&&~~~~~~~~~~~~~~~~~~~~~~~~~~~~~~~~~~~~~~~~~~~~~~~~~~~~~~~~~~~~~~~~~~~~~-\mathcal{T}_{A_{H}} \cos (\kappa  \omega_{2} ))~,
 \end{eqnarray}
 while that for $\alpha_{a_{H}}>1$ case is
 \begin{eqnarray}\label{deltaPAB}
&&\Delta\mathcal{P}_{AB}= -\frac{\alpha_{a_{H}}  a_{H_{B}} \mathcal{T}_{A_{H}}^3 \text{csch}(a_{A_{H}} \kappa ) e^{-\frac{1}{2} (\alpha_{a_{H}} -1) \mathcal{T}_{A_{H}} \omega_{2} }}{16 \kappa  \left(\kappa ^2+\mathcal{T}_{A_{H}}^2\right)}
(\kappa  \sin (\alpha_{a_{H}}  \kappa  \omega_{2} )+\kappa  \sin (\kappa  \omega_{2} )-\mathcal{T}_{A_{H}} \cos (\alpha_{a_{H}}  \kappa  \omega_{2} )\no\\&&~~~~~~~~~~~~~~~~~~~~~~~~~~~~~~~~~~~~~~~~~~~~~~~~~~~~~~~~~~~~~~~~~~~~~+\mathcal{T}_{A_{H}} \cos (\kappa  \omega_{2} ))~. 
 \end{eqnarray}
 Here  $\kappa$ is defined through the relation $\cosh(a_{A_{H}}\kappa)=(a_{A_{H}}/a_{B_{H}}+a_{B_{H}}/a_{A_{H}})/2$. Here it should be pointed out that  for $\alpha_{a}=1$, both expressions of $\mathcal{P}_{AB}$ and $\Delta\mathcal{P}_{B}$, given in (\ref{deltaPABa}), (\ref{deltaPAB}) and (\ref{PBnexpression}) will be reduced to $\Delta\mathcal{P}_{A}$  given in (\ref{PAnexpression}). Thus for $\alpha_{a_{H}}=1$, values of the trace quantities corresponding to work done and heat absorbed by the system are always \textit{zero} and \textit{negative} for initial maximally anti-symmetric and maximally symmetric entangled states, respectively ( due to (\ref{traceFinal}) and (\ref{PAnexpression})). Thus for $\alpha_{a_{H}}=1$, condition (\ref{Tr+}) does not get satisfied for maximally entangled states. \\
 
 For the qualitative analysis of the above results we will need the above expressions in terms of scaled dimensionless parameters.
 If we define the dimensionless quantities as $a_{A_{H}}\mathcal{T}_{A_{H}}(=\mathpzc{A})$, $a_{B_{H}}\mathcal{T}_{A_{H}}(=\mathpzc{B}=\mathpzc{A}/\alpha_{a_{H}})$,  $\omega_{2}\mathcal{T}_{A_{H}}(=\mathpzc{W})$ and $\kappa/\mathcal{T}_{A_{H}}(=\mathpzc{K})$, one can check that  $\mathcal{P}_{A}$, $\mathcal{P}_{B}$ and $\Delta\mathcal{P}_{AB}$ are all dimensionless quantities. In terms of these one finds 
  \begin{eqnarray}\label{PAexpressionD}
&&\mathcal{P}_{A}(\mathpzc{W})=\frac{(\mathpzc{A}/2)^2e^{-\mathpzc{W} }}{16\sin^{2}(\mathpzc{A}/2)}+
\frac{\mathpzc{A}^2 e^{-\frac{2 \pi  \mathpzc{W} }{\mathpzc{A}}} }{64 \pi ^2}
\left(\Phi \left(e^{-\frac{2 \pi \mathpzc{W} }{\mathpzc{A}}},2,1+\frac{\mathpzc{A}}{2 \pi }\right)-\Phi \left(e^{-\frac{2 \pi \mathpzc{W}}{\mathpzc{A}}},2,1-\frac{\mathpzc{A}}{2 \pi }\right)\right)\no\\&&~~~~~~~~+
\frac{\mathpzc{A}\mathpzc{W}  e^{-\frac{2 \pi \mathpzc{W} }{\mathpzc{A}}}}{32 \pi }
\left(\Phi \left(e^{-\frac{2 \pi \mathpzc{W}}{\mathpzc{A}}},1,1+\frac{\mathpzc{A}}{2 \pi }\right)-\Phi \left(e^{-\frac{2 \pi\mathpzc{W} }{\mathpzc{A}}},1,1-\frac{\mathpzc{A}}{2 \pi }\right)\right),
\end{eqnarray}
\begin{eqnarray}\label{PBexpressionD}
&&\mathcal{P}_{B}(\mathpzc{W})=\frac{(\mathpzc{A}/2)^2e^{-\alpha_{a_{H}} \mathpzc{W} }}{16\sin^{2}(\mathpzc{A}/2)}+
\frac{\mathpzc{A}^2 e^{-\frac{2 \pi \mathpzc{W}\alpha_{a_{H}}}{\mathpzc{A}}} }{64 \pi ^2}
\left(\Phi \left(e^{-\frac{2 \pi\mathpzc{W}\alpha_{a_{H}}}{\mathpzc{A}}},2,1+\frac{\mathpzc{A}}{2 \pi }\right)-\Phi \left(e^{-\frac{2 \pi\mathpzc{W}\alpha_{a_{H}}}{\mathpzc{A}}},2,1-\frac{\mathpzc{A}}{2 \pi }\right)\right)\no\\&&~~~~~+
\frac{\mathpzc{A}\mathpzc{W}\alpha_{a_{H}}   e^{-\frac{2 \pi\mathpzc{W} \alpha_{a_{H}}}{\mathpzc{A}}}}{32 \pi }
\left(\Phi \left(e^{-\frac{2 \pi \mathpzc{W}\alpha_{a_{H}}}{\mathpzc{A}}},1,1+\frac{\mathpzc{A}}{2 \pi }\right)-\Phi \left(e^{-\frac{2 \pi \mathpzc{W} \alpha_{a_{H}}}{\mathpzc{A}}},1,1-\frac{\mathpzc{A}}{2 \pi }\right)\right),~~
\end{eqnarray}
 \begin{eqnarray}\label{DPA-Bnexpression}
\mathcal{P}_{B}(-\mathpzc{W})=\frac{\mathpzc{W}\alpha_{a_{H}}}{8}+\mathcal{P}_{B}(\mathpzc{W});~~~~~\mathcal{P}_{A}(-\mathpzc{W})=\frac{\mathpzc{W}}{8}+\mathcal{P}_{A}(\mathpzc{W}),
\end{eqnarray}
 \begin{eqnarray}
\Delta\mathcal{P}_{AB}=-\frac{\alpha_{a_{H}} \mathpzc{B}~\text{csch}(\mathpzc{A}\mathpzc{K} )e^{-\frac{1}{2} (1-\alpha_{a_{H}} ) \mathpzc{W}}}{16 ~\mathpzc{K}~( \mathpzc{K}^{2}+1) }( \mathpzc{K} \sin (\alpha_{a_{H}} \mathpzc{K}\mathpzc{W})+\mathpzc{K} \sin (\mathpzc{K}\mathpzc{W} )+ \cos (\alpha_{a_{H}}\mathpzc{K}\mathpzc{W} )-\cos (\mathpzc{K}\mathpzc{W}))~,
  \end{eqnarray} 
  for $\alpha_{a_{H}}<1$
 and 
 \begin{eqnarray}
\Delta\mathcal{P}_{AB}=-\frac{\alpha_{a_{H}} \mathpzc{B}~\text{csch}(\mathpzc{A}\mathpzc{K} )e^{-\frac{1}{2} (\alpha_{a_{H}} -1) \mathpzc{W}}}{16 ~\mathpzc{K}~( \mathpzc{K}^{2}+1) }( \mathpzc{K} \sin (\alpha_{a_{H}} \mathpzc{K}\mathpzc{W})+\mathpzc{K} \sin (\mathpzc{K}\mathpzc{W} )- \cos (\alpha_{a_{H}}\mathpzc{K}\mathpzc{W} )+\cos (\mathpzc{K}\mathpzc{W}))~,
  \end{eqnarray}  
  for $\alpha_{a_{H}}>1$.
 Using these quantities, the required expression  (\ref{traceFinal}) need to be evaluated to determine the efficiency. Below we will analyse the final expressions for  initial state, $b_{1}|e_{A}g_{B}\ra+b_{2}|g_{A}e_{B}\ra$ with different values of $b_1$ and $b_2$ (under the constraint $b_1^2+b_2^2=1$), separately.
 
\subsubsection{Analysis for maximally entangled symmetrical initial state}
\begin{figure}[h!]
    \centering
    \subfigure[]{\includegraphics[width=0.49\textwidth]{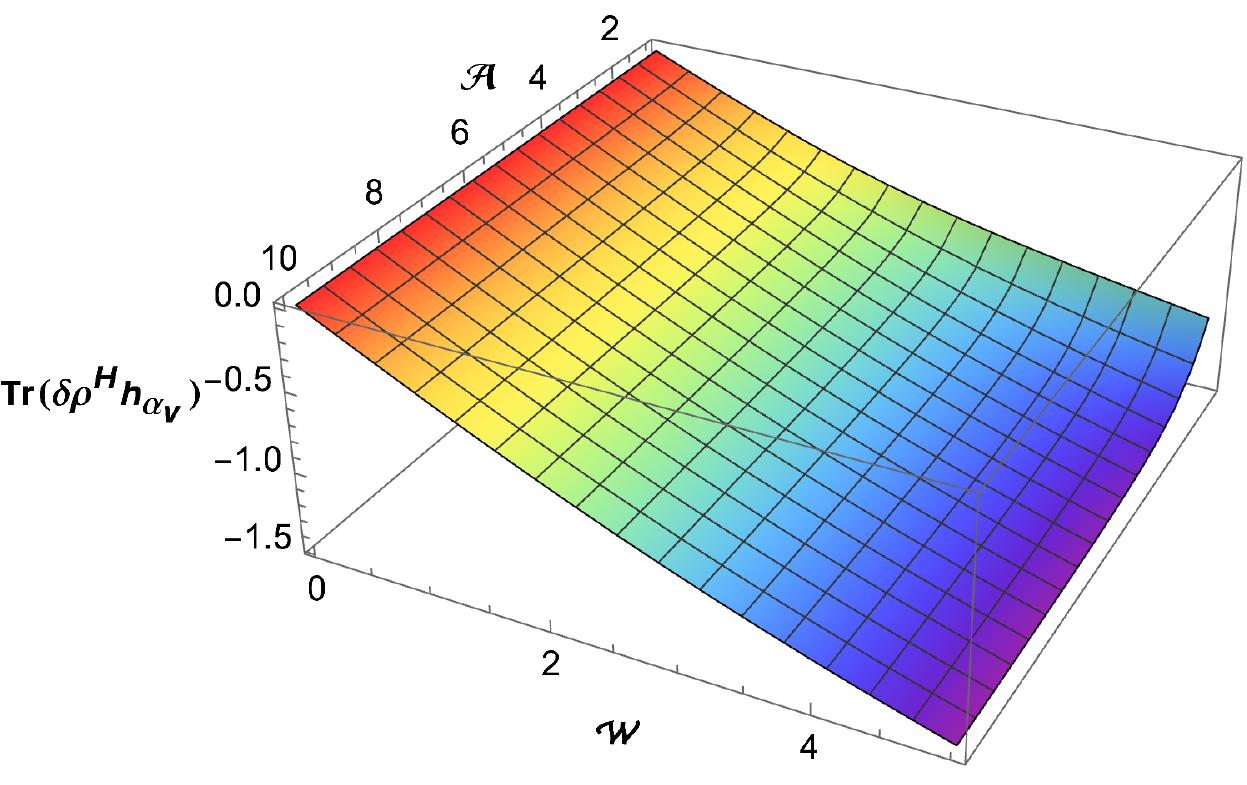}} 
    \subfigure[]{\includegraphics[width=0.49\textwidth]{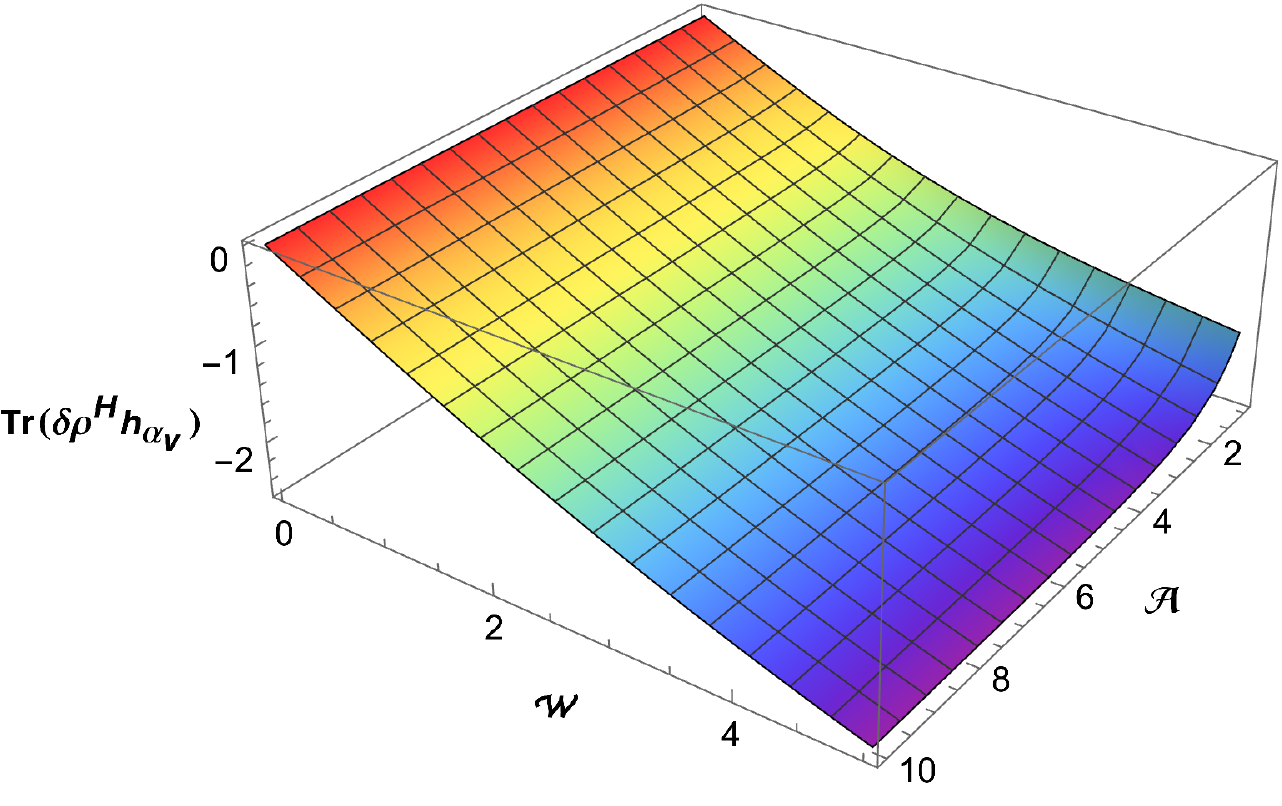}}
  
  \caption{Initial state $|s\ra$ for parallel motion: Plots (a) and (b) show trace corresponding to work done by the system, Tr($ \delta \rho^{H}h_{ \alpha_{v}}$) with respect to dimensionless acceleration of the primary detector ($\mathpzc{A}$) and dimensionless energy gap ($\mathpzc{W}$) of the composite system for $\alpha_{a_{H}}=0.5$ and $\alpha_{a_{H}}=1.5$, respectively.}
\label{fig:PMS-PMSa}
\end{figure}

Here, we have $b_{1}=b_{2}=1/\sqrt{2}$. From (\ref{PAnexpression}) and (\ref{PBnexpression}) we know that \( \Delta\mathcal{P}_{A}(\omega) \) and \(\Delta\mathcal{P}_{B}(\omega)\)  are always negative quantity. Constructing a Otto cycle requires, Tr(\(\delta \rho^{H}h_{ \alpha_{v}}  \)), Tr(\(\delta \rho^{H}h_{ \alpha_{a_{H}}} \)) and Tr(\(\delta \rho^{H}h_{ \alpha_{a_{C}}} \)) need to be positive for some values of $\mathpzc{A},~\mathpzc{B}(=\mathpzc{A}/\alpha_{a_{}}),~\mathpzc{W},~\alpha_{a_{C}}$. We find that Tr(\(\delta \rho^{H}h_{ \alpha_{v}}  \)) is always negative in the parameter space (for instance, see Fig. \ref{fig:PMS-PMSa}) for both the cases ($\alpha_{a}<1$ and $\alpha_{a}>1$). 
 This implies that condition (\ref{Tr+}) is always violated and hence no EUQOE can be made up with two initially symmetric maximally entangled detectors moving in parallel motion.

\subsubsection{Analysis for maximally entangled anti-symmetrical initial state} \label{alphaC}
\begin{figure}[h!]
    \centering
    \subfigure[]{\includegraphics[width=0.49\textwidth]{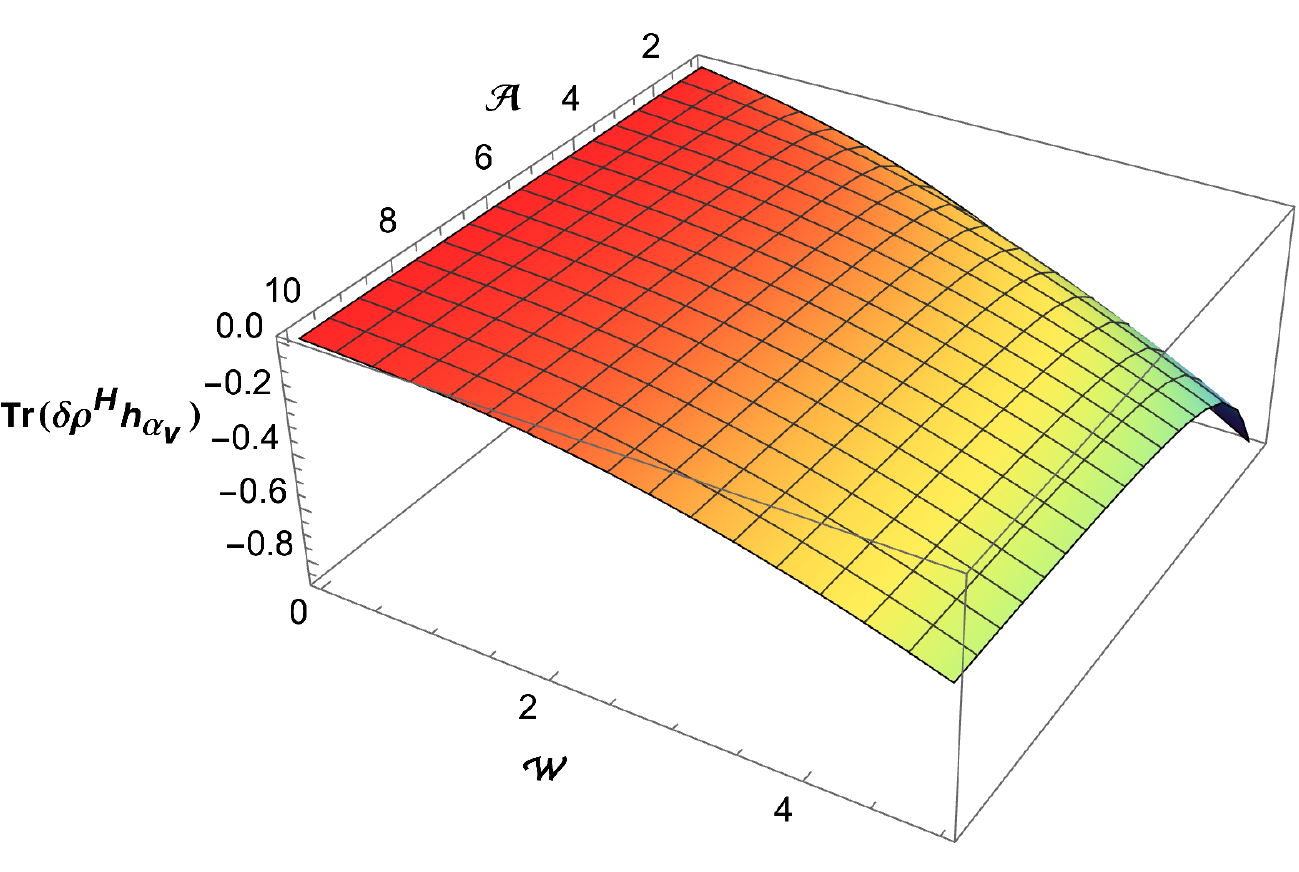}} 
    \subfigure[]{\includegraphics[width=0.49\textwidth]{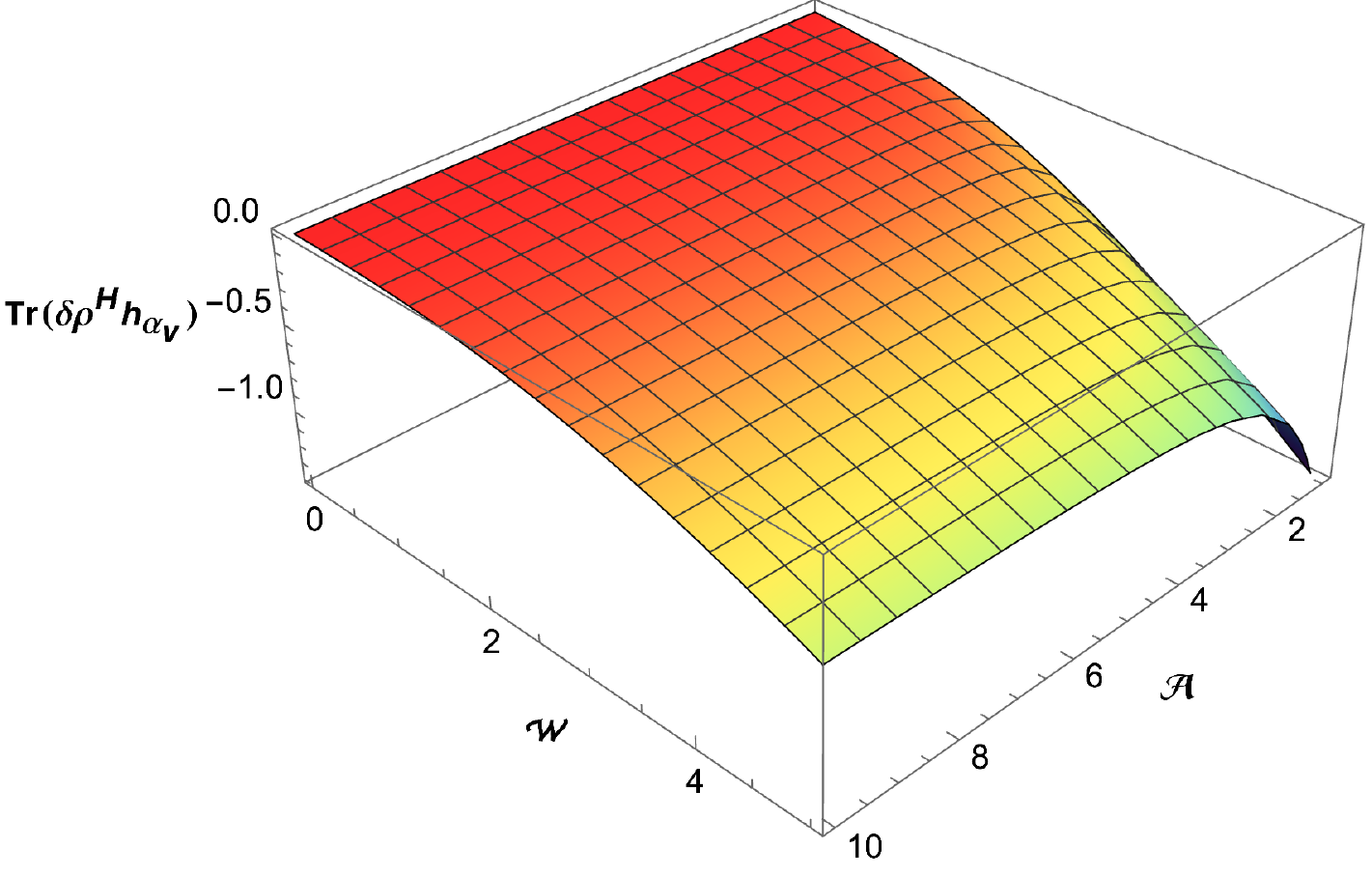}}
 
 \caption{Initial state $|a\ra$  for parallel motion: Plots (a) and (b) show trace corresponding to work done by the system, Tr($ \delta \rho^{H}h_{ \alpha_{v}}$) with respect to dimensionless acceleration of the primary detector ($\mathpzc{A}$) and dimensionless energy gap (\( \mathpzc{W} \)) of the composite system for $\alpha_{a_{H}}<1$ (here $\alpha_{a_{H}} = 0.5$) and $\alpha_{a_{H}}>1$ (here $\alpha_{a_{H}} = 1.5$), repectively.}
\label{fig:PMA-PMAa}
\end{figure}

As mentioned earlier, a successful Otto cycle can be built, provided the condition (\ref{Tr+}) is being satisfied. Therefore, satisfaction of this condition determine the parameter space for the available quantities which are involved in constructing the cycle.  In order to do that we have plotted dimensionless quantity Tr($ \delta \rho^{H}h_{ \alpha_{v}}$) in Fig. \ref{fig:PMA-PMAa}, with respect to dimensionless quantities $\mathpzc{A}$, $\mathpzc{W}$ with fixed $\alpha_{a}$ values ($\alpha_{a}=0.5$ and $\alpha_{a}=1.5$, respectively.). 
 The 3D plots are showing that there are no positive values of Tr($ \delta \rho^{H}h_{ \alpha_{v}}$) for any $\mathpzc{A}$ and $\mathpzc{W}$. Thus making of Entangled Otto cycle is not possible in this case.

\subsubsection{Analysis for non-maximally entangled initial state}
\begin{figure}[h!]
    \centering
    \subfigure[]{\includegraphics[width=0.49\textwidth]{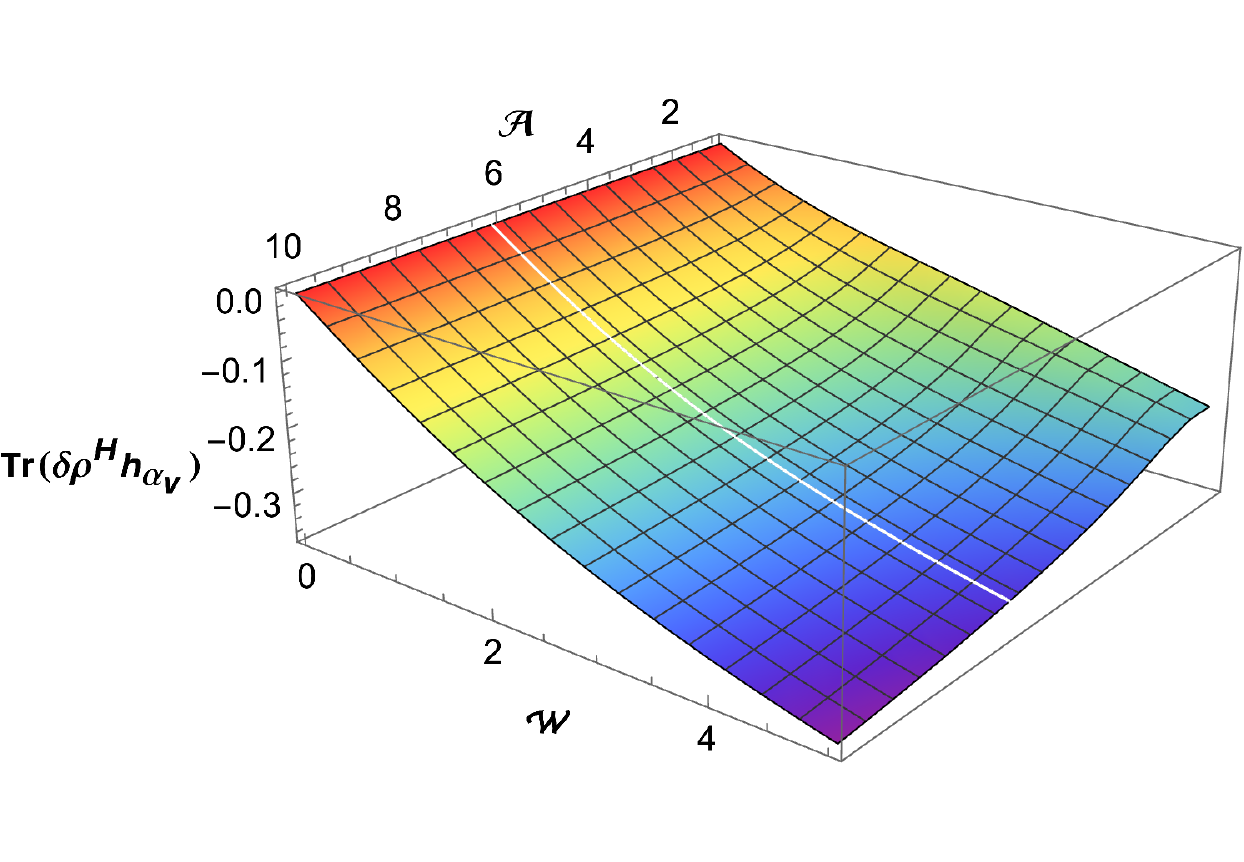}} 
     \subfigure[]{\includegraphics[width=0.45\textwidth]{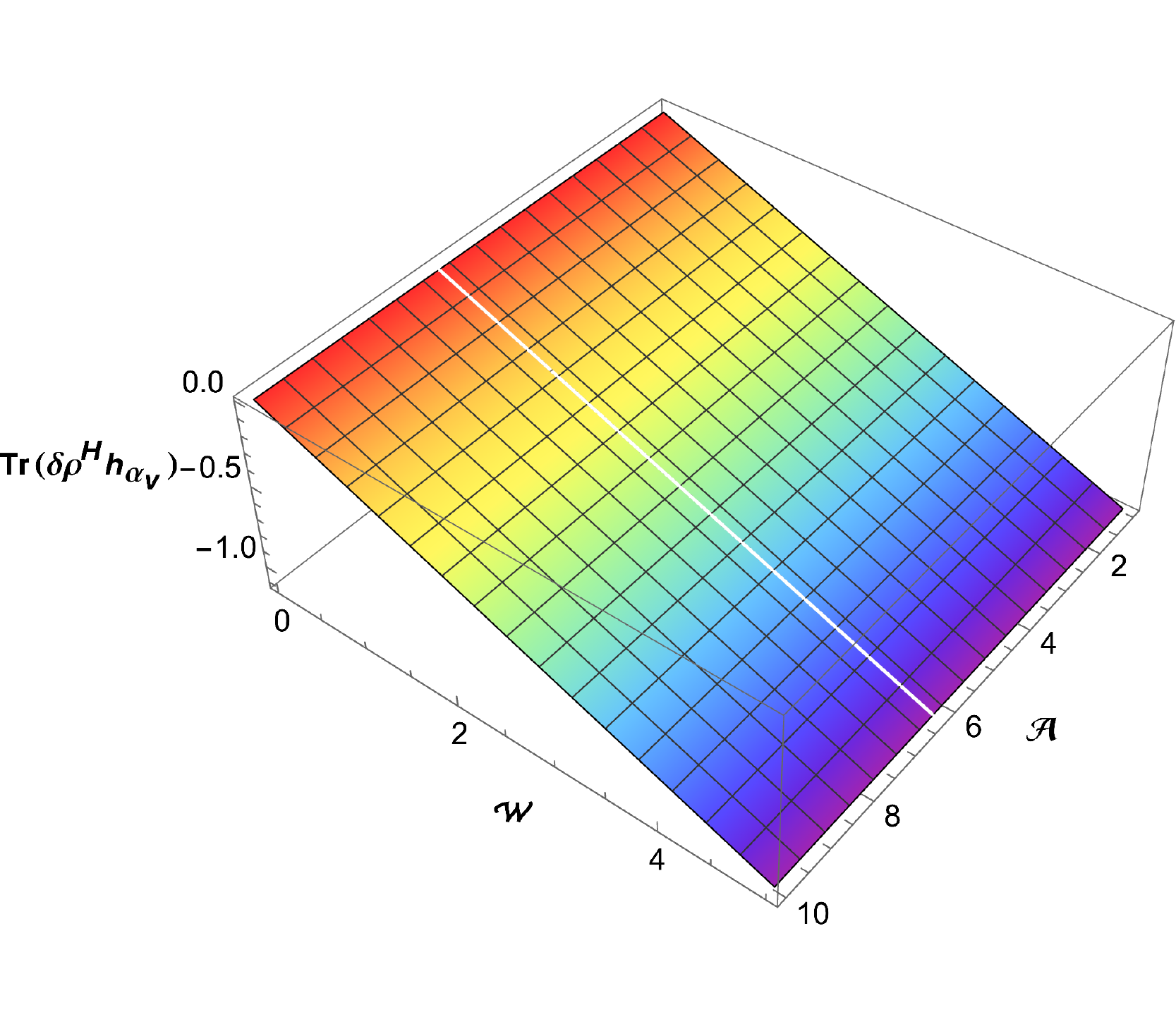}}\\
    \subfigure[]{\includegraphics[width=0.49\textwidth]{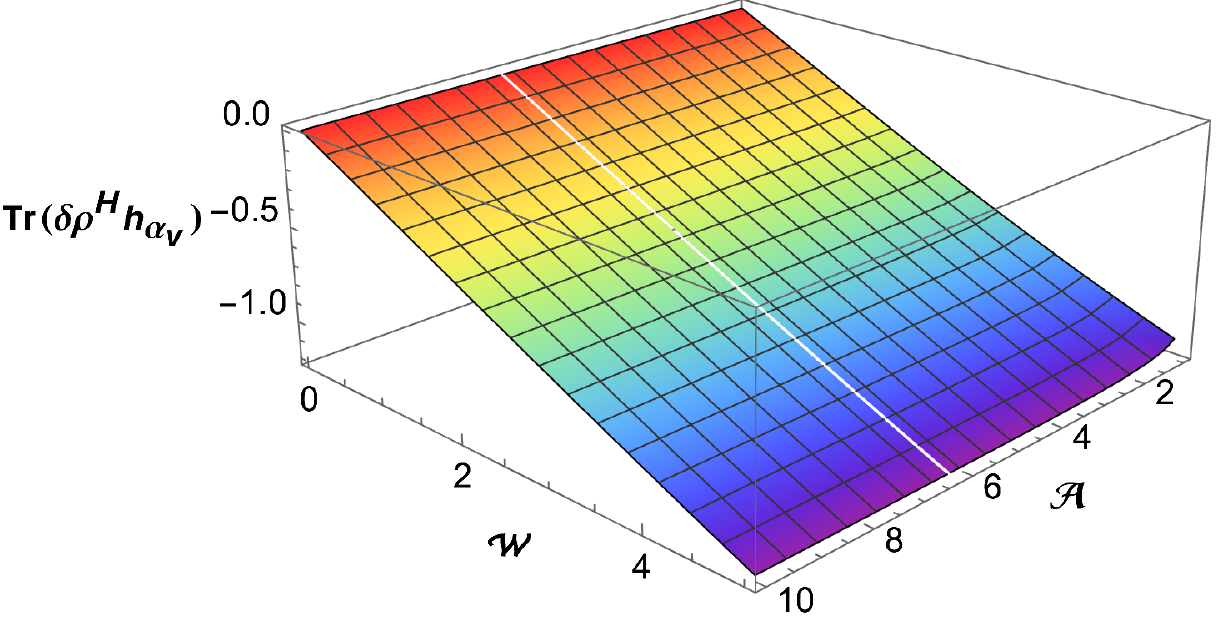}}
     \subfigure[]{\includegraphics[width=0.49\textwidth]{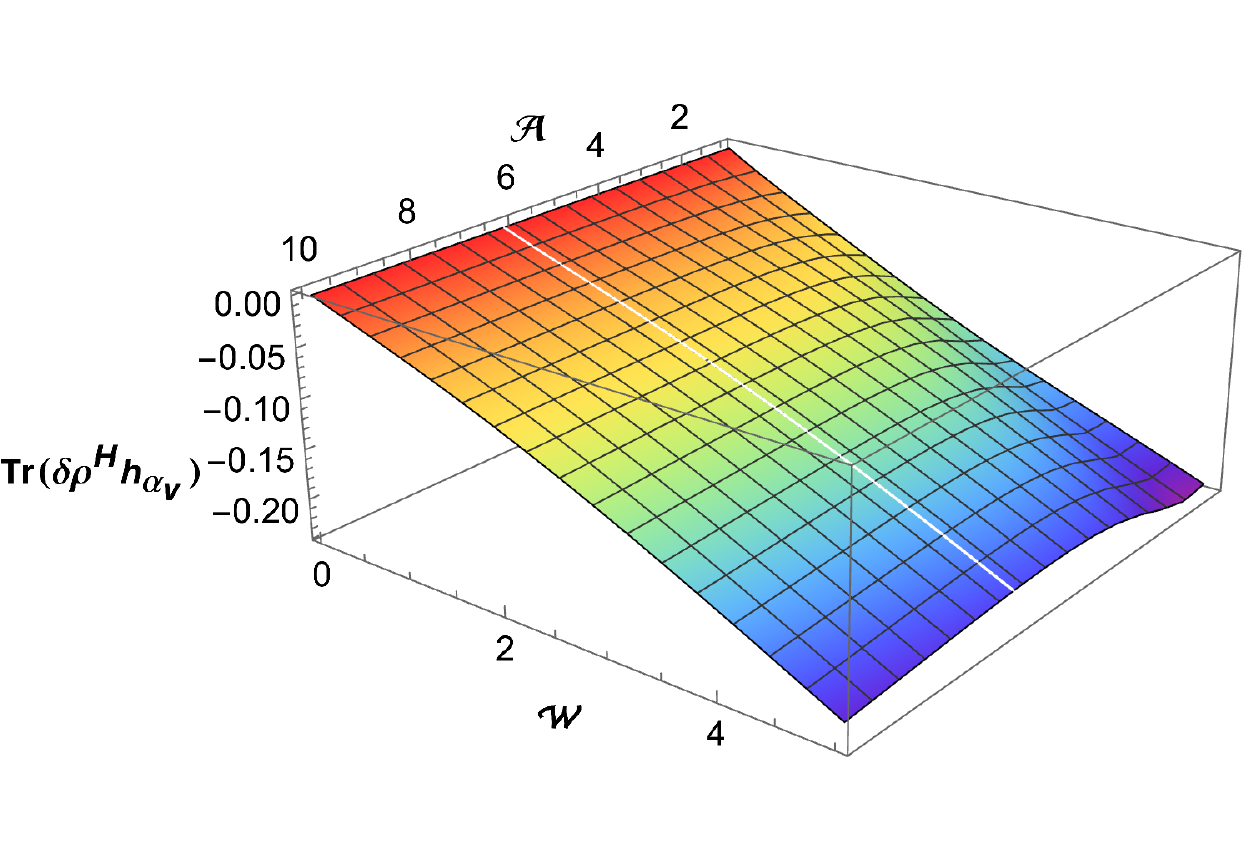}} \\
     \subfigure[]{\includegraphics[width=0.49\textwidth]{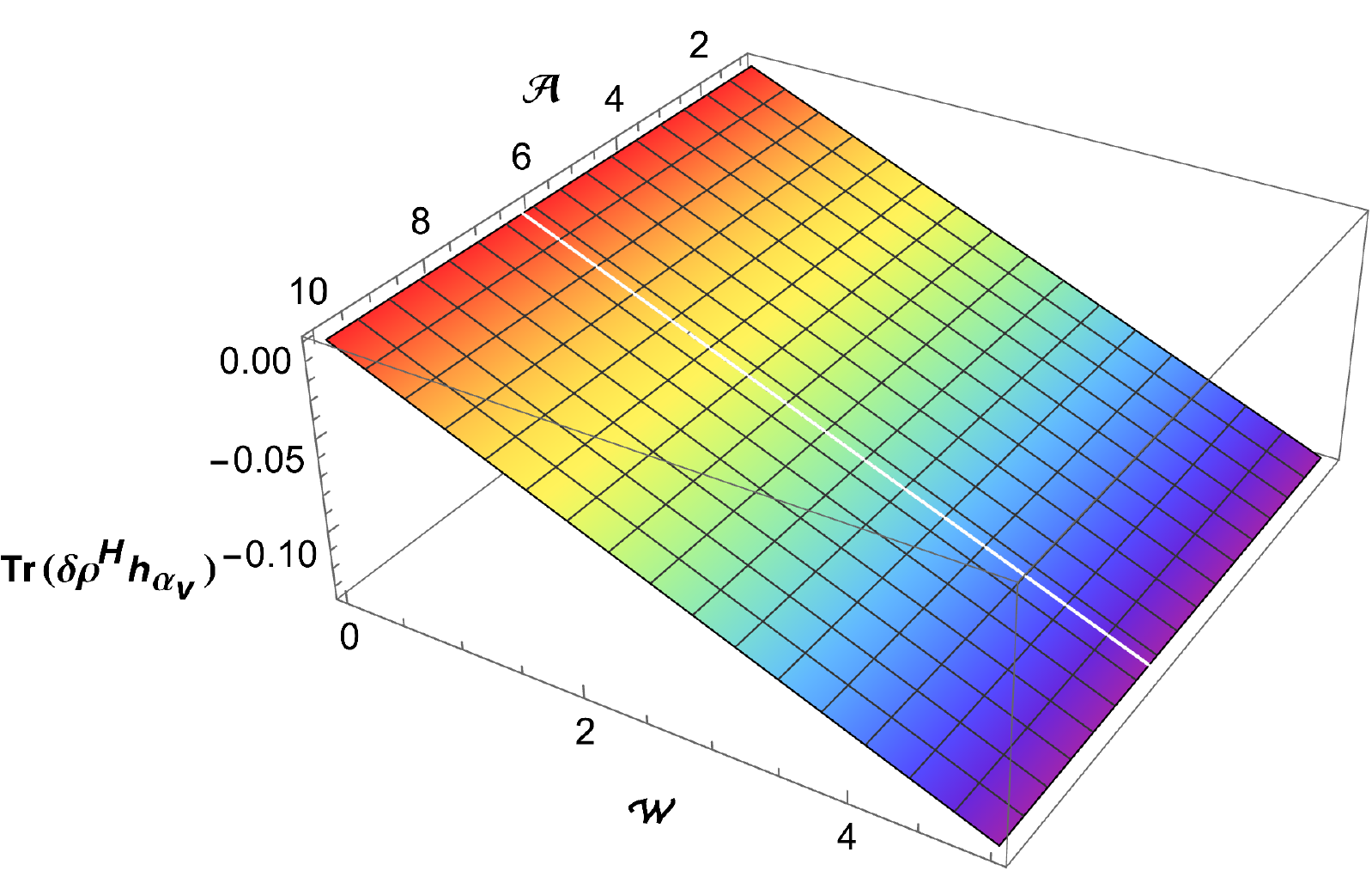}}
    \subfigure[]{\includegraphics[width=0.49\textwidth]{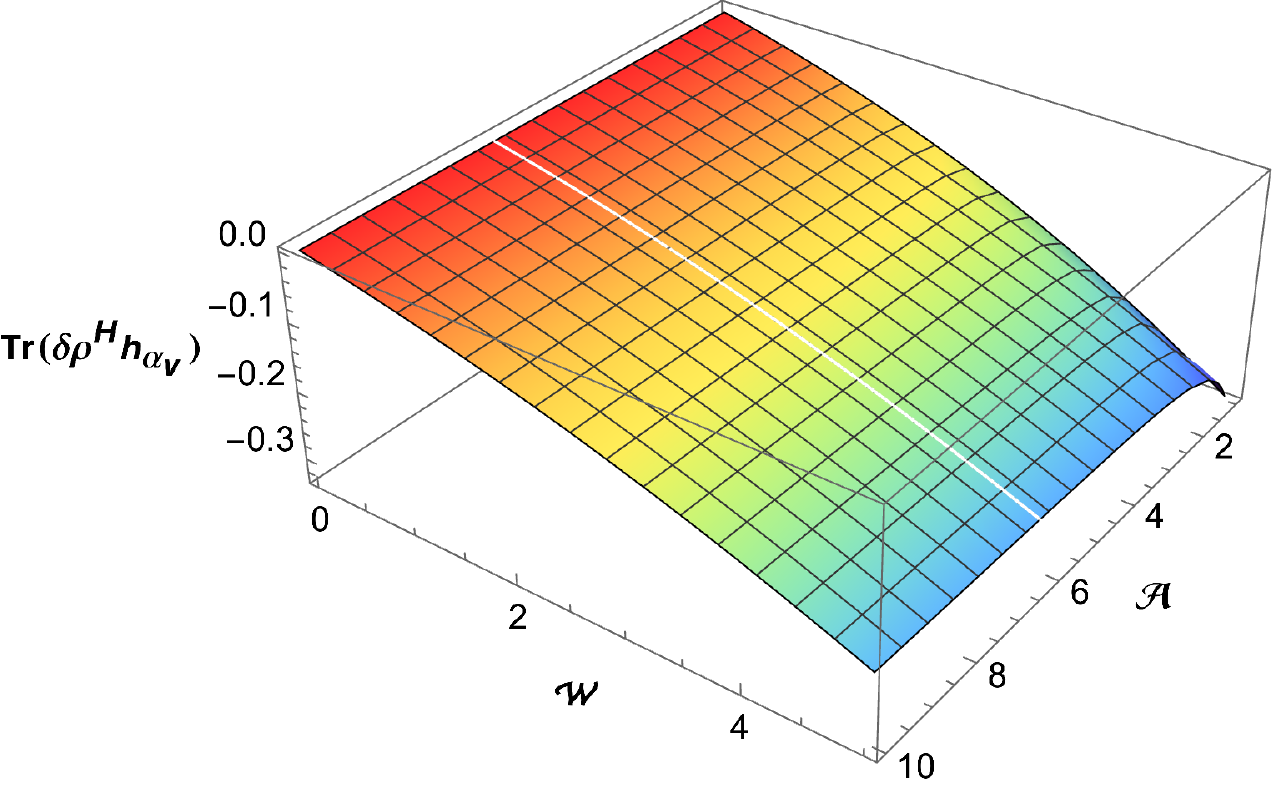}}\\
   \caption{The traces corresponding to work done by the system, Tr($\delta \rho^{H}h_{ \alpha_{v}}$) are plotted with respect to dimensionless acceleration of the primary detector ($\mathpzc{A}$) and dimensionless energy gap (\( \mathpzc{W} \)) of the composite system for detectors' non-maximally entangled initial state in parallel motion: Plots (a), (b) and (c) are showing the same with $b_{1}=b_{2}=0.9$ for $\alpha_{a_{H}}=0.2,~1.0,~1.2$, respectively.  Plots (d), (e) and (f) are showing the same with $b_{1}=-b_{2}=0.9$ for $\alpha_{a_{H}}=0.2,~1.0,~1.2$, respectively. }
   \label{fig:PNMS-PNMSa-PNMA-PNMAa}
\end{figure}

The initial state for $b_{2}\neq1/\sqrt{2}$ with $|b_{1}|^{2}+|b_{2}|^{2}=1$, gives several choices for non-maximally entangled state where $0<b_{2}<1$ (except $b_{2}\neq1/\sqrt{2}$).  In Fig. \ref{fig:PNMS-PNMSa-PNMA-PNMAa}, we have plotted Tr($ \delta \rho^{H}h_{ \alpha_{v}}$) with respect to $\mathpzc{A}$ and$\mathpzc{W}$ for values $\alpha_{a}=0.2,~1.0,~1.2$. We used $|b_{2}|=0.9$ with $b_{1}$ and $b_{2}$ having same or different signs. We have scanned the parameter space for different values of $b_{2}$ and $\alpha_{a}$, but found no region where any of the trace quantities satisfy the condition given by (\ref{Tr+}).
Hence, making of EUQOE is not possible for any entangled initial state, while the detectors are in parallel motion.\\
 
\subsection{The detectors are moving anti-parallely:}\label{sec:calc:anti-para}
In this case detector $A$ is moving in RRW while other one (i.e. $B$) is moving in LRW during the accelerated phase. In this case trajectories of detector $A$ and detector $B$ are given by (\ref{B2}) and (\ref{B3}), respectively. The relation between the proper times of the detectors is given by (from (\ref{tAtB-}))
\begin{equation}
\tau_{B}=-\alpha_{a_{H}}\tau_{A}
\end{equation}
Using this relation, $\mathcal{P}_{A},~\mathcal{P}_{B}$ and $\mathcal{P}_{AB}$ are calculated in terms of the proper time of the detector $A$. Expressions of $\mathcal{P}_{A}$ are already presented in (\ref{PAexpression}) and (\ref{PAnexpression}).
 In Appendix \ref{D}, we obtain expressions of $\mathcal{P}_{B}$ when the detector $B$ is accelerating in LRW.  These are found to be same in value for the detector is accelerating in RRW. Therefore these are given by (\ref{PBexpression}) and (\ref{PBnexpression}). 
 In Appendix \ref{E.2}, we obtain $\Delta\mathcal{P}_{AB}=\mathcal{P}_{AB}(\omega,-\omega)-\mathcal{P}_{AB}(-\omega,\omega)$ for all $\alpha_{a_{}}$ values as
\begin{eqnarray}\label{deltaPABn}
\Delta\mathcal{P}_{AB}=0~.
  \end{eqnarray}
  Therefore, for anti-parallel motion, trace quantities get simplified and expressions are same for any value of $\alpha_{a_{}}$. The relative sign between $b_{1}$ and $b_{2}$ will not matter for anti-parallel motion of the detectors (see (\ref{traceFinal})).

\subsubsection{Analysis for maximally symmetric and anti-symmetric entangled initial state}
For the maximally entangled state {\it i.e.,} $|b_{1}|^{2}=|b_{2}|^{2}=1/2$, we have
 \begin{eqnarray}\label{deltaPAn}
\Delta\mathcal{P}_{A}= -\frac{\mathcal{W}}{8};~~~~
 \Delta\mathcal{P}_{B}=-\frac{\alpha_{a_{H}}\mathcal{W}}{8}=\alpha_{a_{H}} \Delta \mathcal{P}_{A}~,
 \end{eqnarray}
Therefore from (\ref{traceFinal}), we can conclude that work done or heat absorption are same for both symmetric and anti-symmetric initial states. For total work done, $\alpha= \alpha_{v}=1/ \gamma$ is  the inverse Lorentz factor, which is always positive. The trace corresponding to total work done by the system can be calculated from (\ref{traceFinal}), given by
\begin{eqnarray}
\text{Tr}(\delta^{H}h_{\alpha_{v}})=[\Delta\mathcal{P}_{A}+\alpha_{v}\Delta\mathcal{P}_{B}]=(1+\alpha_{v}\alpha_{a_{H}})\Delta\mathcal{P}_{A}
\end{eqnarray}
and the trace corresponding to heat absorbed by the system is
\begin{eqnarray}\label{APMEHeat}
\text{Tr}(\delta^{H}h_{\alpha_{a_{H}}})=[\Delta\mathcal{P}_{A}-\alpha_{a_{H}}\Delta\mathcal{P}_{B}]=(1-\alpha_{a_{H}}^{2})\Delta\mathcal{P}_{A}~.
\end{eqnarray}
Here, we can immediately see that the work done by the system is always negative due to (\ref{deltaPAn}). Therefore, we can not make an Otto cycle in this scenario.

\subsubsection{Analysis for non-maximally entangled initial state}
\begin{figure}[h!]
    \centering
    \subfigure[]{\includegraphics[width=0.49\textwidth]{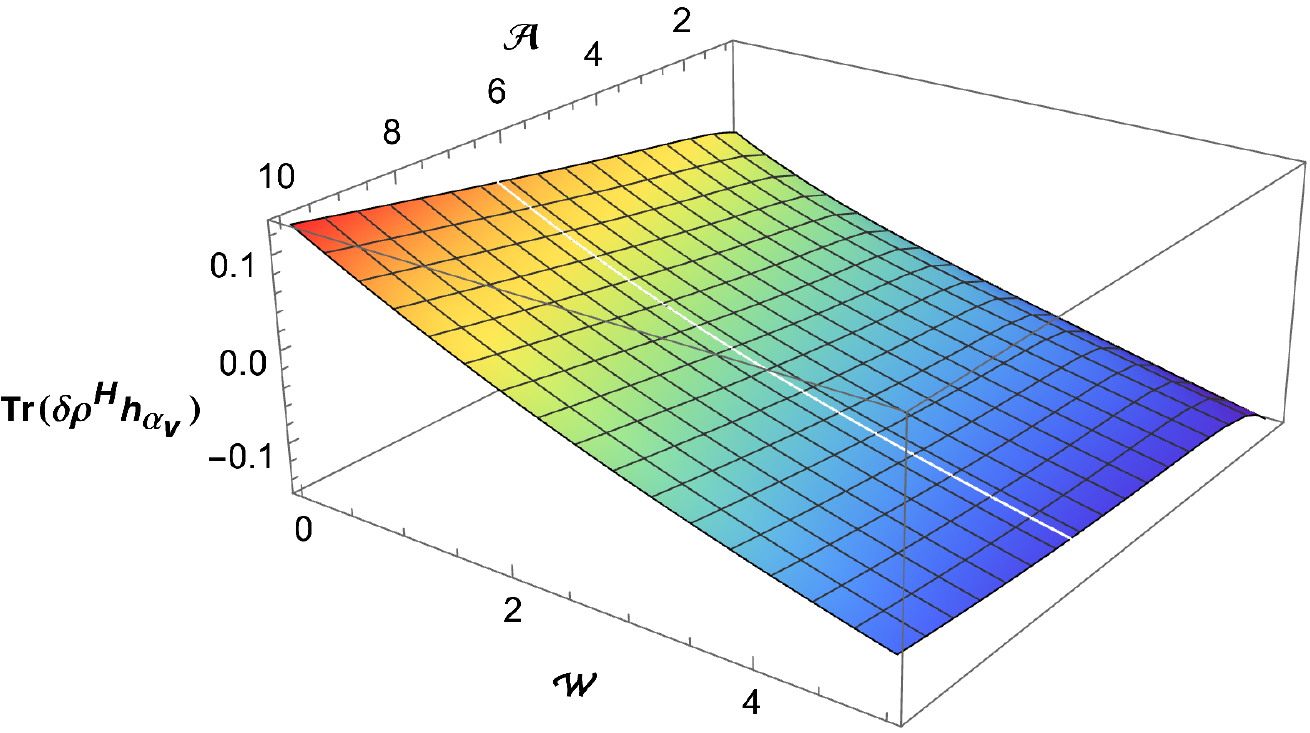}}
    \subfigure[]{\includegraphics[width=0.49\textwidth]{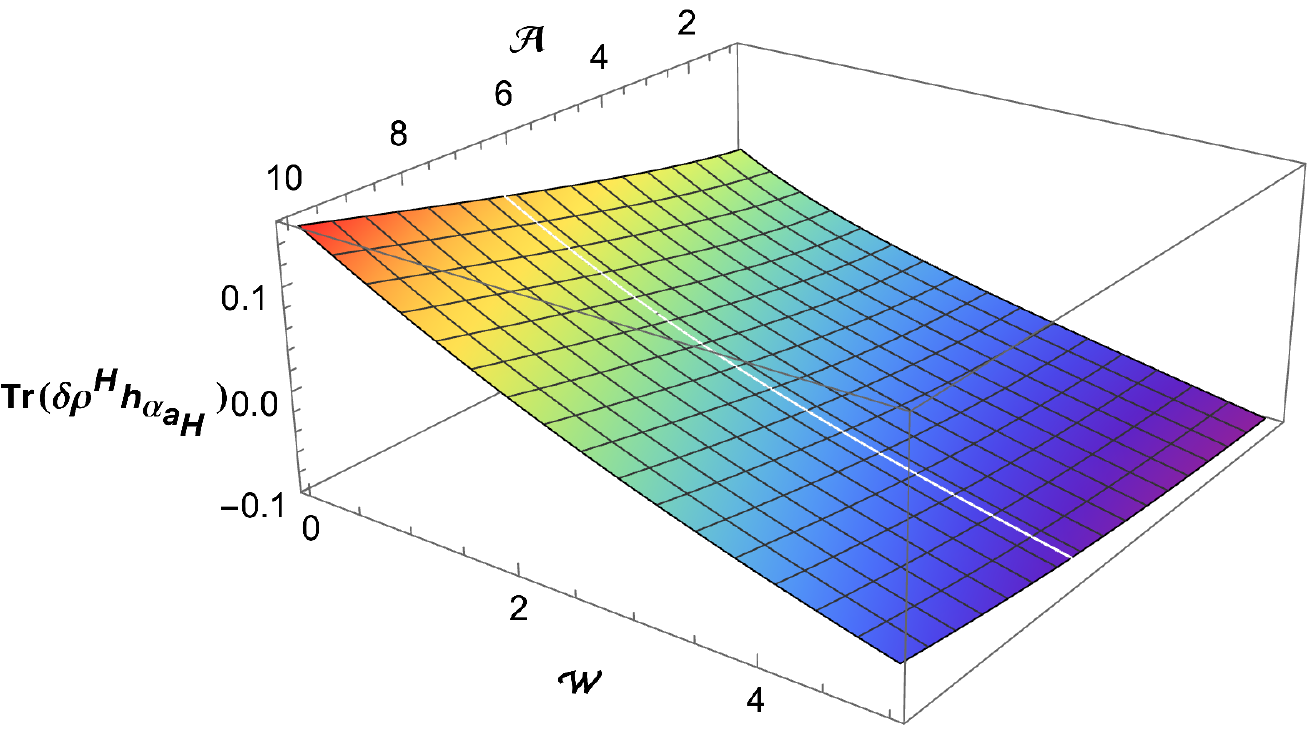}} \\
    \subfigure[]{\includegraphics[width=0.56\textwidth]{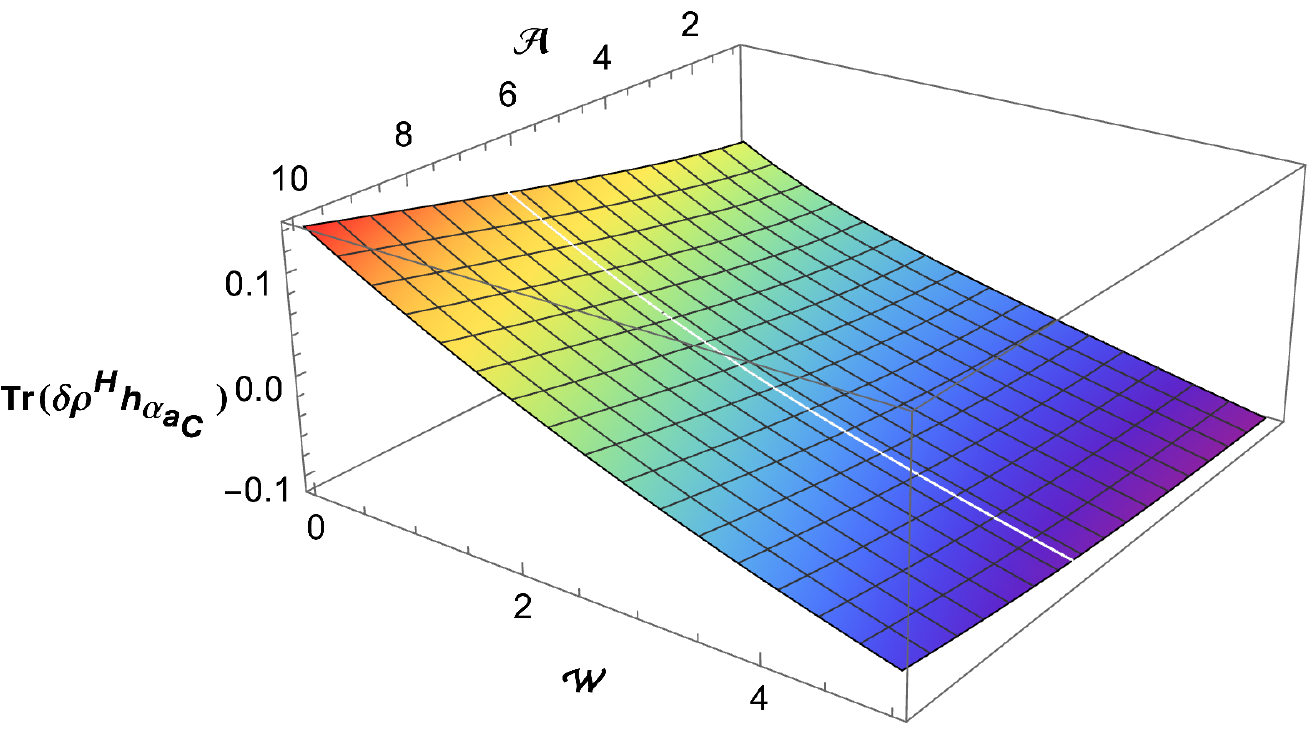}} 
        \subfigure[]{\includegraphics[width=0.4\textwidth]{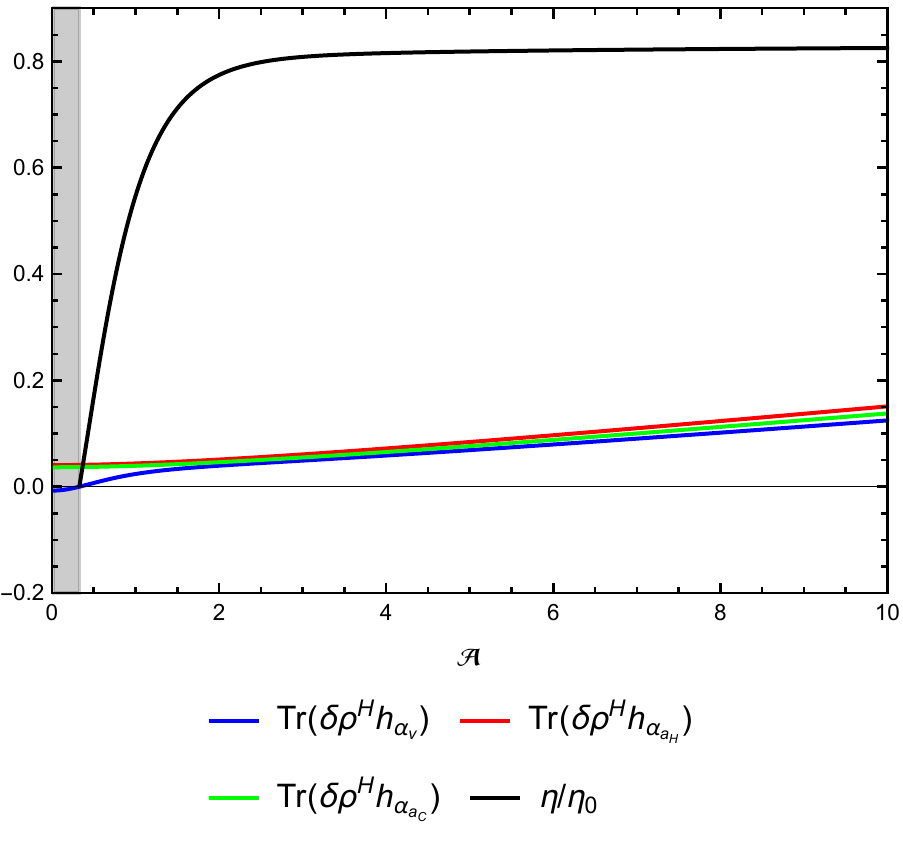}}\\
  \caption{Initially non-maximally entangled state with $|b_{2}|=0.9$ for anti-parallel motion: Plots show traces corresponding to (a) work done by the system, Tr($ \delta \rho^{H}h_{ \alpha_{v}}$), (b) heat absorbed by the system, Tr($ \delta \rho^{H}h_{ \alpha_{a_{H}}}$) with $\alpha_{a_{H}}=0.2$ and (c) heat rejected by the system, Tr($ \delta \rho^{H}h_{ \alpha_{a_{C}}}$) (we chosed $\alpha_{a_{C}}=0.1$)  with respect to dimensionless acceleration of the primary detector ($\mathpzc{A}$) and dimensionless energy gap ($\mathpzc{W}$) of the composite system. Subfigure (d) shows $\eta/\eta_{0}$ in black line, 
  Tr($ \delta \rho^{H}h_{ \alpha_{v}}$) in blue line, Tr($ \delta \rho^{H}h_{ \alpha_{a_{H}}}$) in red line and Tr($ \delta \rho^{H}h_{ \alpha_{a_{C}}}$) in green line  with respect to dimensionless acceleration of the primary detector ($\mathpzc{A}$) for  $\mathpzc{W}=0.2$, $\alpha_{a_{H}}=0.2$ and $\alpha_{a_{C}}=0.1$. The grey region represent the regime where making of the Otto cycle is not possible. }
  \label{fig:APNM}
\end{figure}

In this case (\ref{traceFinal}) reduces to
\begin{eqnarray}\label{traceFinalAP}
\text{Tr}(\delta\rho^H h_{\alpha_{v}})=\{b_{2}^{2}\mathcal{P}_A(\omega)-b_{1}^{2}\mathcal{P}_A(-\omega)\}+\alpha_{v}\{(b_{1}^{2}\mathcal{P}_{B}(\omega)-b_{2}^{2}\mathcal{P}_{B}(-\omega)\}~.
\end{eqnarray}
\begin{eqnarray}\label{traceFinalAP1}
\text{Tr}(\delta\rho^H h_{\alpha_{a_{H}}})=\{b_{2}^{2}\mathcal{P}_A(\omega)-b_{1}^{2}\mathcal{P}_A(-\omega)\}-\alpha_{a_{H}}\{(b_{1}^{2}\mathcal{P}_{B}(\omega)-b_{2}^{2}\mathcal{P}_{B}(-\omega)\}~.
\end{eqnarray}
 Here $\alpha=\alpha_{v}=\sqrt{1-v_{rel}^{2}}$ for work done, where by using (\ref{B2}) and (\ref{B3}), one can find $v_{rel}=-2\tanh(\mathpzc{A})/(1+\tanh^{2}(\mathpzc{A}))$. For heating stage, we have $\alpha=-\alpha_{a_{H}}$. In sub-figure (a) and (b) of Fig. \ref{fig:APNM}, we find that for some fixed $b_{2}$ and $\alpha_{a_{H}}$, Tr($ \delta \rho^{H}h_{ \alpha_{a_{H}}}$) and Tr($ \delta \rho^{H}h_{ \alpha_{v}}$) can be positive within a range of values of $\mathpzc{A}$ and $\mathpzc{W}$.
 We also need to ensure that Tr($\delta\rho^{H}h_{\alpha_{a_{C}}}$) is positive, which is necessary in order to happening of the heat rejection by the cycle in the cooling process. Otherwise, the efficiency of the cycle will be greater than one. This can be checked from (\ref{E-conserv}), which states that for Tr($\delta\rho^{H}h_{\alpha_{a_{C}}}$)$<0$, $(\omega_{2}-\omega_{1})\text{Tr}(\delta\rho^{H}h_{\alpha_{v}})$ will be greater than $\omega_{2}\text{Tr}(\delta\rho^{H}h_{\alpha_{a_{H}}})$, and in that case one will have $\eta_{E}>1$ (see, expression of $\eta_{E}$ in (\ref{eta-E})). As stated earlier, such is not physically possible. We know that velocity of $j$-th detector $(j=A,B)$ at the start and end of the stage II ($-\mathcal{T}_{j_{H}}/2$ and $\mathcal{T}_{j_{H}}/2$) is same as that at the end and start of the stage IV ($\mathcal{T}_{j_{C}}/2$ and $-\mathcal{T}_{j_{C}}/2$).  Using this fact, the following relations can be obtained:
 \begin{eqnarray}\label{heat-cool-A}
 a_{A_{C}}\mathcal{T}_{A_{C}}=-a_{A_{H}}\mathcal{T}_{A_{H}}~;
 \end{eqnarray}
\begin{eqnarray}\label{heat-cool-B}
 a_{B_{C}}\mathcal{T}_{B_{C}}=-a_{B_{H}}\mathcal{T}_{B_{H}}~;
 \end{eqnarray}
and \begin{eqnarray}
 \alpha_{a_{C}}\mathcal{T}_{A_{C}}= \mathcal{T}_{B_{C}}~.
 \label{B10}
  \end{eqnarray}
Here $\alpha_{a_{C}}$ is given by $a_{A_{C}}/a_{B_{C}}$.  In (\ref{heat-cool-A}) and (\ref{heat-cool-B}) as the parameters in the heating process (stage II) are already fixed, we can always freely choose $a_{j_{C}}$. Note that Eq. (\ref{B10}) can be obtained by taking ratio between (\ref{heat-cool-A}) and (\ref{heat-cool-B}) and then using (\ref{B5}). Therefore, any two parameters can be chosen freely and and then the third parameter is determined. Here we fix $\alpha_{a_{C}}$ and $\mathcal{T}_{A_{C}}$ ( $\mathcal{T}_{B_{C}}$ will be automatically determined) such that Tr($\delta\rho^{H}h_{\alpha_{a_{C}}})>0$. Now a sufficient condition for satisfaction of energy conservation ( from (\ref{E-conserv})) can be taken as
 \begin{eqnarray}
 \frac{\mathpzc{W}}{\omega_{1}\mathcal{T}_{A_{H}}}(\alpha_{a_{H}}-1)=(\alpha_{a_{C}}-1)~.
 \label{B11}
 \end{eqnarray}

 Since the values of $\mathpzc{W}$ and $\alpha_{a_{H}}$ are already determined by considering positivity of Tr($ \delta \rho^{H}h_{ \alpha_{v}}$) and Tr($ \delta \rho^{H}h_{ \alpha_{a_{H}}}$), while $\alpha_{a_{C}}$ in (\ref{B10}) is chosen to be free, the above equation can fix the value of dimensionless energy gap $\omega_{1}\mathcal{T}_{A_{H}}$ for $\alpha_{a_{C}}\neq1$. In this connection remember that $\omega_{1}$ is always need to be less than $\omega_{2}$. So we have to choose $\alpha_{a_{C}}$ such that $\omega_{1}<\omega_{2}$ and (\ref{B11}) are satisfied simultaneously.  One can notice that $\omega_{1}$ is positive, if we choose $\alpha_{a_{C}}>1~(\alpha_{a_{C}}<1)$ for $\alpha_{a_{H}}>1~(\alpha_{a_{H}}<1).$ It should be pointed out that for $\alpha_{a_{H}}=1$, we obtain $\omega_{1}=0~($ for $\alpha_{a_{C}}\neq1)$ or any value of $\omega_{1}~($ for $\alpha_{a_{C}}=1)$. However, we will not consider $\omega_{1}=0$ case, as this implies that at the beginning of stage I, energy gap of the system is zero. If we choose $\alpha_{a_{H}}=1$ then we also need to choose $\alpha_{a_{C}}=1$. For other values of $\alpha_{a_{H}}$ in detector's anti-parallel motion with non-maximally entangled case, we can always choose $\alpha_{a_{C}}$ such that $\omega_{1}<\omega_{2}$ and Tr($ \delta \rho^{H}h_{ \alpha_{a_{C}}})>0$ along with the energy conservation relation satisfied. In sub-figure (c) of Fig. \ref{fig:APNM} we find that Tr($ \delta \rho^{H}h_{ \alpha_{a_{C}}}$) has positive values within a range of values of $\mathpzc{A}$ and $\mathpzc{W}$. In this case since we took $\alpha_{a_{C}}=0.1$, Eq. (\ref{B11}) implies $\omega_{1}<\omega_{2}$ (as we took $\alpha_{a_{H}}=0.2$). In sub-figure (d) of Fig. \ref{fig:APNM}, we have plotted  Tr($ \delta \rho^{H}h_{ \alpha_{v}}$),  Tr($ \delta \rho^{H}h_{ \alpha_{a_{H}}}$),  Tr($ \delta \rho^{H}h_{ \alpha_{a_{C}}}$) and  $\eta/\eta_{0}$ as a function of $\mathpzc{A}$ with $\mathpzc{W}=0.2$, $\alpha_{a_{H}}=0.2$, $\alpha_{a_{C}}=0.1$ and $|b_{2}|=0.9$. One can observe that for $\mathpzc{A}<0.33$ (in the gray region), our Tr($ \delta \rho^{H}h_{ \alpha_{v}}$) has negative values  for these fixed values of parameters. Therefore the  Otto cycle is only possible for $\mathpzc{A}>0.33$ for this case. One can also notice that $\eta/\eta_{0}$ varies with the detectors' acceleration ($\mathpzc{A}$) for fixed values of other parameters and always less than one. Thus even though construction of the entangled Otto cycle is possible, enhancement of efficiency of the cycle is not possible.

 \begin{figure}[h!]
    \centering
       \subfigure[]{\includegraphics[width=0.49\textwidth]{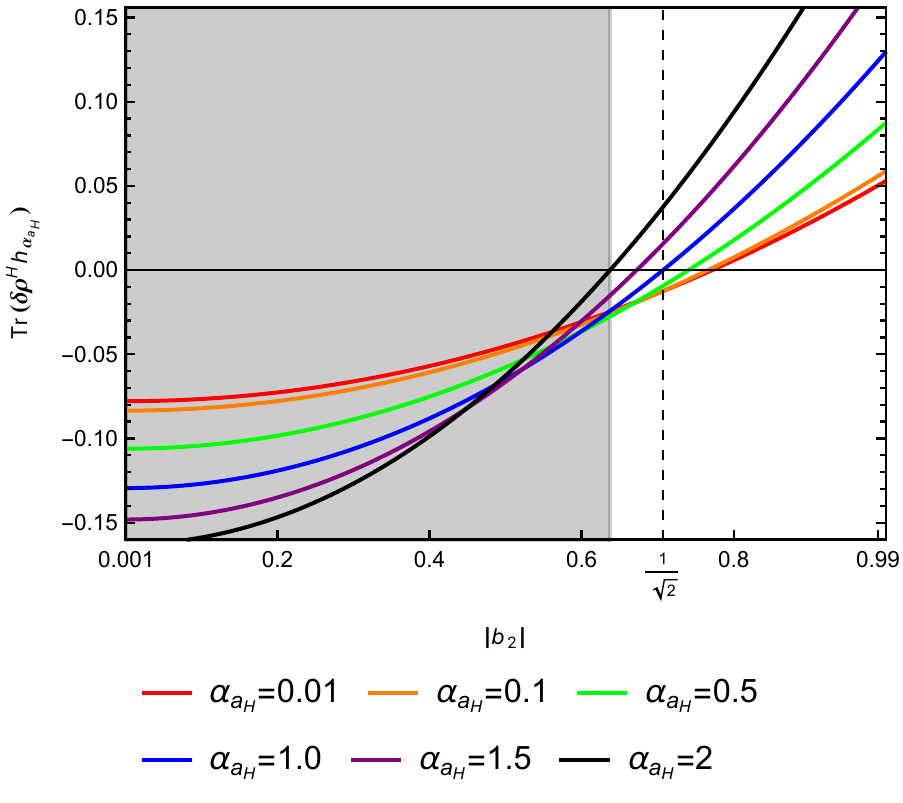}} 
     \subfigure[]{\includegraphics[width=0.49\textwidth]{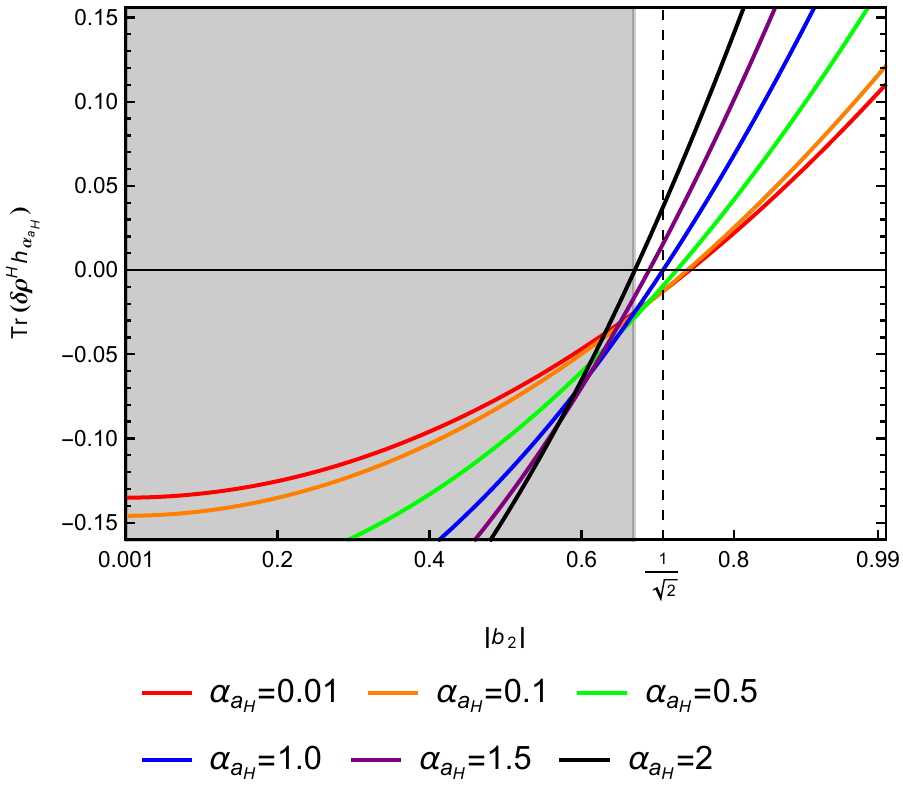}} \\
     \subfigure[]{\includegraphics[width=0.49\textwidth]{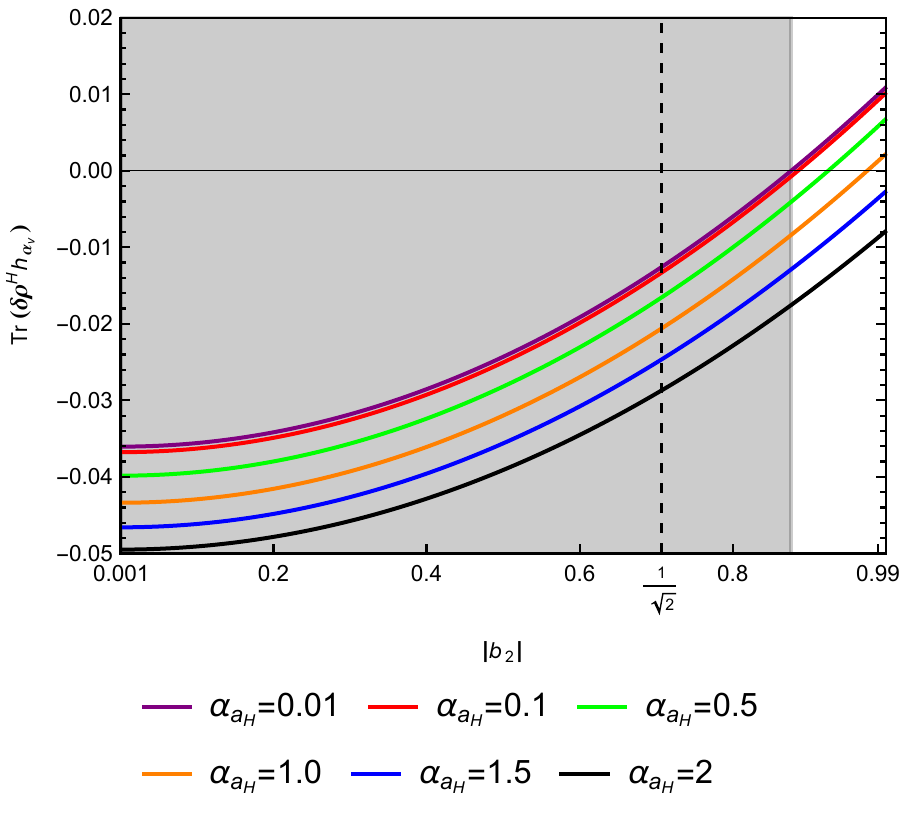}}
    \subfigure[]{\includegraphics[width=0.49\textwidth]{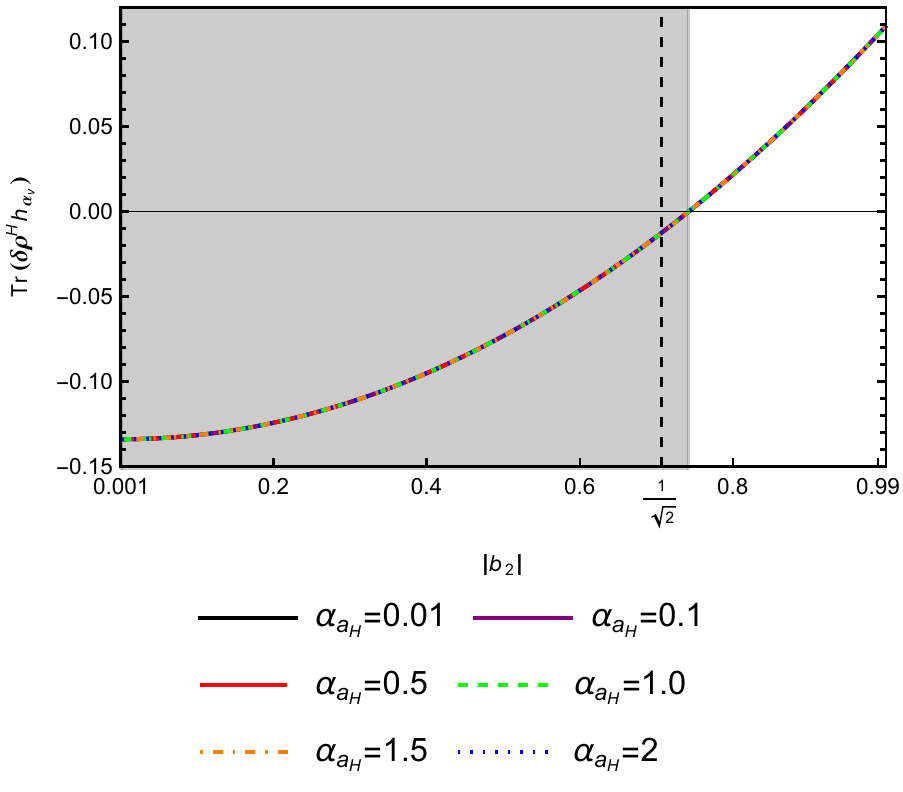}}
  \caption{Initially non-maximally entangled state for anti-parallel motion (with fixed  $\mathpzc{W}=0.2$): Plots (a) and (b) are showing traces corresponding to heat absorbed by the system, Tr($ \delta \rho^{H}h_{ \alpha_{a_{H}}}$)   with respect to $|b_{2}|$ for different $\alpha_{a_{H}}$ for $\mathpzc{A}=0.5$ and $5.0$, respectively. Plots (c) and (d) are showing traces corresponding to work done by the system, Tr($ \delta \rho^{H}h_{ \alpha_{v}}$)  with respect to $|b_{2}|$ for different $\alpha_{a_{H}}$ for $\mathpzc{A}=0.5$ and $5.0$, respectively. The grey region in the plots roughly represent the regime with negative traces for all given $\alpha_{a_{H}}$ values.}
  \label{fig:APNM-al}
\end{figure}

 \begin{figure}[h!]
    \centering
    \subfigure[]{\includegraphics[width=0.49\textwidth]{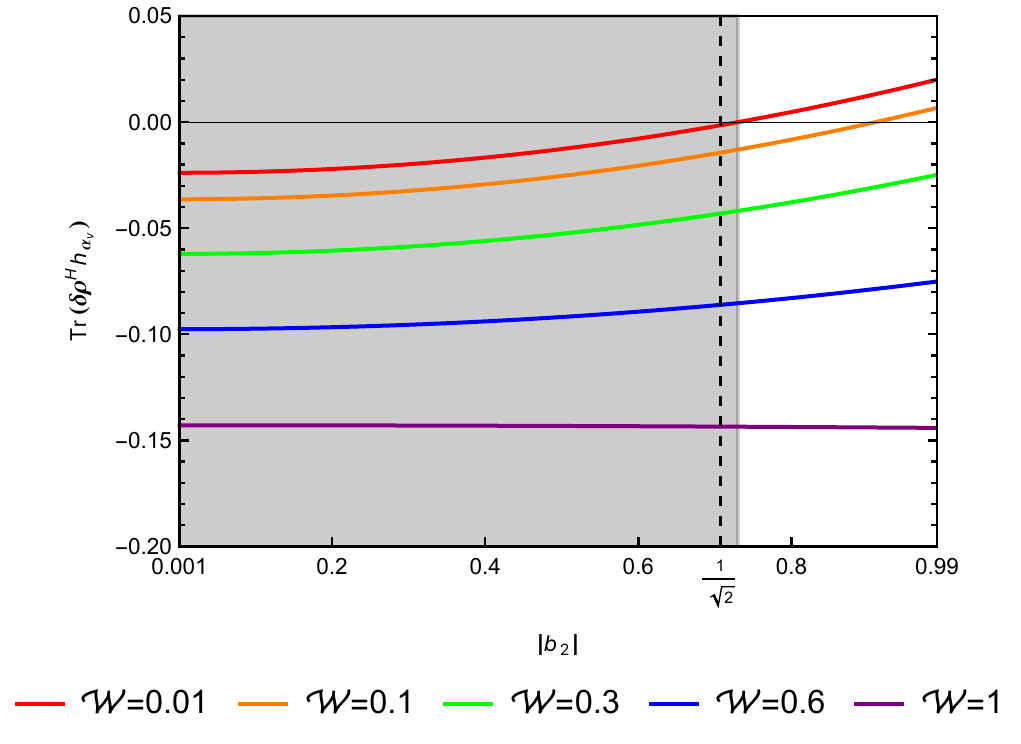}}
    \subfigure[]{\includegraphics[width=0.49\textwidth]{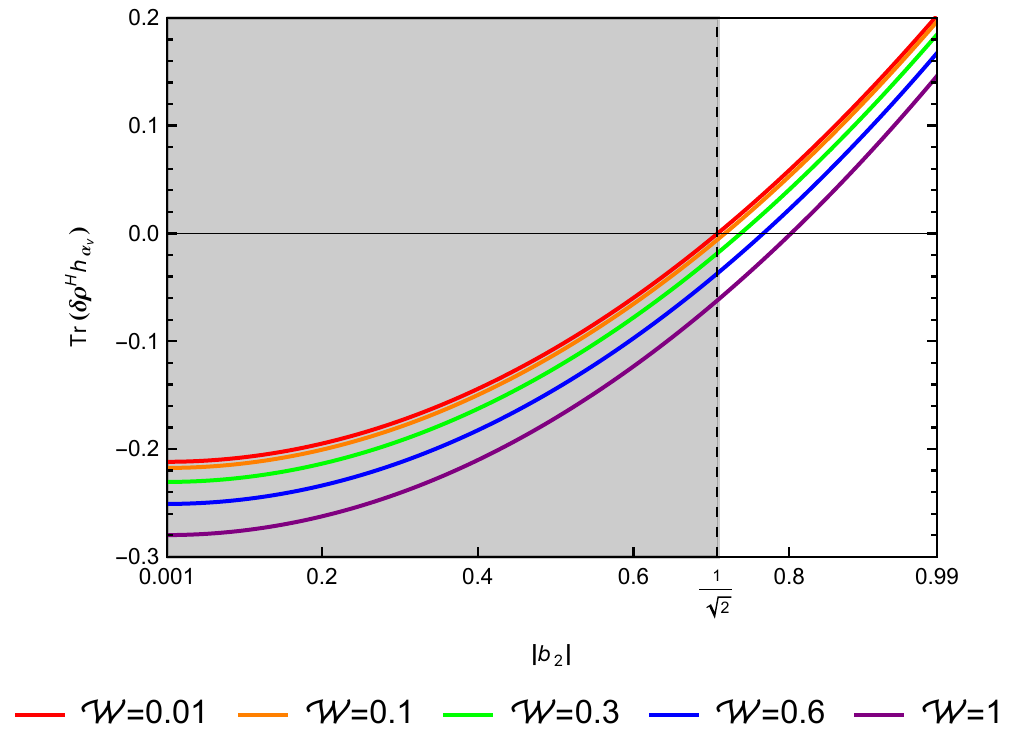}}\\
     \subfigure[]{\includegraphics[width=0.49\textwidth]{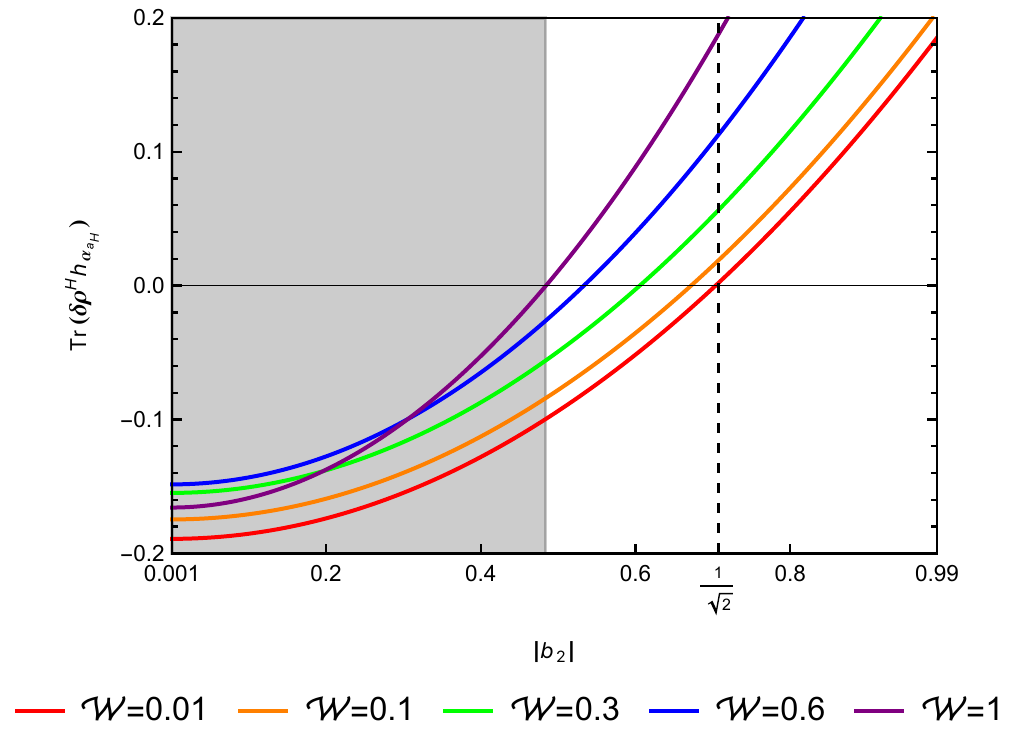}}
    \subfigure[]{\includegraphics[width=0.49\textwidth]{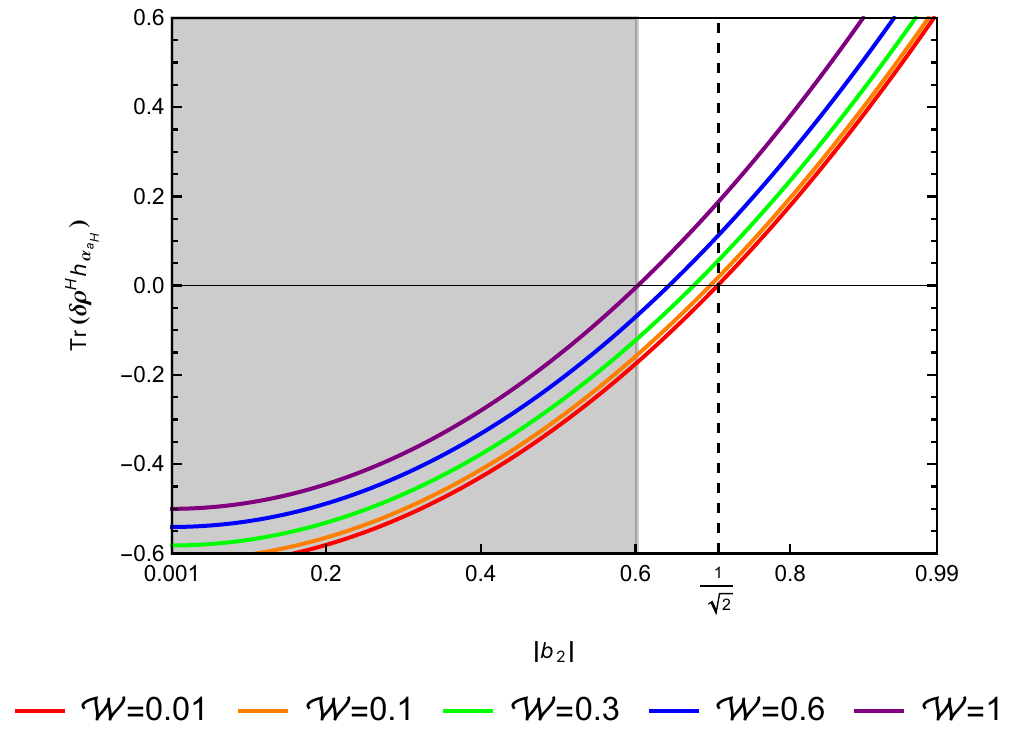}}
  \caption{Initially non-maximally entangled state for anti-parallel motion (with fixed $\alpha_{a_{H}}=2.0$
): Plots  (a) and (b) are showing traces corresponding to work done by the system, Tr($ \delta \rho^{H}h_{ \alpha_{v}}$)  with respect to $|b_{2}|$ for different dimensionless energy gap ($\mathpzc{W}$) of the composite system for $\mathpzc{A}=0.5$ and $\mathpzc{A}=10.0$, respectively. Plots  (c) and (d) are showing traces corresponding to heat absorbed by the system, Tr($ \delta \rho^{H}h_{ \alpha_{a_{H}}}$)  with respect to $|b_{2}|$ for different dimensionless energy gap ($\mathpzc{W}$) of the composite system for $\mathpzc{A}=0.5$ and $\mathpzc{A}=10.0$, respectively. The grey region in the plots roughly represent the regime with negative traces for given $\mathpzc{W}$ values.}
  \label{fig:APNM-b2}
\end{figure}

 \begin{figure}[h!]
    \centering
    \subfigure[]{\includegraphics[width=0.49\textwidth]{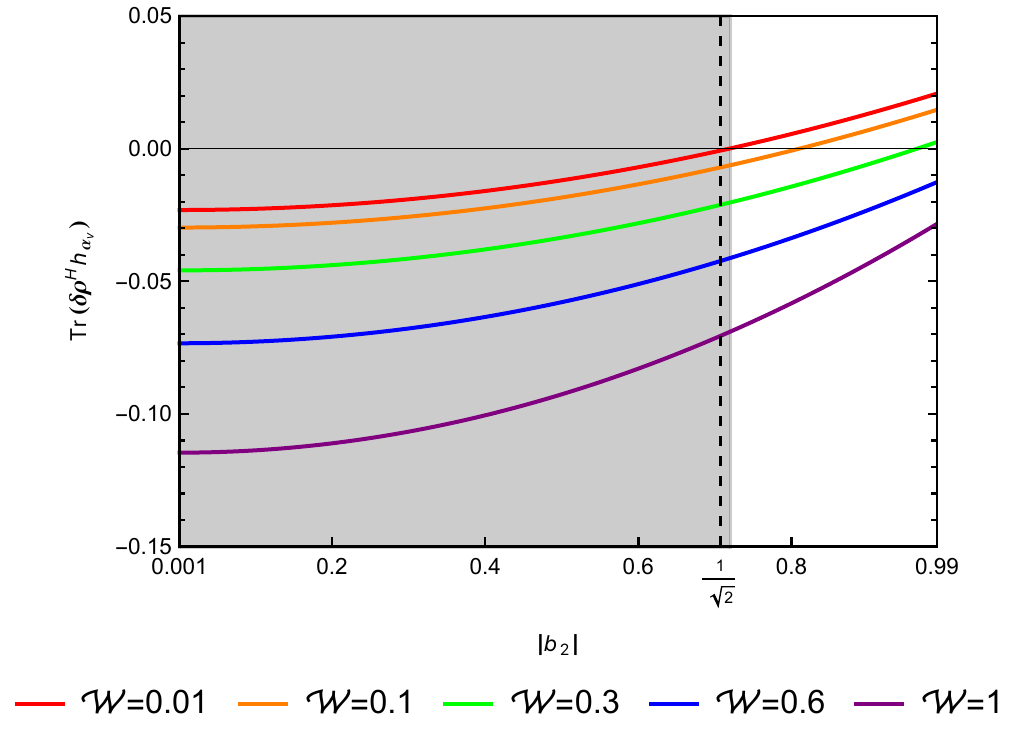}}
    \subfigure[]{\includegraphics[width=0.49\textwidth]{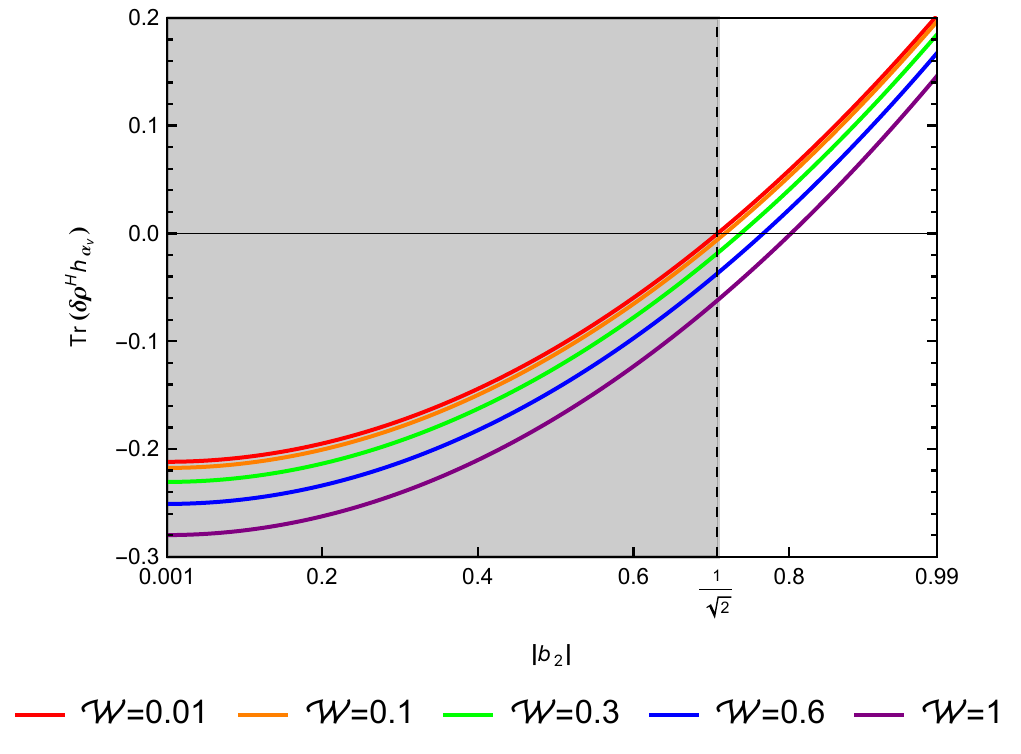}}\\
     \subfigure[]{\includegraphics[width=0.49\textwidth]{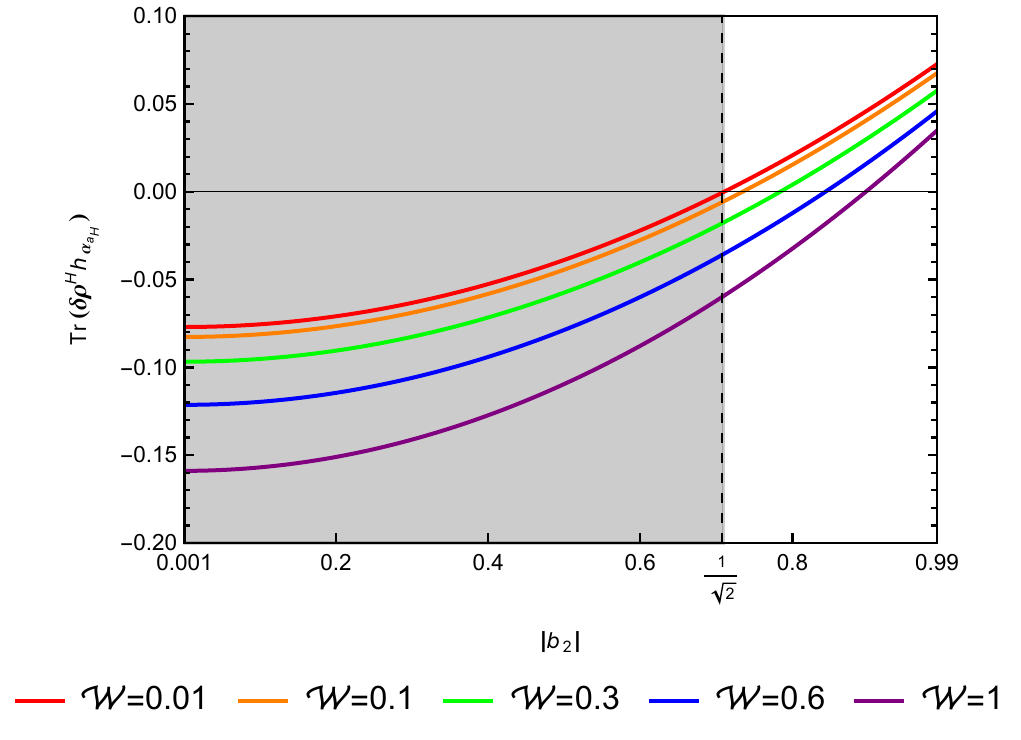}}
    \subfigure[]{\includegraphics[width=0.49\textwidth]{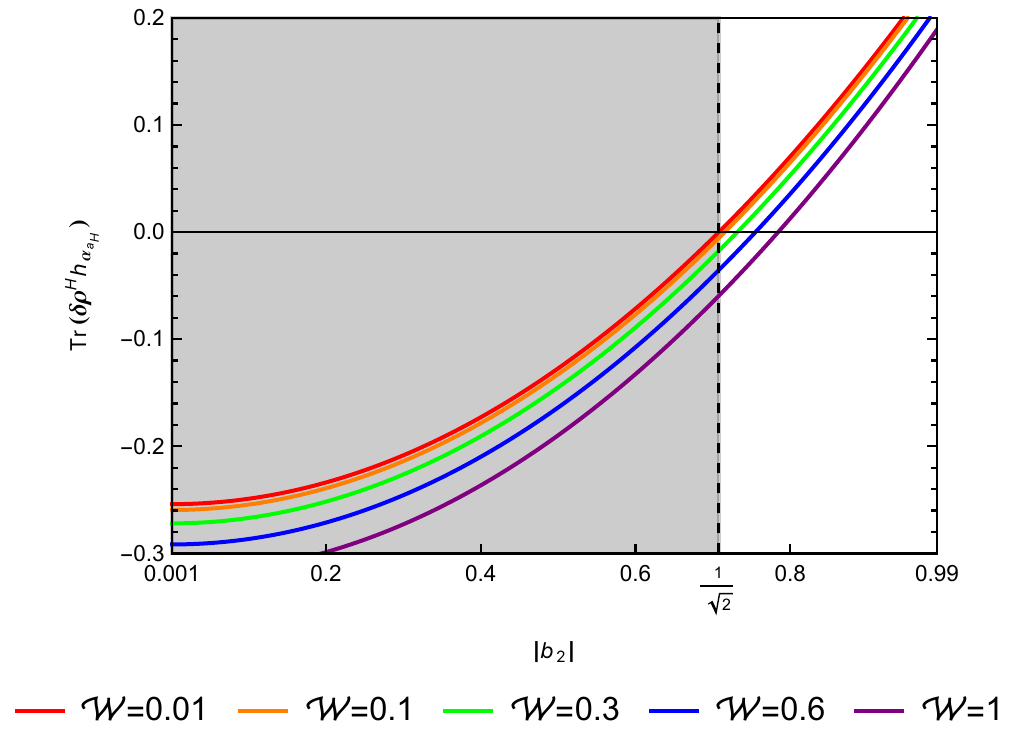}}
  \caption{Initially non-maximally entangled state for anti-parallel motion (with fixed $\alpha_{a_{H}}=0.2$
): Plots  (a) and (b) are showing traces corresponding to work done by the system, Tr($ \delta \rho^{H}h_{ \alpha_{v}}$)  with respect to $|b_{2}|$ for different values of dimensionless energy gap ($\mathpzc{W}$) of the composite system for $\mathpzc{A}=0.5$ and $\mathpzc{A}=10.0$, respectively. Plots  (c) and (d) are showing traces corresponding to heat absorbed by the system, Tr($ \delta \rho^{H}h_{ \alpha_{a_{H}}}$)  with respect to $|b_{2}|$ for different values of dimensionless energy gap ($\mathpzc{W}$) of the composite system for $\mathpzc{A}=0.5$ and $\mathpzc{A}=10.0$, respectively. The grey region in the plots roughly represent the regime with negative traces for given $\mathpzc{W}$ values.}
  \label{fig:APNM-b2a}
\end{figure}
Now we aim to investigate how much deviation from maximally entangled state is needed in order to construct a EUQOE.
 In Fig. \ref{fig:APNM-al} we have plotted the trace quantities with respect to $b_{2}$ for different values of $\alpha_{a_{H}}$. The positivity of the trace quantity related to heat absorption by the system depend on $\alpha_{a_{H}}$ as indicated by (\ref{APMEHeat}). The grey region in the plots represent region with negative traces for given parameter values. 
 In sub-figure (a) (corresponding to $\mathpzc{A}=0.5$) and (b) (for $\mathpzc{A}=5.0$),  it is shown that the trace quantity related to heat absorbed by the system is $\alpha_{a_{H}}$ dependent.  Higher the $\alpha_{a_{H}}$ values, larger the range of $b_{2}$ values for which this trace is positive. It should be mentioned that when $\alpha_{a_{H}}=1$, Tr($ \delta \rho^{H}h_{ \alpha_{a_{H}}}$) is always $zero$ for $|b_{2}|=1/\sqrt{2}$ (as eq. (\ref{traceFinalAP1}) reduces to $(b_{2}^{2}- b_{1}^{2})[\mathcal{P}_A(\omega)+\mathcal{P}_A(-\omega)]$). For $\alpha_{a_{H}}>1$, this quantity also can be positive for $|b_{2}|<1/\sqrt{2}$. Same nature will be obtained for Tr($ \delta \rho^{H}h_{ \alpha_{a_{C}}}$)  along with satisfaction of (\ref{B11}), as only $\alpha_{a_{H}}$ will replaced by $\alpha_{a_{C}}$.
  However this will not be useful unless all trace quantities are not positive simultaneously.
  In sub-figure (c) the trace quantity related to work done by the system is plotted for  $\mathpzc{A}=0.5$, which is $\alpha_{a_{H}}$ dependent, as expected. Higher the $\alpha_{a_{H}}$ values, smaller the range of $b_{2}$ values for which this trace is positive.  In sub-figure (d) it is shown that it is almost independent of $\alpha_{a_{H}}$ for $\mathpzc{A}=5.0$. It is due to the fact that $\alpha_{v}=\sqrt{1-v_{rel}^{2}}$ (with $v_{rel}=-2\tanh(\mathpzc{A})/(1+\tanh^{2}(\mathpzc{A}))$ get suppressed with higher values of $\mathpzc{A}$.

In sub-figures (a) and (b) of Fig. \ref{fig:APNM-b2}, we have shown the traces corresponding to work done by the system with respect to $b_{2}$ for different values of $\mathpzc{W}$ (for $\mathpzc{A}=0.5$ and $\mathpzc{A}=10.0$ respectively). Sub-figure (b) shows that for higher $\mathpzc{A}$, Tr($ \delta \rho^{H}h_{ \alpha_{v}}$) is $positive$ for $b_{2}$ is very close to $1/\sqrt{2}$. In sub-figures (c) and (d) of Fig. \ref{fig:APNM-b2}, showing the traces corresponding to heat absorption by the system with respect to $b_{2}$ for different values of $\mathpzc{W}$ (with $\mathpzc{A}=0.5$ and $\mathpzc{A}=10.0$ respectively). It is visible that Tr($\delta \rho^{H}h_{ \alpha_{a_{H}}}$) is $positive$ for $|b_{2}|<1/\sqrt{2}$ for $\alpha_{a_{H}}>1$.  In all the plots in Fig. \ref{fig:APNM-b2}, we choose $\alpha_{a_{H}}=2$.
 Same is drawn for $\alpha_{a_{H}}=0.2$ in Fig. \ref{fig:APNM-b2a}. All plots are showing that larger the value of $\mathpzc{W}$, smaller the range of $b_{2}$, for which the traces corresponding to work done by the system are positive. Additionally when $\alpha_{a_{H}}>1$, the region of $b_2$ for which the traces corresponding to heat absorbed by the system are positive will be larger as $\mathpzc{W}$ increases. Whereas for $\alpha_{a_{H}}<1$ the same will be smaller as $\mathpzc{W}$ increases. Howerever, since making of EUOE is only possible for a parameter space where all trace quantities are positive simultaneously, therefore we will consider a common parameter space from these figures where all the traces are positive.  Here we can see that even though the construction of Otto cycle is not possible for maximally entangled state, but a small deviation form the maximally entangled state can allow us to construct an Otto cycle.

\begin{figure}[h!]
    \centering
  \subfigure[]{\includegraphics[width=0.49\textwidth]{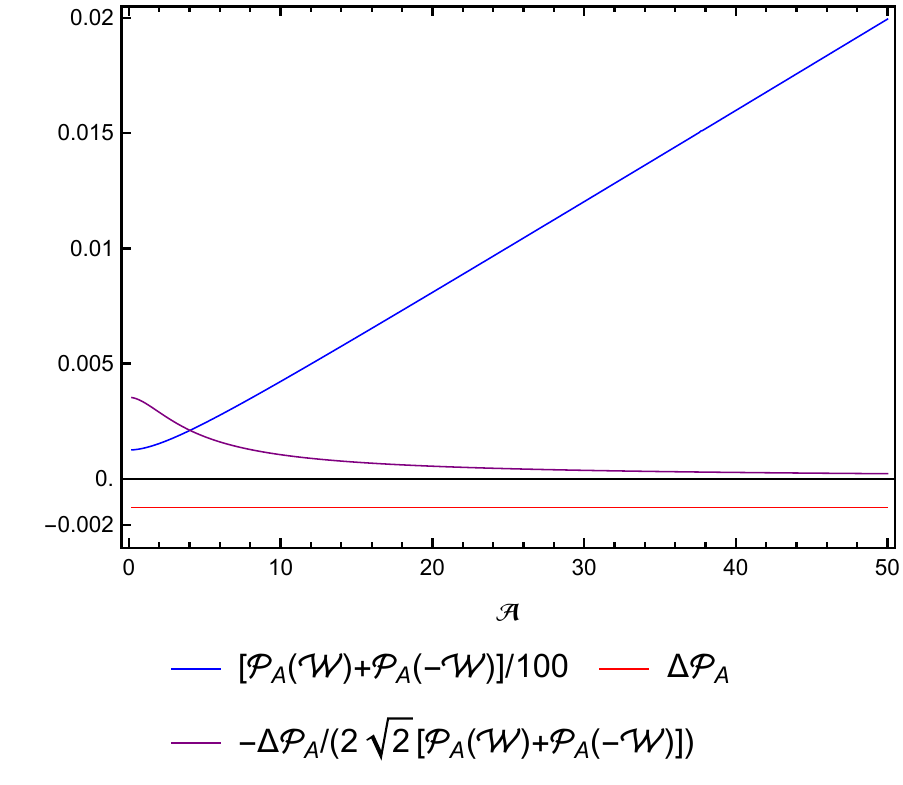}}
   \subfigure[]{\includegraphics[width=0.49\textwidth]{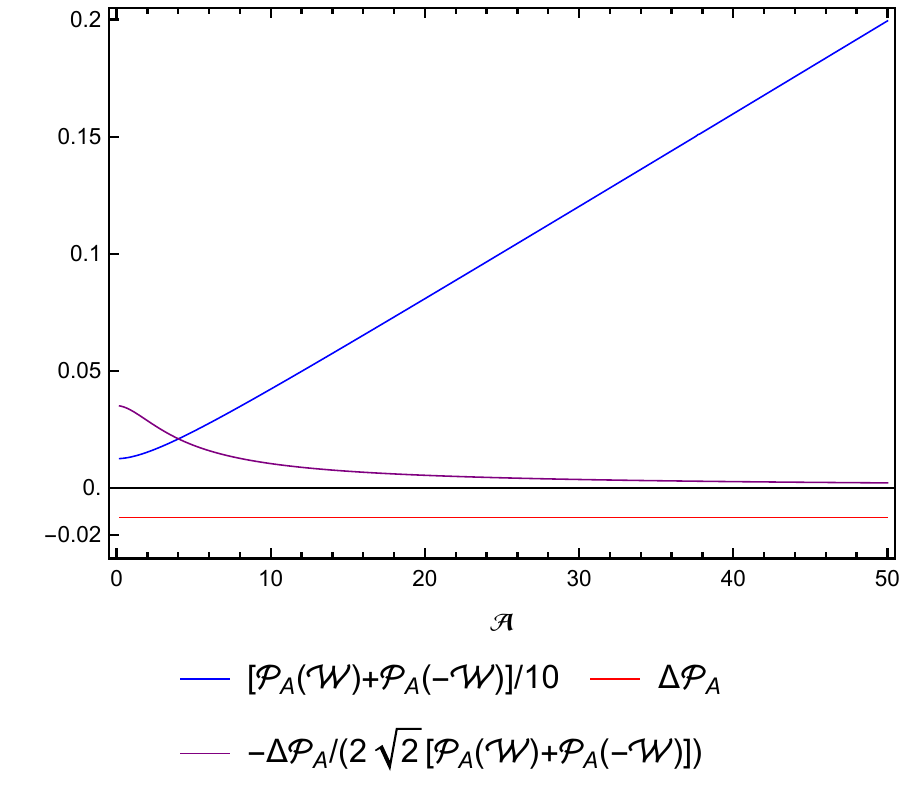}}
      \caption{Estimation of $\epsilon_{0}$, for (a) $\mathpzc{W}=0.01$ and (b)  $\mathpzc{W}=0.1$ (independent of $\alpha_{a_{H}}$).}
  \label{fig:APNM-ep}
\end{figure}

 As shown in Figs. \ref{fig:APNM-b2} and \ref{fig:APNM-b2a}, depending upon the given parameters $\textrm{Tr}(\delta\rho^H h_{\alpha_{a_{H}}})$ can be $positive$ for $b_{2}<1/\sqrt{2}$ or $b_{2}$ slightly {\it greater than} $1/\sqrt{2}$. However $\textrm{Tr}(\delta\rho^H h_{\alpha_{v}})$ is $positive$ for $b_{2}$ is {\it greater than} $1/\sqrt{2}$. Therefore for these given parameters the making of Otto cycle is posssible only when one is away from maximally entangled state. But it may be noticed that larger the value of $\mathpzc{A}$ the right side boundary of the shaded region approching towards $b_2=1/\sqrt{2}$ value. Therefore it may appear that a near-maximally entangled state is capable of yielding a Otto cycle. Here we want to estimate such closeness. For higher $\mathpzc{A}$, only the terms in first  braces of  (\ref{traceFinalAP}) will dominate.  In this situation, ${\textrm{Tr}}(\delta\rho^H h_{\alpha_v})$ can be positive for small deviation of $b_{2}$ from the maximally entangled case (consider $|b_{2}|=1/\sqrt{2}+\epsilon$, with $\epsilon~ (>0)$ very small). Then considering upto order $\epsilon^2$ term one finds
\begin{eqnarray}\label{traceFinalAP2}
\begin{aligned}
\text{Tr}(\delta\rho^H h_{\alpha_{v}})=&\left(\frac{1}{\sqrt{2}}+\epsilon\right)^{2}\mathcal{P}_A(\mathpzc{W})-\left[1-\left(\frac{1}{\sqrt{2}}+\epsilon\right)^{2}\right]\mathcal{P}_A(-\mathpzc{W}))\\\approx&\frac{1}{2}\Delta\mathcal{P}_{A}+\sqrt{2}\epsilon[\mathcal{P}_{A}(\mathpzc{W})+\mathcal{P}_{A}(-\mathpzc{W})]+\epsilon^{2}\mathcal{P}_{A}(-\mathpzc{W})~.
\end{aligned}•
\end{eqnarray}
Here the first quantity is $negative$ and independent of $\mathpzc{A}$ (see (\ref{DPA-Bnexpression})), but second and third quantities are $positive$ (see (\ref{PAexpressionD}) and (\ref{DPA-Bnexpression}) for expressions of $\mathcal{P}_{A}(\mathpzc{W})$ and $\mathcal{P}_{A}(-\mathpzc{W})$; which are individually positive quantity and increases with $\mathpzc{A}$). Therefore if at least we have $\epsilon=-\Delta\mathcal{P}_{A}/(2\sqrt{2}[\mathcal{P}_{A}(\mathpzc{W})+\mathcal{P}_{A}(-\mathpzc{W})])\equiv\epsilon_{0}$, then this trace will be positive. For $\epsilon=\epsilon_{0}$, first two terms in (\ref{traceFinalAP2})  cancel each other. The remaining term be the third term, which is in the order of $\epsilon^{2}$. In Fig. \ref{fig:APNM-ep}, we have plotted $\mathcal{P}_{A}(\mathpzc{W})+\mathcal{P}_{A}(-\mathpzc{W})$, $\Delta\mathcal{P}_{A}$ and $\epsilon_{0}$ with respect to $\mathpzc{A}$. These plots are independent of $\alpha_{a_{H}}$. The plot of $\mathcal{P}_{A}(\mathpzc{W})+\mathcal{P}_{A}(-\mathpzc{W})$ is divided by a numerical factor just for plotting convenience.
We can see that $\epsilon_{0}$ is indeed very small (with $\mathpzc{W}=0.01$ and $0.1$, respectively), and decreases more for the higher values of $\mathpzc{A}$ where the above estimation can be trusted. Thus, very  small $\epsilon$ is required for small $\mathpzc{W}$ and large $\mathpzc{A}$ to make this trace positive. We add a table (Table \ref{table:ep}) for values of $\epsilon_{0}$ and corresponding values of Tr$(\delta\rho^{H}h_{\alpha_{v}})$ from Fig. \ref{fig:APNM-ep}. As the value of $\mathpzc{A}$ is increasing, $\epsilon_{0}$ is getting more smaller. However in this case the values of Tr$(\delta\rho^{H}h_{\alpha_{v}})$ are becoming almost negligible as it is of the order $\epsilon^2$ in this range. The order of magnitute of the numerical values of Tr$(\delta\rho^{H}h_{\alpha_{v}})$ are becoming so small that these fall into the range of noise in the numerical analysis and hence these can not be trusted. Therefore $\epsilon$ can not be arbitrarily small in order to satisfy the condition (\ref{Tr+}) and so the initial entangled stated must be ``considerable'' away (at least in the range of numerical perceverence) from the maximally entangled state in order to make an Otto cycle.

\begin{figure}[h!]
    \centering
    \subfigure[]{\includegraphics[width=0.49\textwidth]{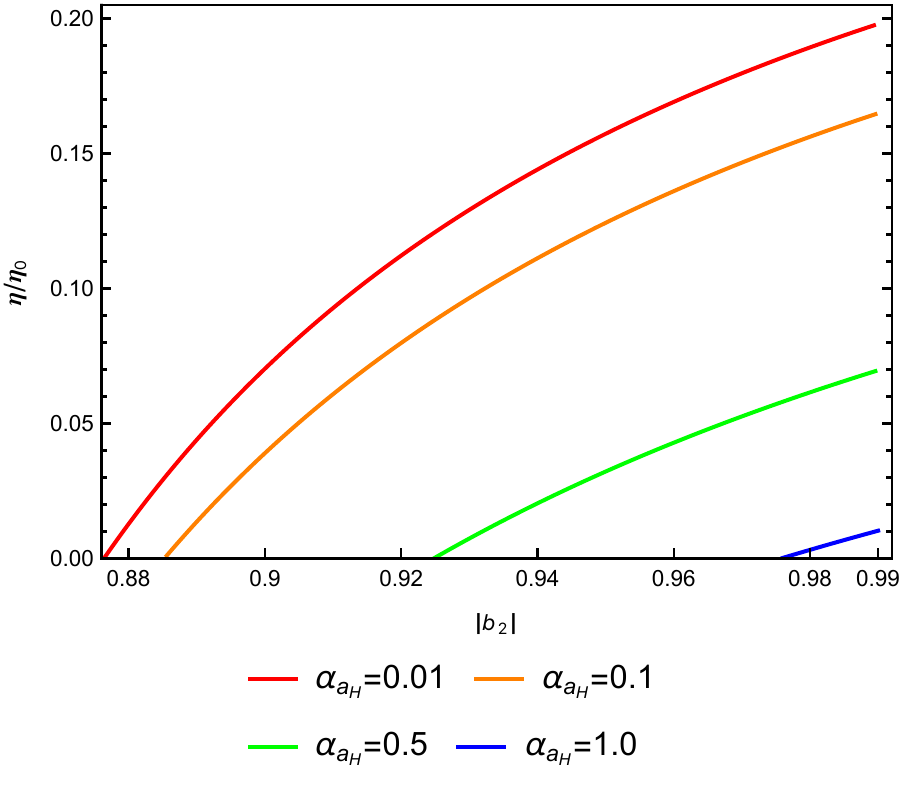}}
    \subfigure[]{\includegraphics[width=0.49\textwidth]{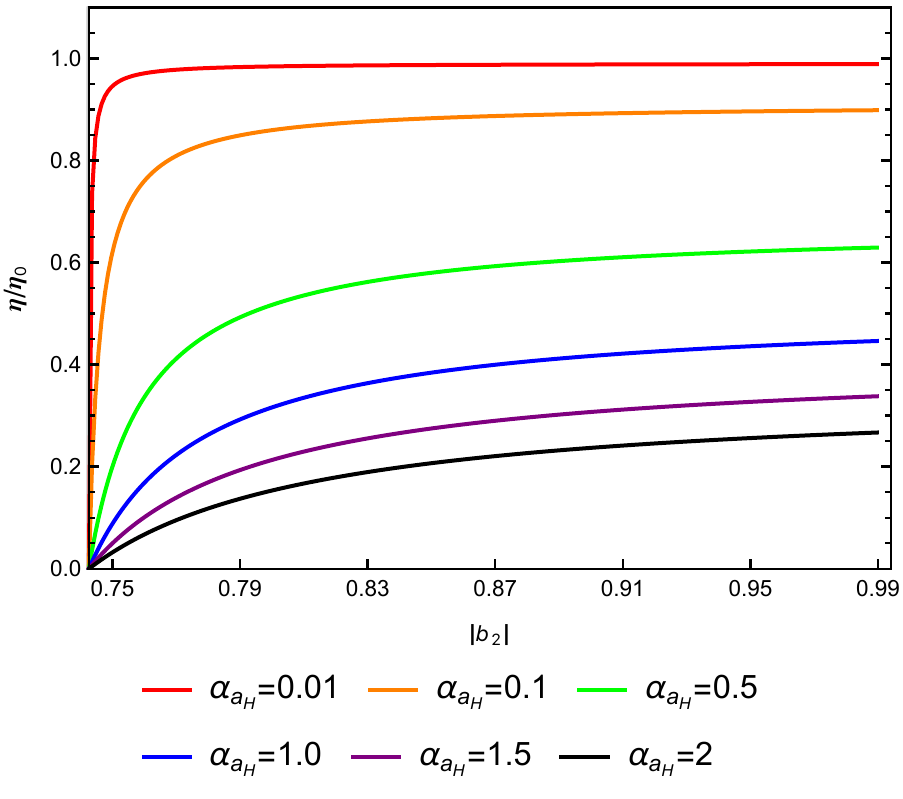}}      \caption{$\eta/\eta_{0}$ is plotted with respect to $|b_{2}|$ for different values of $\alpha_{a_{H}}$ with $\mathpzc{W}=0.2$. Plots (a) and (b) are for $\mathpzc{A}=0.5$ and $\mathpzc{A}=5.0$, respectively. }
  \label{fig:AP-NM-eta}
\end{figure}

\begin{figure}[h!]
    \centering
         \subfigure[]{\includegraphics[width=0.46\textwidth]{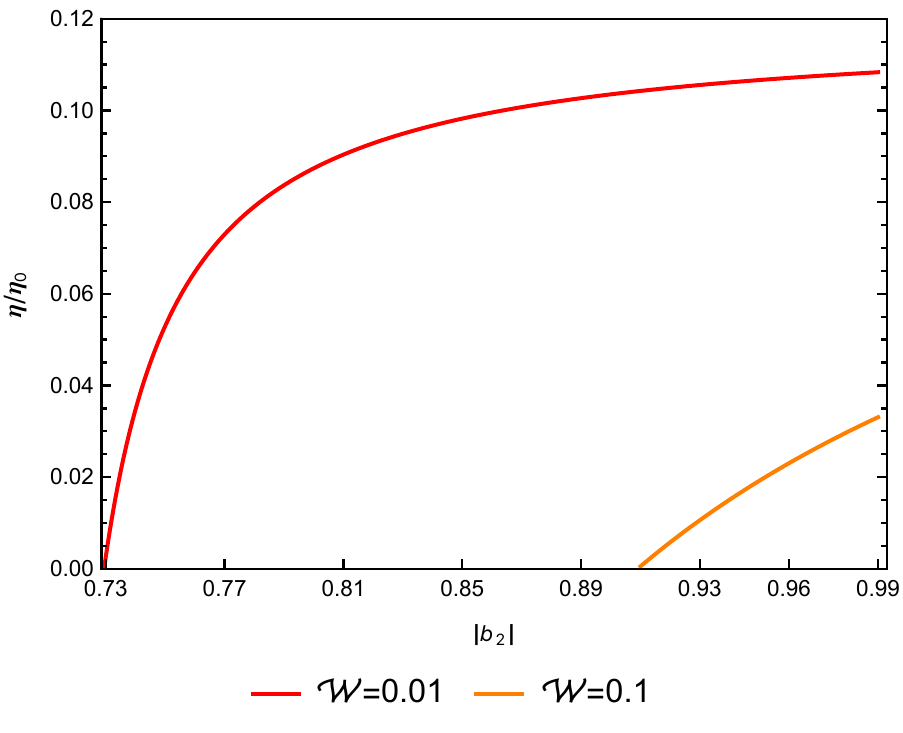}}
    \subfigure[]{\includegraphics[width=0.5\textwidth]{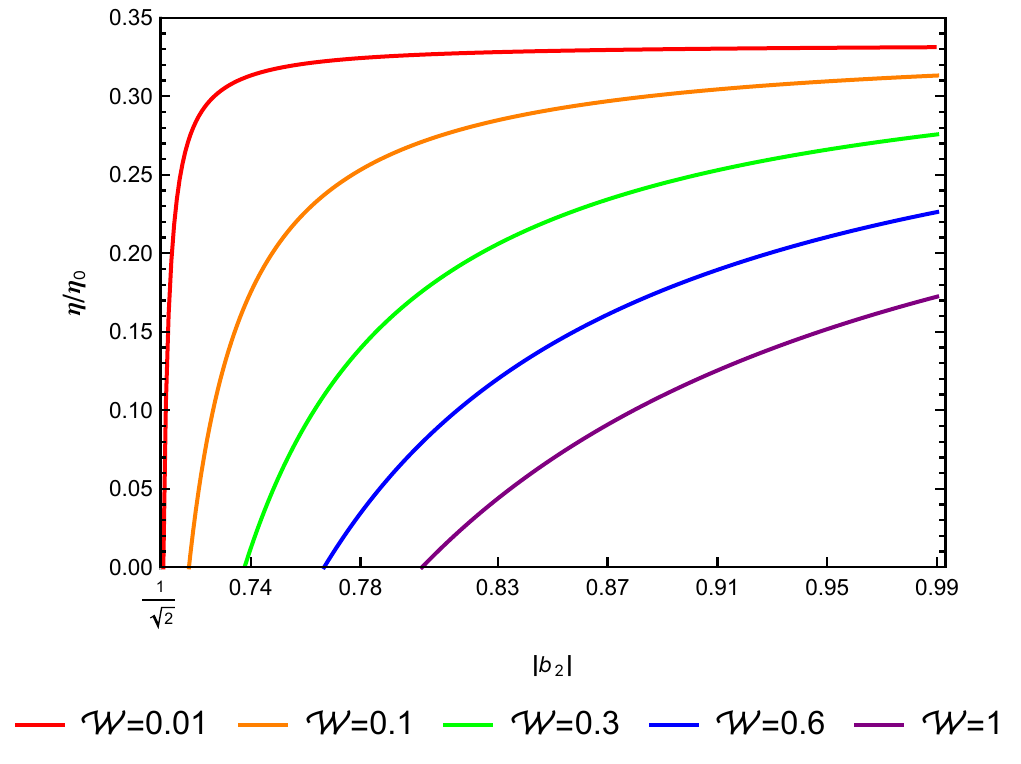}}\\
    \subfigure[]{\includegraphics[width=0.46\textwidth]{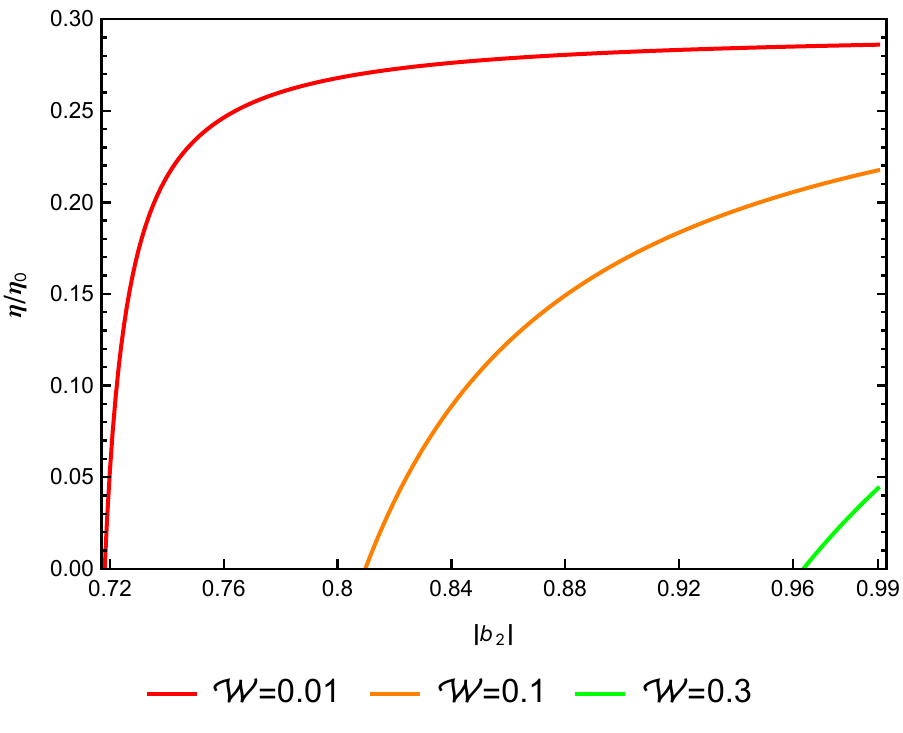}}
    \subfigure[]{\includegraphics[width=0.5\textwidth]{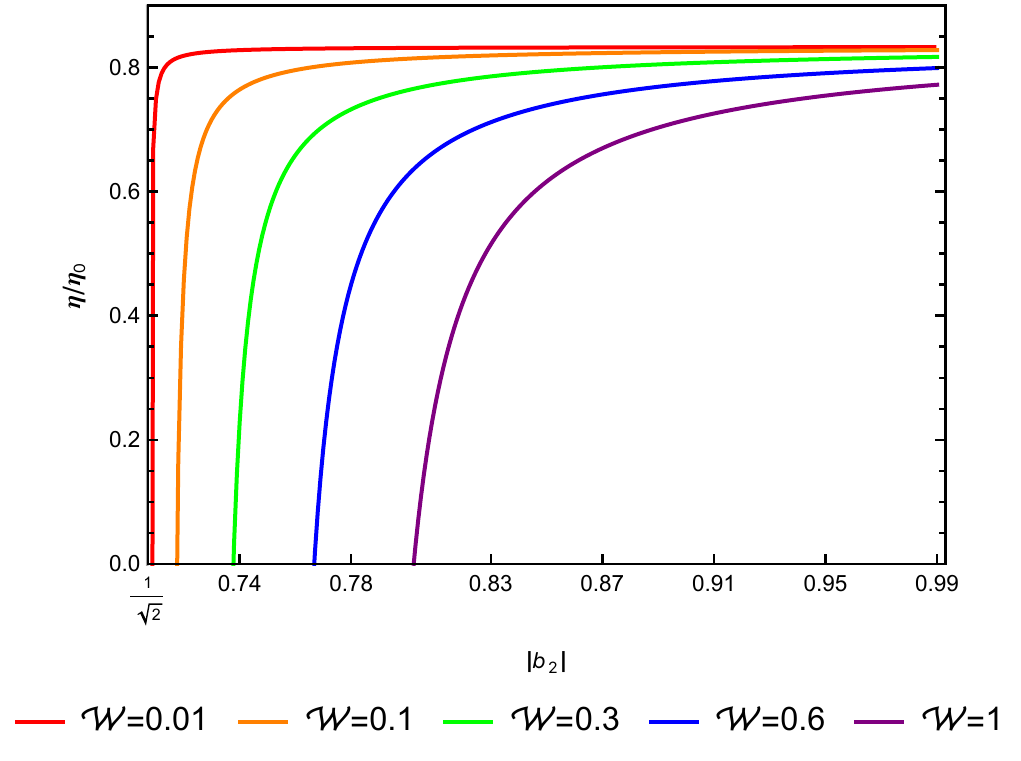}}
  \caption{$\eta/\eta_{0}$ is plotted with respect to $|b_{2}|$ for different values of $\mathpzc{W}$ when the detectors' in anti-parallel motion: For $\alpha_{a_{H}}=2.0$, plots (a) and (b) are showing this for $\mathpzc{A}=0.5$  and $\mathpzc{A}=10.0$, respectively. For $\alpha_{a_{H}}=0.2$, plots (c) and (d) are showing the same for $\mathpzc{A}=0.5$ and $\mathpzc{A}=10.0$, respectively.}
  \label{fig:AP-NM-eta2}
\end{figure}

The above analysis, through the figures  \ref{fig:APNM-al}, \ref{fig:APNM-b2} and \ref{fig:APNM-b2a}, provides the space of parameters where our Otto cycle is possible. We will now investigate the nature of efficiency with respect to strength of initial entanglement within these available values of parameters.  
 For this in Fig. \ref{fig:AP-NM-eta}, we have shown variation of the quantity $\eta/\eta_{0}$ with respect to $b_{2}$ for different values of $\alpha_{a_{H}}$ with a fixed $\mathpzc{W}=0.2$. Here in sub-figures (a) and (b), we have used $\mathpzc{A}=0.5$ and $\mathpzc{A}=5.0$, respectively. Similarly in Fig. \ref{fig:AP-NM-eta2}, the variation of $\eta/\eta_{0}$ with respect to $b_{2}$ is shown for different values of $\mathpzc{W}$ with fixed $\alpha_{a_{H}}$ values.
 Here sub-figures (a) and (b) are showing the same with $\alpha_{a_{H}}=2.0$ for $\mathpzc{A}=0.5$ and $\mathpzc{A}=10.0$, respectively. Whereas sub-figures (c) and (d) are corresponding to $\alpha_{a_{H}}=0.2$ for $\mathpzc{A}=0.5$ and $\mathpzc{A}=10.0$, respectively. All these plots show that the value of $\eta/\eta_{0}$ is increasing as $b_{2}$ is moving away from $b_{2}=1/\sqrt{2}$, but never exceeding $one$. This implies that the efficiency of our EUQOE is always less than that of UQOE. Moreover, this efficiency increases if the initial composite state moves away from the maximally entangled state. Therefore it is likely that ``the entanglement suppresses the efficiency of the Otto cycle and as initial entanglement decreases, the efficiency gradually grows''. In addition it is to be observed in Fig. \ref{fig:AP-NM-eta} that for a given value of $\mathpzc{W}$ the engine is becoming less efficient as relative acceleration $\alpha_{a_{H}}$ between the detectors increases. Similarly Fig. \ref{fig:AP-NM-eta2} indicates that for a given $\alpha_{a_{H}}$ the engine losses its efficiency with the increase of $\mathpzc{W}$. 
\begin{table}[h]
\begin{center}
\caption{The values of $\epsilon_{0}$ and corresponding values of Tr$(\delta\rho^{H}h_{\alpha_{v}})$ from Fig. \ref{fig:APNM-ep}.}
\label{table:ep}
\begin{tabular}{|c|>{\centering\arraybackslash}m{2.cm}|>{\centering\arraybackslash}m{2.3cm}|>{\centering\arraybackslash}m{3cm}|>{\centering\arraybackslash}m{3cm}|}
\hline\hline 
  $\mathpzc{W}$ & $\alpha_{a_{H}}$ & $\mathpzc{A }$ &  $\epsilon_{0}$ & Tr$(\delta\rho^{H}h_{\alpha_{v}})$\\ [1ex] 
\hline
		&	2	&	10&		0.0104596	&	0.0000462253		\\[.8ex]
  	0.1	&	2	&	20&		0.00546493	&	0.0000241518		\\[.8ex]
		&	2	&	30&		0.00367563	&	0.0000162441		\\[.8ex]
		&	2	&	40&		0.00276543	&	0.0000122216		\\[.8ex]
		&	2	&	50&		0.0022156	&	9.79166×$10^{-6}$	\\[.8ex]
 \hline
		&	2	&	10	&	0.00104626	&	 4.62368×$10^{-7}$	\\[.8ex]
	0.01	&	2	&	20	&	0.000546536	&	2.41537 ×$ 10^{-7}$	\\[.8ex]
		&	2	&	30	&	0.000367576	&	1.62447×$10^{-7}$	\\[.8ex]
		&	2	&	40	&	0.000276548	&	1.22218×$10^{-7}$	\\[.8ex]
		&	2	&	50	&	0.000221563	&	9.7918×$10^{-8}$	\\[.8ex]
\hline
\end{tabular}
\end{center}
\end{table}

\section{Conclusion}\label{sec:conclu}
We have investigated the possibility of construction of a EUQOE with different initial entangled states between two qubits (taken as monopole detectors), which was recently initiated in \cite{PhysRevD.104.L041701}. The thermal bath has been mimicked by uniformly accelerating the detectors, either in parallel or in anti-parallel motion. As mentioned in the introduction, there were few limitations and incompleteness in the initial attempt. The present discussion raised above those limitations and provided a complete study of constructing a EUQOE. 

We found that making of EUQOE is not possible for all cases. There are only one situation when all necessary conditions for making an Otto engine is satisfied. This is when the detectors are in anti-parallel motion with the initial detector's states are non-maximally entangled states. In this case we got some continuous range of values of acceleration of first detector  $(\mathpzc{A})$, for some fixed values of the system's energy gap and ratio of the accelerations of the detectors. In the allowed  $\mathpzc{A}$ range, the efficiency first increasing and then approaching a constant value with respect to $\mathpzc{A}$, however the efficiency never goes beyond $\eta_0$. Thus entangled Otto engines are not fruitful in terms of efficiency of the engine. In other words, the entanglement causes suppression of efficiency. In fact, as the detectors' initial state is approaching towards the maximally entangled state (in allowed $b_{2}$ range), efficiency of the cycle deteriorates sharply and becomes $zero$. For all other cases making of EUQOE is not possible. We summarise these outcomes in Table \ref{table:cases}. 
\begin{table}[h]
\begin{center}
\caption{Summery of possible scenarios of EUQOE.}
\label{table:cases}
\begin{tabular}{|c|>{\centering\arraybackslash}m{6cm}|>{\centering\arraybackslash}m{2.3cm}|>{\centering\arraybackslash}m{3cm}|}
\hline\hline 
  Motion & Initial States & Satisfying all criteria? & Can $\eta/\eta_{0}$ be greater than one?\\ [1ex] 
\hline
				&	Maximally entangled, Symmetric		&	No	&	-	\\[.8ex]
  	Parallel		&	Maximally entangled, Anti-symmetric 	&	No	&	-	\\[.8ex]
				&	Non-maximally entangled				&	No	&	-	\\[.8ex]
 \hline
				&	Maximally entangled, Symmetric 		&	No	&	- 	\\[.8ex]
	Anti-parallel	&	Maximally entangled, Anti-symmetric		&	No	&	-	\\[.8ex]
				&	Non-maximally entangled				&	Yes	&	No	\\[.8ex]
\hline
\end{tabular}
\end{center}
\end{table}

Couple of comments are in order. In the original QOE, one provides a real thermal bath and therefore the existence of this bath is classical in nature. While our EUQOE (also UQOE, proposed in \cite{Arias:2018wc,Gray:2018uw}) is a pure quantum mechanical in construction in the sense that the thermal bath itself is provided through a pure quantum (plus relativistic) effect(s). Finally, it need to be mentioned that the efficiency here can be regulated by changing the observer's acceleration, ratio of detector $A$ to detector $B$'s acceleration and the energy gap of the detectors' system. Such a unique feature of EUQOE can be very important for experimental verification of the Unruh effect as this phenomenon is the main input in our analysis. In this study we discussed the possible scenario to setup an Otto cycle. We also provided an example range of parameter values where making of Otto cycle is possible. We hope that with the use of those values of the system parameters one will be able to construct an experimental apparatus which will also be able to verify the existence  of Unruh phenomenon.

\vskip 4mm
\noindent
{\bf Acknowledgments:}
DB would like to acknowledge Ministry of Education, Government of India for providing financial support for his research via the PMRF May 2021 scheme. The research of BRM is supported by a START-UP RESEARCH 
GRANT (No. SG/PHY/P/BRM/01) from the Indian Institute of Technology Guwahati, 
India and by a Core Research Grant (File no. CRG/2020/000616) from Science and Engineering Research Board (SERB), Department of Science $\&$ Technology (DST), Government of India.



\begin{appendix}
\section*{Appendices}
\numberwithin{equation}{section}

\section{Relation between the proper times of the detectors}\label{A}

The uniformly accelerated detectors follow hyperbolic paths on the Minkowski spacetime. These trajectories for different accelerations never meet each other except at null past $(t=-x)$ and null future $(t=x)$. Whereas the $(t/x) =$ constant lines are straight lines on the $x-t$ plane, passing through the origine, and one such line meets all hyperbola trajectories. Therefore all the accelerated frames will asign same value of $(t/x)$ ($= C_0$, say) where a particular $(t/x) = C_0$ line meets all the corresponding trajectories. On the other hand the proper times of our present detectors are extented from $-\infty$ to $+\infty$ and so they starts simultaneouly from the null past horizon ($t/x=-1$) and reach at future null horizon ($t/x=1$) again simultaneouly. Also if both the detectors starts from same $t/x = $ constant line, then at each moment they will measure the same $t/x$, but different from the initial value. This simple setup is very useful to find a relation between the indivitual proper times of the frames. Moreover usually the relation between proper times of two different frames are done by using a quantity which is ``numerically'' same for both the frames (like the infinitesimal spacetime distant element $dS^2$). The above discussion implies that a particular $t/x = $ constant ($C_0$, say) line therefore can be used to connect the two hyperbolic trajectories, in particular to find the relation between the proper times of these frames. The same idea and simpliest situation was adopted in \cite{RODRIGUEZCAMARGO2018266} (see the start of the section $2$ of this reference) as well. Here also we will seek the relation by considering the same idea.

Two detectors $A$ and $B$ that are uniformly accelerating in the same Rindler wedge (namely RRW) with accelerations $a_{A}$ and $a_{B}$, respectively. Their trajectories in terms of their proper times are determined by the relation between the Minkowski coordinates and accelerated frame proper time:
\begin{eqnarray}\label{accelerationParallel}
&t_{A}=\frac{1}{a_{A}}\sinh(a_{A}\tau_{A});~~~x_{A}=\frac{1}{a_{A}}\cosh(a_{A}\tau_A)~,\no\\&t_{B}=\frac{1}{a_{B}}\sinh(a_{B}\tau_{B});~~~x_{B}=\frac{1}{a_{B}}\cosh(a_{B}\tau_{B})~.
\end{eqnarray}
Now using the fact that $t/x$ is same for both these frames yields a relation between the proper times of the detectors as
\begin{equation}\label{tAtB}
\tau_{B}=\alpha_{a}\tau_{A}~.
\end{equation}
This yields the required relation between the proper times of the detectors as (\ref{B5}) with $\alpha = \alpha_a = (a_{A}/a_{B})$.

For anti-parallelly accelerating detectors $A$ and $B$, with accelerations $a_{A}$ and $-a_{B}$, respectively,  then their trajectories in terms of their proper times are given by
\begin{eqnarray}\label{accelerationAntiParallel}
&t_{A}=\frac{1}{a_{A}}\sinh(a_{A}\tau_{A});~~~x_{A}=\frac{1}{a_{A}}\cosh(a_{A}\tau_A)~,\no\\&t_{B}=\frac{1}{a_{B}}\sinh(a_{B}\tau_{B});~~~x_{B}=-\frac{1}{a_{B}}\cosh(a_{B}\tau_{B})~.
\end{eqnarray}
For detectors to be described by constant $t/x$ or $\eta$, we obtain the following relation between the detectors' proper times
\begin{equation}\label{tAtB-}
\tau_{B}=-\alpha_{a}\tau_{A}~.
\end{equation}
In this case, therefore we have $\tau_B = \alpha \tau_a$ where $\alpha = -\alpha_{a}=-(a_{A}/a_{B})$. 

If the detectors are moving with constant velocities and have relative velocity $v_{rel}$. Then we can have $\tau_{B}=\alpha_{v}\tau_{A}$, where
\begin{eqnarray}
\alpha_{v}=\sqrt{1-v_{rel}^{2}}~.
\end{eqnarray}
 For the detectors are moving with same constant velocity, then $v_{rel}=0$ and hence $\alpha_{v}=1$.

\section{Calculation of the late time density matrix}\label{B}
In this  Appendix, we will briefly calculate the change in the density matrix ($\delta\rho^{}$) of two detectors in stage II or stage IV, where two detectors uniformly accelerate and interacts with the background massless real scalar quantum field through monopole coupling. This calculation is for both heating process ($H$) and cooling process ($C$). Therefore we drop the subscripts $H$ or $C$, used in the main text and keep it as a general discussion. Also the energy gap of the combined system is taken as $\omega$ (instead of $\omega_{1}$ or $\omega_{2}$ used in the main text). This interaction is governed by the action given by
\begin{equation}
		S_{int}=\int c_{A}\chi_A(\tau_A)m_A(\tau_A){\phi}(\bar{x}_A)d\tau_A+\int c_{B}\chi_B(\tau_B)m_B(\tau_B){\phi}(\bar{x}_B)d\tau_B.
	\end{equation}
We choose the initial state of the combined system in the asymptotic past as
	\begin{equation}
		|in\rangle=|0\rangle|D\rangle=|0\rangle[b_1|e_{A}g_{B}\rangle+b_2|g_{A}e_{B}\rangle],
	\end{equation}
	where $|0\rangle$ is the vacuum of the quantum field in tensor product with the entangled state of the detectors. In the above we will choose $b_1=(1/\sqrt{2}) = b_2$ for $|s\ra$,  $b_1=(1/\sqrt{2})$, $b_2 = -(1/\sqrt{2})$ for $| a\ra$ and $b_{1}\neq1/\sqrt{2}~(0<b_{1}<1),~b_{2}=\pm\sqrt{1-b_{1}^{2}}$ for non-maximally entangled states in the main analysis of our paper. At the asymptotic future, the evolved quantum state can be given by \cite{book:Peskin}
	\begin{equation}\label{densityMFinal}
		|out\rangle=Te^{iS_{int}}|in\rangle~.
	\end{equation}
Here $T$ denotes the time order product of the operators.	

The reduced density matrix of the detectors can be obtained by tracing out the field part of the final density matrix \cite{PhysRevA.97.062338}
		\begin{equation}\label{densityMFinalR}
		\rho_{AB}=\textrm{Tr}_{\phi}|out\rangle\langle out|.
	\end{equation}
The elements of the reduced density matrix are evaluated from the  Dyson series expansion of (\ref{densityMFinal}) upto second order in $c_{j}$ inside (\ref{densityMFinalR}):
	\begin{eqnarray}
			&&\langle {{n}_A} {{n}_B}| \rho_{AB}|{\hat{n}_A\hat{n}_B}\rangle\no\\&=&\textrm{Tr}_{\phi}\langle {{n}_A}{{n}_B}| 
			[1+iS_{int}-T(S_{int}S'_{int}/2)]
			[b_1|e_{A}g_{B}\rangle+b_2|g_{A}e_{B}\rangle]|0\rangle
			\langle0|[b_{1}^{\star}\langle e_{A}g_{B}|+b_{2}^{\star}\langle g_{A}e_{B}|]\no\\&&
			[1-iS_{int}-T(S_{int}S'_{int})/2]
			|{\hat{n}_A}{\hat{n}_B}\rangle\no\\
			&=&\langle {{n}_A}{{n}_B}|[b_{1}|e_{A}g_{B}\rangle+b_{2}|g_{A}e_{B}\rangle][b_{1}^{\star}\langle e_{A}g_{B}|+b_{2}^{\star}\langle g_{A}e_{B}|]	|{\hat{n}_A}{\hat{n}_B}\rangle	\no\\&&
			+\textrm{Tr}_{\phi}\langle {{n}_A} {{n}_B}|S_{int}	|0\rangle[b_{1}|e_{A}g_{B}\rangle+b_{2}|g_{A}e_{B}\rangle]
		[b_{1}^{\star}\langle e_{A}g_{B}|+b_{2}^{\star}\langle g_{A}e_{B}|]\langle0|S_{int}	|{\hat{n}_A}{\hat{n}_B}\rangle\no\\&&
				-\textrm{Tr}_{\phi}\langle {{n}_A} {{n}_B}|T[S_{int}S'_{int}/2]	|0\rangle\langle0|[b_{1}|e_{A}g_{B}\rangle+b_{2}|g_{A}e_{B}\rangle]
		[b_{1}^{\star}\langle e_{A}g_{B}|+b_{2}^{\star}\langle g_{A}e_{B}|]	|{\hat{n}_A}{\hat{n}_B}\rangle\no\\&&
				-\textrm{Tr}_{\phi}\langle {n}_{A} {n}_{B}| [b_{1}|e_{A}g_{B}\rangle+b_{2}|g_{A}e_{B}\rangle]
			[b_{1}^{\star}\langle e_{A}g_{B}|+b_{2}^{\star}\langle g_{A}e_{B}|]|0\rangle\langle0|T[S_{int}S_{int}^{\p}/2]	|{\hat{n}_A}{\hat{n}_B}\rangle\no\\&=& \rho_{0}+{\underbrace{[R_{n_An_B,\hat{n}_A\hat{n}_B}^{(1)}+R_{n_An_B,\hat{n}_A\hat{n}_B}^{(2)}+R_{\hat{n}_A\hat{n}_B,n_An_B}^{(2)\star}]}_{O(c^{2})}}+O(c^3),
	\end{eqnarray}
where $\rho_{0}$ is the initial density matrix in the bases of $|e_{A}e_{B}\ra$, $|e_{A}g_{B}\ra$, $|g_{A}e_{B}\ra$ and $|g_{A}g_{B}\ra$, given by
\begin{equation}
	\rho_{0}=\begin{pmatrix}
		0&0&0&0\\
		0&|b_{1}|^2&b_{1}b_{2}^{\star}&0\\
		0&b_{2} b_{1}^{\star}&|b_{2}|^2&0\\
		0&0&0&0
	\end{pmatrix}.
\end{equation}
Terms first order in $c_{j}$ vanishes individually due to the trace operation $(\la0_{M}|\phi(\bar{x}_{i})|0_{M}\ra=0)$ and the terms second order in $c_{j}$ are calculated as follows: 
\begin{eqnarray}
R_{n_An_B,\hat{n}_A\hat{n}_B}^{(1)}&=&\textrm{Tr}_{\phi}\langle {{n}_A} {{n}_B}|S_{int}	[b_{1}|e_{A}g_{B}\rangle+b_{2}|g_{A}e_{B}\rangle]|0\rangle
		\langle0|[b_{1}^{\star}\langle e_{A}g_{B}|+b_{2}^{\star}\langle g_{A}e_{B}|]S_{int}	|{\hat{n}_A}{\hat{n}_B}\rangle\no\\&&=
\textrm{Tr}_{\phi}\langle {{n}_A} {{n}_B}|\Big(\int c_A\chi_A(\tau_A)m_A(\tau_A)\phi(\bar{x}_A)d\tau_A+\int c_B\chi_B(\tau_B)m_B(\tau_B)\phi(\bar{x}_B)d\tau_B\Big)\no\\&&
		[b_{1}|e_{A}g_{B}\rangle+b_{2}|g_{A}e_{B}\rangle]	|0\rangle
			\langle0|[b_{1}^{\star}\langle e_{A}g_{B}|+b_{2}^{\star}\langle g_{A}e_{B}|]\no\\&&\Big(\int c_A\chi_A(\tau_A)m_A(\tau_A)\phi(\bar{x}_A)d\tau_A+\int c_B\chi_B(\tau_B)m_B(\tau_B)\phi(\bar{x}_B)d\tau_B\Big)|{\hat{n}_A}{\hat{n}_B}\rangle\no\\&=&
\textrm{Tr}_{\phi}\Big(\int d\tau_{A}c_A \chi_{A}(\tau_{A})\langle {{n}_A} {{n}_B}|m_A(\tau_{A})[b_{1}|e_{A}g_{B}\rangle+b_{2}|g_{A}e_{B}\rangle]\phi(\bar{x}_A)|0\rangle\no
\\&&+\int d\tau_{B}c_B \chi_{B}(\tau_{B})\langle {{n}_A} {{n}_B}|m_B(\tau_{B})[b_{1}|e_{A}g_{B}\rangle+b_{2}|g_{A}e_{B}\rangle] \phi(\bar{x}_B)|0\rangle\Big)\no\\&&
\Big( \int d\tau_{A}^{\prime}c_A\chi_A(\tau_A^{\prime})[b_{1}^{\star}\langle e_{A}g_{B}|+b_{2}^{\star}\langle g_{A}e_{B}|]m_A(\tau_A^{\prime})|{\hat{n}_A}{\hat{n}_B}\rangle\langle0|\phi(\bar{x}_A^{\prime})\no\\&&
+\int d\tau_{B}^{\prime}c_{B}\chi_B(\tau_B^{\prime})[b_{1}^{\star}\langle e_{A}g_{B}|+b_{2}^{\star}\langle g_{A}e_{B}|]m_B(\tau_B^{\prime})|{\hat{n}_A}{\hat{n}_B}\rangle	\langle0|\phi(\bar{x}_B^{\prime})
\Big)\no\\&=&
\textrm{Tr}_{\phi}\Big[\int d\tau_{A}c_A \chi_{A}(\tau_{A})	[b_{1}e^{i(g_{A}-e_{A})\tau_{A}}\delta_{n_{A},g_{A}} \delta_{n_{B},g_{B}}+b_{2}e^{i(e_{A}-g_{A})\tau_{A}}\delta_{n_{A},e_{A}} \delta_{n_{B},e_{B}}] \phi(\bar{x}_A)|0\rangle
\no\\&&+\int d\tau_{B}c_B \chi_{B}(\tau_{B})[b_{1}e^{i(e_{B}-g_{B})\tau_{B}}\delta_{n_{A},e_{A}} \delta_{n_{B},e_{B}}+b_{2}e^{i(g_{B}-e_{B})\tau_{B}}\delta_{n_{A},g_{A}} \delta_{n_{B},g_{B}}
] \phi(\bar{x}_B)|0\rangle\Big]\no\\&&
\Big[ \int d\tau_{A}^{\prime}c_A\chi_A(\tau_A')[b_{1}^{\star}e^{i(e_{A}-g_{A})\tau_A'}\delta_{\hat{n}_{A},g_{A}}\delta_{\hat{n}_{B},g_{B}} +b_{2}^{\star}e^{i(g_{A}-e_{A})\tau_A'}\delta_{\hat{n}_{A},e_{A}}\delta_{\hat{n}_{B},e_{B}} 
]\langle0|\bar{\phi}(\bar{x}_A')\no\\&&
+\int d\tau_{B}^{\prime}c_B\chi_B(\tau_B')[b_{1}^{\star}e^{i(g_{B}-e_{B})\tau_B'}\delta_{\hat{n}_{A},e_{A}}\delta_{\hat{n}_{B},e_{B}} 
 +b_{2}^{\star}e^{i(e_{B}-g_{B})\tau_B'}\delta_{\hat{n}_{A},g_{A}}\delta_{\hat{n}_{B},g_{B}} 
]\langle0|\bar{\phi}(\bar{x}_B')
\Big]\no\\&&\equiv AA_{n_An_B,\hat{n}_A\hat{n}_B}+AB_{n_An_B,\hat{n}_A\hat{n}_B}+BB_{n_An_B,\hat{n}_A\hat{n}_B}+BA_{n_An_B,\hat{n}_A\hat{n}_B},
\end{eqnarray}
where $e_{j}=\omega/2$ and $g_{j}=-\omega/2$, are the energy levels of the individual detectors.  The positive frequency Wightman function, $G^{+}(\tau_{A}, \tau_{B})$ is obtained as
\begin{eqnarray}
G^{+}(\tau_{A}, \tau_{B})&=&\textrm{Tr}_{\phi}\Big[\phi(\bar{x}_{B})|0\ra\la0|\phi(\bar{x}_{A})\Big]\no\\&=&
\sum_{\textrm{all field states}(n)} \la{n}|\phi(\bar{x}_{B})|0\ra\la0|\phi(\bar{x}_{A})|n\ra \no\\&=&
\la0|\phi(\bar{x}_{A})\Big(\sum_{n}|n\ra\la{n}|\Big)\phi(\bar{x}_{B})|0\ra\no\\&=&\la0|\phi(\bar{x}_{A})\phi(\bar{x}_{B})|0\ra
\end{eqnarray}
where {\it completeness} identity of the field states has been used in the last equality.
We have
\begin{eqnarray}\label{R1-AA1}
	AA&=&\int \int c_A^2 \chi_{A}(\tau_{A}) \chi_{A}(\tau_{A}')d\tau_{A}d\tau_{A}' G^{+}(\bar{x}_A',\bar{x}_A)\begin{pmatrix}
	|b_{2}|^2e^{i\omega(\tau_{A}-\tau_{A}')}&0&0&b_{2} b_{1}^{\star}e^{i\omega(\tau_{A}+\tau_{A}')}\\
	0&0&0&0\\
	0&0&0&0\\
	b_{1}b_{2}^{\star}e^{i\omega(-\tau_{A}-\tau_{A}')}&0&0&|b_{1}|^2e^{i\omega(-\tau_{A}+\tau_{A}')}\\
	\end{pmatrix}\no\\
	&=&\begin{pmatrix}
	|b_{2}|^2\mathcal{P}_A(\omega)&0&0&b_{2}b_{1}^{\star}\mathcal{P}_{AA}(\omega)\\ 0 &0&0&0\\0&0&0&0\\
	b_{1}b_{2}^{\star}\mathcal{P}_{AA}(-\omega)&0&0&|b_{1}|^2\mathcal{P}_{A}(-\omega)\\
	\end{pmatrix},
\end{eqnarray}
\begin{eqnarray}\label{R1-BB1}
BB&=&\int \int c_B^2 \chi_{B}(\tau_{B}) \chi_{B}(\tau_{B}')d\tau_{B}d\tau_{B}' G^{+}(\bar{x}_B',\bar{x}_B)\begin{pmatrix}
	|b_{1}|^2e^{i\omega(\tau_{B}-\tau_{B}')}&0&0&b_{1} b_{2}^{\star} e^{i\omega(\tau_{B}+\tau_{B}')}\\
	0&0&0&0\\0&0&0&0\\
	b_{2}b_{1}^{\star}e^{i\omega(-\tau_{B}-\tau_{B}')}&0&0& |b_{2}|^2 e^{i\omega(-\tau_{B}+\tau_{B}')}
\end{pmatrix}\no\\&=&\begin{pmatrix}
|b_{1}|^2\mathcal{P}_B(\omega)&0&0&b_{1}b_{2}^{\star}\mathcal{P}_{BB}(\omega)\\0&0&0&0\\0&0&0&0\\
b_{2}b_{1}^{\star} \mathcal{P}_{BB}(-\omega)&0&0&|b_{2}|^2 \mathcal{P}_{B}(-\omega)\\
\end{pmatrix},
\end{eqnarray}
\begin{eqnarray}\label{R1-AB1}
AB&=&\int\int c_Ac_B \chi_{A}(\tau_{A}) \chi_{B}(\tau_{B}')d\tau_{A}d\tau_{B}'G^{+}(\bar{x}_B',\bar{x}_A)\begin{pmatrix}
		b_{2} b_{1}^{{\star}} e^{i\omega(\tau_{A}-\tau_{B}')}&0&0&|b_{2}|^2 e^{i\omega(\tau_{A}+\tau_{B}')}\\
		0&0&0&0\\
		0&0&0&0\\
		|b_{1}|^2 e^{-i\omega(\tau_{A}+\tau_{B}')}&0&0&b_{1} b_{2}^{{\star}} e^{-i\omega(\tau_{A}-\tau_{B}')}
	\end{pmatrix}\no\\&=&\begin{pmatrix}
		b_{2} b_{1}^{\star}	\mathcal{P}_{AB}(\omega,-\omega)&0&0&|b_{2}|^2 \mathcal{P}_{AB}(\omega,\omega)\\
		0&0&0&0\\
		0&0&0&0\\
		|b_{1}|^2\mathcal{P}_{AB}(-\omega,-\omega)&0&0&b_{1}b_{2}^{\star} \mathcal{P}_{AB}(-\omega,\omega)
	\end{pmatrix},
\end{eqnarray}
\begin{eqnarray}\label{R1-BA1}
BA&=&	\int\int c_Ac_B \chi_{A}(\tau_{A}') \chi_{B}(\tau_{B})d\tau_{A}'d\tau_{B} G^{+}(\bar{x}_A',\bar{x}_B)\begin{pmatrix}
b_{1} b_{2}^{\star} e^{-i \omega(\tau_{A}'-\tau_{B})}&0&0& |b_{1}|^2 e^{i \omega(\tau_{A}'+\tau_{B})}\\
0&0&0&0\\
0 &0&0&0\\
|b_{2}|^2 e^{-i \omega(\tau_{A}'+\tau_{B})}&0&0&b_{2} b_{1}^{\star} e^{i \omega(\tau_{A}'-\tau_{B})}\\
		\end{pmatrix}\no\\&=&\begin{pmatrix}
b_{1}b_{2}^{\star}\mathcal{P}_{AB}^{\star}(\omega,- \omega)&0&0&|b_{1}|^2\mathcal{P}_{AB}^{\star}(- \omega,- \omega)\\
0&0&0&0\\
0&0& 0& 0\\
|b_{2}|^2\mathcal{P}_{AB}^{\star}(\omega, \omega)&0&0&b_{2}b_{1}^{\star}\mathcal{P}_{AB}^{\star}(- \omega, \omega)
		\end{pmatrix},
\end{eqnarray}
where $\mathcal{P}_j(\pm \omega)$ is defined by (\ref{Pi-Def}). We denote 
 \begin{equation}\label{Pii}
 \mathcal{P}_{jj}(\pm \omega)=c_j^2\int \int \chi_{j}(\tau_{j}) \chi_{j}(\tau_{j}')d\tau_{j}d\tau_{j}' G^{+}(\bar{x}_j',\bar{x}_j)e^{\pm i\omega(\tau_j+\tau'_j)},
 \end{equation}
 \begin{equation}\label{PAB}
\mathcal{P}_{AB}(\pm \omega,\pm \omega)=c_Ac_B\int\int{d\tau_A}{d\tau_B}e^{i(\pm \omega\tau_A\pm \omega\tau_B')}G^{+}(\bar{x}_B',\bar{x}_A).
\end{equation}
Using these definitions, we can identify that the matrix $BA$ as hermitian conjugate of matrix $AB$. \\

 The positive frequency Wightman function $G^{+}(\bar{x}^{\p},\bar{x})$ can be written as $G^{+}(\tau^{\p},\tau)$, satisfies the following properties
\begin{eqnarray}\label{greenPN}
G^{+}(-\tau_{i},- \tau_{j})=G^{+}(\tau_{j},\tau_{i}); ~~~~~G^{+}(\tau_{i}, \tau_{j})=[G^{+}(\tau_{j},\tau_{i})]^{\star}.
\end{eqnarray}
Using the above properties of the positive frequency Wightman function, we can check that $\mathcal{P}_{AB}$ (similarly, $\mathcal{P}_{j}$) is a real quantity, as
 \begin{eqnarray}\label{PABstr}
\mathcal{P}_{AB}(\pm \omega,\pm \omega)&=&c_Ac_B\int\int{d\tau_A}{d\tau_B}e^{i(\pm \omega\tau_A\pm \omega\tau_B')}G^{+}(\tau_B',\tau_A)\no\\&=&c_Ac_B\int\int{d\tau_A}{d\tau_B}e^{-i(\pm \omega\tau_A\pm \omega\tau_B')}G^{+}(-\tau_B',-\tau_A)\no\\&=&c_Ac_B\int\int{d\tau_A}{d\tau_B}e^{-i(\pm \omega\tau_A\pm \omega\tau_B')}[G^{+}(\tau_B^{\prime},\tau_A)]^{\star}\no\\&=&\mathcal{P}_{AB}^{\star}(\pm \omega,\pm \omega),
\end{eqnarray}
where after the second equality, we have changed $\tau_{j}\to-\tau_{j}$. Then, we have used the property of the Wightman function, given in (\ref{greenPN}).\\

The final expression of $R^{(1)}$ can be obtained by adding (\ref{R1-AA1}), (\ref{R1-BB1}), (\ref{R1-AB1}) and (\ref{R1-BA1}), as
\begin{eqnarray}\label{R1matrix}
R^{(1)}=\begin{pmatrix}
	|b_{2}|^2\mathcal{P}_A(\omega)+|b_{1}|^2\mathcal{P}_B(\omega)&0&0&b_{2}b_{1}^{\star}\mathcal{P}_{AA}(\omega)+b_{1}b_{2}^{\star}\mathcal{P}_{BB}(\omega)\\
	+b_{2} b_{1}^{\star}	\mathcal{P}_{AB}(\omega,-\omega)+b_{1}b_{2}^{\star}\mathcal{P}_{AB}(\omega,- \omega)&&&+|b_{2}|^2 \mathcal{P}_{AB}(\omega,\omega)+|b_{1}|^2\mathcal{P}_{AB}(- \omega,- \omega)\\
	 0 &0&0&0\\0&0&0&0\\
	b_{1}b_{2}^{\star}\mathcal{P}_{AA}(-\omega)+b_{2}b_{1}^{\star} \mathcal{P}_{BB}(-\omega)&0&0&|b_{1}|^2\mathcal{P}_{A}(-\omega)+|b_{2}|^2 \mathcal{P}_{B}(-\omega)\\
	+|b_{1}|^2\mathcal{P}_{AB}(-\omega,-\omega)+|b_{2}|^2\mathcal{P}_{AB}(\omega, \omega)&&&+b_{1}b_{2}^{\star} \mathcal{P}_{AB}(-\omega,\omega)+b_{2}b_{1}^{\star}\mathcal{P}_{AB}(- \omega, \omega)
\end{pmatrix},
\end{eqnarray}
where we have used the identity (\ref{PABstr}).

Next we calculate the following term:
\begin{eqnarray}
&&R_{n_An_B,\hat{n}_A\hat{n}_B}^{(2)}\no\\&&=-\textrm{Tr}_{\phi}\langle {{n}_A} {{n}_B}|T(S_{int}S_{int}^{\p}/2) |0\rangle\langle0|[b_{1}|e_{A}g_{B}\rangle+b_{2}|g_{A}e_{B}\rangle][b_{1}^{\star}\langle e_{A}g_{B}|+b_{2}^{\star}\langle g_{A}e_{B}|]|{\hat{n}_A}{\hat{n}_B}\rangle\no\\
			&&=-\langle {{n}_A} {{n}_B}|\langle0|T[\left(\int {}d\tau_{A}c_A\chi_{A}(\tau_{A})m_A(\tau_{A})\phi(\bar{x}_A)+\int{}d\tau_{B}c_B\chi_{B}(\tau_{B})m_B(\tau_{B})\phi(\bar{x}_B)\right)
			\Bigg(\int{}d\tau_{A}^{\p}c_A\chi_{A}(\tau_{A}^{\p})\no\\&&m_A(\tau_{A}^{\p})\phi(\bar{x}_A^{\p})+\int{}d\tau_{B}^{\p}c_B\chi_{B}(\tau_{B}^{\p})m_B(\tau_{B}^{\p})\phi(\bar{x}_B^{\p})\Bigg)/2]|0\rangle\scalebox{0.97}{$[b_1|e_{A}g_{B}\rangle+b_2|g_{A}e_{B}\rangle][b_1^{\star}\langle e_{A}g_{B}|+b_2^{\star}\langle g_{A}e_{B}|]|{\hat{n}_A}{\hat{n}_B}\rangle$}
			\no\\&&=-\langle {{n}_A} {{n}_B}|\langle0|T[\Bigg(c_A^2\int\int{}d\tau_{A}d\tau_{A}^{\p}\chi_{A}(\tau_{A})\chi_{A}(\tau_{A}^{\p})m_A(\tau_{A})m_A(\tau_{A}^{\p})\phi(x_A) \phi(\bar{x}_A^{\p})\no\\&&
			+c_B^2\int\int{}d\tau_{B}d\tau_{B}^{\p}\chi_{B}(\tau_{B})
			\chi_{B}(\tau_{B}^{\p})m_B(\tau_{B})m_B(\tau_{B}^{\p})\phi(x_B)\phi(\bar{x}_B^{\p})+
			c_Ac_B\int\int{}d\tau_{B}d\tau_{A}^{\p}\chi_{B}(\tau_{B})\chi_{A}(\tau_{A}^{\p})\no\\&&m_B(\tau_{B})m_A(\tau_{A}^{\p})\phi(\bar{x}_B)\phi(\bar{x}_A^{\p})+c_Ac_B
			\int\int{d\tau_{A}d\tau_{B}^{\p}m_A(\tau_{A})m_B(\tau_{B}^{\p})\chi_{A}(\tau_{A})\chi_{B}(\tau_{B}^{\p})\phi(\bar{x}_A)\phi(\bar{x}_B^{\p})}\Bigg)/2]|0\rangle\no\\&&[b_1|e_{A}g_{B}\rangle+b_2|g_{A}e_{B}\rangle][b_1^{\star}\langle e_{A}g_{B}|+b_2^{\star}\langle g_{A}e_{B}|]|{\hat{n}_A}{\hat{n}_B}\rangle
			\no\\&&[\text{Now assuming non-primed proper times are greater than primed proper times, i.e., $\tau>\tau'$}, \text{see} \cite{book:Peskin} ]
			\no\\&&=-\langle {{n}_A} {{n}_B}|[\left(c_A^2\int\int{d\tau_{A}d\tau_{A}^{\p}}\chi_{A}(\tau_{A})\chi_{A}(\tau_{A}^{\p})m_A(\tau_{A})m_A(\tau_{A}^{\p})\Theta(\tau_{A}-\tau_{A}^{\p}) G_W(\bar{x}_A,\bar{x}_A^{\p})\right.\no\\&&\left.
			+c_B^2\int\int{d\tau_{B}d\tau_{B}^{\p}}\chi_{B}(\tau_{B})\chi_{B}(\tau_{B}^{\p})m_B(\tau_{B})m_B(\tau_{B}^{\p})\Theta(\tau_{B}-\tau_{B}^{\p})G_W(\bar{x}_B,\bar{x}_B^{\p})\right.\no\\&&+\left.
			c_Ac_B\int\int{d\tau_{B}d\tau_{A}^{\p}}\chi_{B}(\tau_{B})\chi_{A}(\tau_{A}^{\p})m_B(\tau_{B})m_A(\tau_{A}^{\p})\Theta(\tau_{B}-\tau_{A}^{\p})G_W(\bar{x}_B,\bar{x}_A^{\p})\right.\no\\&&+\left.
			c_Ac_B\int\int{d\tau_{A}d\tau_{B}^{\p}}\chi_{A}(\tau_{A})\chi_{B}(\tau_{B}^{\p})m_A(\tau_{A})m_B(\tau_{B}^{\p})\Theta(\tau_{A}-\tau_{B}^{\p})G_W(\bar{x}_A,\bar{x}_B^{\p})\right)][b_1|e_{A}g_{B}\rangle+b_2|g_{A}e_{B}\rangle]\no\\&&[b_1^{\star}\delta_{\hat{n}_A,e_{A}}\delta_{\hat{n}_B,g_{B}}+b_2^{\star}\delta_{\hat{n}_A,g_{A}}\delta_{\hat{n}_B,e_{B}}]\no\\&&=
			-\langle {{n}_A} {{n}_B}|\left(c_A^2\int\int{d\tau_{A}d\tau_{A}^{\p}}\chi_{A}(\tau_{A})\chi_{A}(\tau_{A}^{\p})m_A(\tau_{A})m_A(\tau_{A}^{\p})[b_1|e_{A}g_{B}\rangle+b_2|g_{A}e_{B}\rangle]\Theta(\tau_{A}-\tau_{A}^{\p})G_W(\bar{x}_A,\bar{x}_A^{\p})\right.\no\\&&\left.
			+c_B^2\int\int{d\tau_{B}d\tau_{B}^{\p}}\chi_{B}(\tau_{B})\chi_{B}(\tau_{B}^{\p})m_B(\tau_{B})m_B(\tau_{B}^{\p})[b_1|e_{A}g_{B}\rangle+b_2|g_{A}e_{B}\rangle]\Theta(\tau_{B}-\tau_{B}^{\p})G_W(\bar{x}_B,\bar{x}_B^{\p})\right.\no\\&&+\left.
			c_Ac_B\int\int{d\tau_{B}d\tau_{A}^{\p}}\chi_{B}(\tau_{B})\chi_{A}(\tau_{A}^{\p})m_B(\tau_{B})m_A(\tau_{A}^{\p})[b_1|e_{A}g_{B}\rangle+b_2|g_{A}e_{B}\rangle]\Theta(\tau_{B}-\tau_{A}^{\p})G_W(\bar{x}_B,\bar{x}_A^{\p})\right.\no\\&&+\left.
			c_Ac_B\int\int{d\tau_{A}d\tau_{B}^{\p}}\chi_{A}(\tau_{A})\chi_{B}(\tau_{B}^{\p})m_A(\tau_{A})m_B(\tau_{B}^{\p})[b_1|e_{A}g_{B}\rangle+b_2|g_{A}e_{B}\rangle]\Theta(\tau_{A}-\tau_{B}^{\p})G_W(\bar{x}_A,\bar{x}_B^{\p})\right)\no\\&&[b_1^{\star}\delta_{\hat{n}_A,e_{A}}\delta_{\hat{n}_B,g_{B}}+b_2^{\star}\delta_{\hat{n}_A,g_{A}}\delta_{\hat{n}_B,e_{B}}]\no\\&&
			\scalebox{0.93}{$=-\Big(c_A^2\int\int{d\tau_{A}d\tau_{A}^{\p}}\chi_{A}(\tau_{A})\chi_{A}(\tau_{A}^{\p})[b_1e^{i\omega\tau_A}e^{-i\omega\tau_A^{\p}}\delta_{n_A,e_{A}}\delta_{n_B,g_{B}}+b_2e^{-i\omega\tau_A}e^{i\omega\tau_A^{\p}}\delta_{n_A,g_{A}}\delta_{n_B,e_{B}}]\Theta(\tau_{A}-\tau_{A}^{\p})G_W(\bar{x}_A,\bar{x}_A^{\p})$}\no\\&&+
			\scalebox{0.95}{$c_B^2\int\int{d\tau_{B}d\tau_{B}^{\p}}\chi_{B}(\tau_{B})\chi_{B}(\tau_{B}^{\p})[b_1e^{-i\omega\tau_B}e^{i\omega\tau_B^{\p}}\delta_{n_A,e_{A}}\delta_{n_B,g_{B}}+b_2 e^{i\omega\tau_B}e^{-i\omega\tau_B^{\p}}\delta_{n_A,g_{A}}\delta_{n_B,e_{B}}]\Theta(\tau_{B}-\tau_{B}^{\p})G_W(\bar{x}_B,\bar{x}_B^{\p})$}\no\\&&+
			\scalebox{0.95}{$c_Ac_B\int\int{d\tau_{B}d\tau_{A}^{\p}}\chi_{B}(\tau_{B})\chi_{A}(\tau_{A}^{\p})[b_1e^{i\omega\tau_B-i\omega\tau_A^{\p}}\delta_{n_A,g_{A}}\delta_{n_B,e_{B}}+b_2e^{-i\omega\tau_{B}+i\omega\tau_{A}^{\p}}\delta_{n_A,e_{A}}\delta_{n_B,g_{B}}]\Theta(\tau_{B}-\tau_{A}^{\p})G_W(\bar{x}_B,\bar{x}_A^{\p})$}\no\\&&
			\scalebox{0.94}{$+c_Ac_B\int\int{d\tau_{A}d\tau_{B}^{\p}}\chi_{A}(\tau_{A})\chi_{B}(\tau_{B}^{\p})[b_1e^{i\omega\tau_{B}^{\p}-i\omega\tau_{A}}\delta_{n_A,g_{A}}\delta_{n_B,e_{B}}+b_2e^{-i\omega\tau_{B}^{\p}+i\omega\tau_{A}}\delta_{n_A,e_{A}}\delta_{n_B,g_{B}}]\Theta(\tau_{A}-\tau_{B}^{\p})G_W(\bar{x}_A,\bar{x}_B^{\p})\Big)$}\no\\&&[b_{1}^{\star}\delta_{\hat{n}_A,e_{A}}\delta_{\hat{n}_B,g_{B}}+b_{2}^{\star}\delta_{\hat{n}_A,g_{A}}\delta_{\hat{n}_B,e_{B}}]\no\\
			&&
			\scalebox{0.95}{$=-\Big(c_{A}^{2}\int\int{d\tau_{A}d\tau_{A}^{\p}}\chi_{A}(\tau_{A})\chi_{A}(\tau_{A}^{\p})[b_{1}e^{i\omega(\tau_A-\tau_A^{\p})}\delta_{n_A,e_{A}}\delta_{n_B,g_{B}}+b_{2}e^{-i\omega(\tau_A-\tau_A^{\p})}\delta_{n_A,g_{A}}\delta_{n_B,e_{B}}] \Theta(\tau_{A}-\tau_{A}^{\p})G_W(\bar{x}_A,\bar{x}_A^{\p})$}
			\no\\&&
			+c_{B}^{2}\int\int{d\tau_{B}d\tau_{B}^{\p}}\chi_{B}(\tau_{B})\chi_{B}(\tau_{B}^{\p})\scalebox{0.95}{$[b_{1}e^{-i\omega(\tau_B-\tau_B^{\p})}\delta_{n_{A},e_{A}}\delta_{n_{B},g_{B}}+b_{2}e^{i\omega(\tau_B-\tau_B^{\p})}\delta_{n_A,g_{A}}\delta_{n_B,e_{B}}]\Theta(\tau_{B}-\tau_{B}^{\p})G_W(\bar{x}_B,\bar{x}_B^{\p})$}\no\\&&
			+c_{A}c_{B}
			\int\int{d\tau_{A}d\tau_{B}^{\p}}\chi_{A}(\tau_{A})\chi_{B}(\tau_{B}^{\p})\scalebox{0.95}{$[b_1e^{-i\omega(\tau_{A}-\tau_{B}^{\p})}\delta_{n_A,g_{A}}\delta_{n_B,e_{B}}+b_{2}e^{i\omega(\tau_{A}-\tau_{B}^{\p})}\delta_{n_A,e_{A}}\delta_{n_B,g_{B}}][iG_F(\bar{x}_A,\bar{x}_B^{\p})]$}\Big)\no\\&&[b_{1}^{\star}\delta_{\hat{n}_A,e_{A}}\delta_{\hat{n}_B,g_{B}}+b_{2}^{\star}\delta_{\hat{n}_A,g_{A}}\delta_{\hat{n}_B,e_{B}}],
\end{eqnarray}
where in the third term after the second last equality, we have changed $\tau_{A}^{\p}\to\tau_{A}$ and $\tau_{B}\to\tau_{B}^{\p}$. Then we used  $\Theta(\tau_{A}-\tau_{B}^{\p})+\Theta(\tau_{B}^{\p}-\tau_{A})=1$, to obtain the final expression. \\

This $R^{(2)}$ can be written in a matrix form, given by
\begin{eqnarray}\label{R2matrix}
R^{(2)}=-\begin{pmatrix}
		0&0&0&0\\
		0&b_{1}^{{\star}}(b_1F_{A}(\omega)+b_1F_{B}(-\omega)+b_2\mathcal{E}(\omega))&b_2^{{\star}}(b_1F_{A}(\omega)+b_1F_{B}(-\omega)+b_2\mathcal{E}(\omega))&0\\
	0&b_1^{{\star}}(b_2F_{A}(-\omega)+b_2F_{B}(\omega)+b_1\mathcal{E}(-\omega))&b_2^{{\star}}(b_2F_{A}(-\omega)+b_2F_{B}(\omega)+b_1\mathcal{E}(-\omega))&0\\
0&0&0&0\\
	\end{pmatrix}	,
\end{eqnarray}
where we defined
\begin{equation}\label{FA_FB}
\begin{aligned}
	F_{j}(\pm \omega)=	c_j^2\int\int{d\tau_{j}d\tau_{j}^{\p}}\chi_{j}(\tau_{j})\chi_{j}(\tau_{j}^{\p})e^{\pm i\omega(\tau_{j}-\tau_{j}^{\p})}\Theta(\tau_{j}-\tau_{j}^{\p})G^{+}(\tau_j,\tau_j^{\p}),
\end{aligned}\end{equation}
\begin{equation}\label{epsilonAB}
	\mathcal{E}(\pm\omega)=c_Ac_B\int\int{d\tau_{A}d\tau_{B}^{\p}}\chi_{A}(\tau_{A})\chi_{B}(\tau_{B}^{\p})e^{\pm i\omega(\tau_{A}- \tau_{B}^{\p})}[iG_F(\tau_A,\tau_B^{\p})].
\end{equation}
One can see that $F_{j}( \pm\omega)$ satisfies the following relations,
\begin{equation}\label{Fi++Fi+str}
\begin{aligned}
&F_{j}( \pm\omega)+F_{j}^{\star}(\pm\omega)\\&=c_j^2\int\int{d\tau_{j}d\tau_{j}^{\p}}\chi_{j}(\tau_{j})\chi_{j}(\tau_{j}^{\p})\Big[e^{\pm i\omega(\tau_{j}-\tau_{j}^{\p})}\Theta(\tau_{j}-\tau_{j}^{\p})G^{+}(\tau_j,\tau_j^{\p})+e^{\mp i \omega(\tau_{j}-\tau_{j}^{\p})}\Theta(\tau_{j}-\tau_{j}^{\p})G^{+}(\tau_j^{\p},\tau_j)\Big]\\&=c_j^2\int\int{d\tau_{j}d\tau_{j}^{\p}}\chi_{j}(\tau_{j})\chi_{j}(\tau_{j}^{\p})e^{\mp i\omega(\tau_{j}-\tau_{j}^{\p})}G^{+}(\tau_j^{\p},\tau_j)\\&=\mathcal{P}_j(\mp\omega),
\end{aligned}
\end{equation}
where in the first term of the second line we interchanged  $\tau_j$ and $\tau_j^{\p}$. Also, one finds
\begin{equation}\label{Fi++Fi-str}
\begin{aligned}
&F_{j}( \pm\omega)+F_{j}^{\star}(\mp\omega)\\&=c_j^2\int\int{d\tau_{j}d\tau_{j}^{\p}}\chi_{j}(\tau_{j})\chi_{j}(\tau_{j}^{\p})\Big[e^{\pm i\omega(\tau_{j}-\tau_{j}^{\p})}\Theta(\tau_{j}-\tau_{j}^{\p})G^{+}(\tau_j,\tau_j^{\p})+e^{\pm i\omega(\tau_{j}-\tau_{j}^{\p})}\Theta(\tau_{j}-\tau_{j}^{\p})G^{+}(\tau_j^{\p},\tau_j)\Big]\\&=c_j^2
\int\int{d\tau_{j}d\tau_{j}^{\p}}\chi_{j}(\tau_{j})\chi_{j}(\tau_{j}^{\p})e^{\pm i\omega(\tau_{j}-\tau_{j}^{\p})}\Theta(\tau_{j}-\tau_{j}^{\p})(G^{+}(\tau_j,\tau_j^{\p})+G^{+}(\tau_j^{\p},\tau_j))\\&\equiv Q_{j}(\pm\omega)=Q_{j}^{\star}(\mp\omega).
\end{aligned}
\end{equation}
 $\mathcal{E}(\omega)$ satisfies the following relation
\begin{eqnarray}\label{mathcalErelation1}
&&\mathcal{E}(\omega)+\mathcal{E}^{\star}(\omega)\no\\&&=c_Ac_B\int\int{d\tau_{A}d\tau_{B}^{\p}}\chi_{A}(\tau_{A})\chi_{B}(\tau_{B}^{\p})e^{i\omega(\tau_{A}-\tau_{B}^{\p})} [\Theta(\tau_A-\tau_B^{\p})G^+(\tau_A,\tau_B^{\p})+\Theta(\tau_B^{\p}-\tau_A)G^+(\tau_B^{\p},\tau_A)]\no\\&&+c_Ac_B\int\int{d\tau_{A}d\tau_{B}^{\p}}\chi_{A}(\tau_{A})\chi_{B}(\tau_{B}^{\p})e^{-i\omega(\tau_{B}^{\p}-\tau_{A})}[\Theta(\tau_B^{\p}-\tau_A)G^+(\tau_A,\tau_B^{\p})+\Theta(\tau_A-\tau_B^{\p})G^+(\tau_B^{\p},\tau_A)]\no\\&&=c_Ac_B\int\int{d\tau_{A}d\tau_{B}^{\p}}\chi_{A}(\tau_{A})\chi_{B}(\tau_{B}^{\p})e^{i\omega(\tau_{A}-\tau_{B}^{\p})}[G^+(\tau_A,\tau_B^{\p})+G^+(\tau_B^{\p},\tau_A)]\no\\&&=\mathcal{P}_{AB}^{\star}(-\omega,\omega)+\mathcal{P}_{AB}(\omega,-\omega)\no\\&&=\mathcal{P}_{AB}(-\omega,\omega)+\mathcal{P}_{AB}(\omega,-\omega),
\end{eqnarray}
where we interchanged $\tau_A$ and $\tau_B'$ in the second line. To arrive in the last equality, we have used the fact that $\mathcal{P}_{AB}$ is a real quantity, see (\ref{PABstr}). 
Now adding (\ref{R2matrix}) with it's hermitian conjugate, and using (\ref{Fi++Fi+str}), (\ref{Fi++Fi-str}) and (\ref{mathcalErelation1}), we obtain
\begin{eqnarray}\label{R2+R2hcDensityMatrix}
R^{(2)}+R^{(2){\dagger}}=-\begin{pmatrix}
0&0&0&0\\
0&b_1^2(\mathcal{P}_A(-\omega)+\mathcal{P}_B(\omega))+&b_1b_2(Q_A(\omega)+Q_B(-\omega))&0\\
&b_1b_2(\mathcal{P}_{AB}(-\omega,\omega)+\mathcal{P}_{AB}(\omega,-\omega))&+b_2^2\mathcal{E}(\omega)+b_1^2\mathcal{E}^{\star}(-\omega)&\\0&b_1b_2(Q_A^{\star}(\omega)+Q_B^{\star}(-\omega))&b_2^2(\mathcal{P}_A(\omega)+\mathcal{P}_B(-\omega))+&0\\
&+b_2^2\mathcal{E}^{\star}(\omega)+b_1^2\mathcal{E}(-\omega)&b_1b_2(\mathcal{P}_{AB}(-\omega,\omega)+\mathcal{P}_{AB}(\omega,-\omega))&\\0&0&0&0
\end{pmatrix}.~~
\end{eqnarray}

Finally, by adding (\ref{R1matrix}) and (\ref{R2+R2hcDensityMatrix}), we obtain the variation of the density matrix (\( \delta \rho \)) upto second order in $c_{i}$, as
\begin{equation}\label{deltarho}
\delta \rho=R^{(1)}+R^{(2)}+R^{(2){\dagger}}.
\end{equation}

\section{Calculation of $\text{Tr}(\delta \rho^{H}h_{\alpha_{k}})$}\label{C}

To evaluate the trace (\ref{traceFinal}), we need to add the diagonal terms of $\delta\rho^{H}h_{\alpha_{k}}~(k=v,~\alpha_{a_{H}},~\alpha_{a_{C}})$. This is given by multiplication of matrices given by (\ref{deltarho}) and (\ref{H0-hamiltonian}). We consider the coefficients of the initial detector state, $b_{1}$, $b_{2}$ are to be real numbers, satisfying $|b_{1}|^{2}+|b_{2}|^{2}=1$. The diagonal terms are:
\begin{eqnarray}
(\delta\rho^{H}h_{\alpha_{k}})_{e_{A}e_{B},e_{A}e_{B}}&=&[b_2^2\mathcal{P}_A(\omega_{})+b_1^2\mathcal{P}_B(\omega_{})+2b_1b_2\mathcal{P}_{AB}(\omega_{},-\omega_{})] \frac{1+\alpha_{k}}{2};\\
(\delta\rho^{H}h_{\alpha_{k}})_{e_{A}g_{B},e_{A}g_{B}}&=&-[b_1^2(\mathcal{P}_A(-\omega_{})+\mathcal{P}_B(\omega_{}))+b_1b_2(\mathcal{P}_{AB}(\omega_{},-\omega_{})+\mathcal{P}_{AB}(-\omega_{},\omega_{}))] \frac{1-\alpha_{k}}{2};\\
(\delta\rho^{H}h_{\alpha_{k}})_{g_{A}e_{B},g_{A}e_{B}}&=&-[b_2^2(\mathcal{P}_A(\omega_{})+\mathcal{P}_B(-\omega_{}))+b_1b_2(\mathcal{P}_{AB}(\omega_{},-\omega_{})+\mathcal{P}_{AB}(-\omega_{},\omega_{}))] \frac{-1+\alpha_{k}}{2};~~~~\\
(\delta\rho^{H}h_{\alpha_{k}})_{g_{A}g_{B},g_{A}g_{B}}&=&[b_1^2\mathcal{P}_A(-\omega_{})+b_2^2\mathcal{P}_B(-\omega_{})+2b_1b_2\mathcal{P}_{AB}(-\omega_{},\omega_{})]\frac{-1-\alpha_{k}}{2}.
\end{eqnarray}

Trace of $\delta\rho^{H}h_{\alpha_{k}}$ is obtained as,
\begin{eqnarray}\label{EffAppenC}
\textrm{Tr}[\delta\rho^{H} h_{\alpha_{k}}]=\{b_2^2\mathcal{P}_A(\omega_{})-b_1^2\mathcal{P}_A(-\omega_{})+b_1b_2[\mathcal{P}_{AB}(\omega_{},-\omega_{})-\mathcal{P}_{AB}(-\omega_{},\omega_{})]\}\nonumber\\+{\alpha_{k}}\{b_1^2\mathcal{P}_B(\omega_{})-b_2^2\mathcal{P}_B(-\omega_{})+b_1b_2[\mathcal{P}_{AB}(\omega_{},-\omega_{})-\mathcal{P}_{AB}(-\omega_{},\omega_{})]\}.
\end{eqnarray}
This leads to (\ref{traceFinal}) for any general initially entangled states.

\section{Calculation of $\mathcal{P}_{j}(\pm\omega)$}\label{D}

\begin{figure}[h]
    \centering
    \subfigure[]{\includegraphics[width=0.48\textwidth]{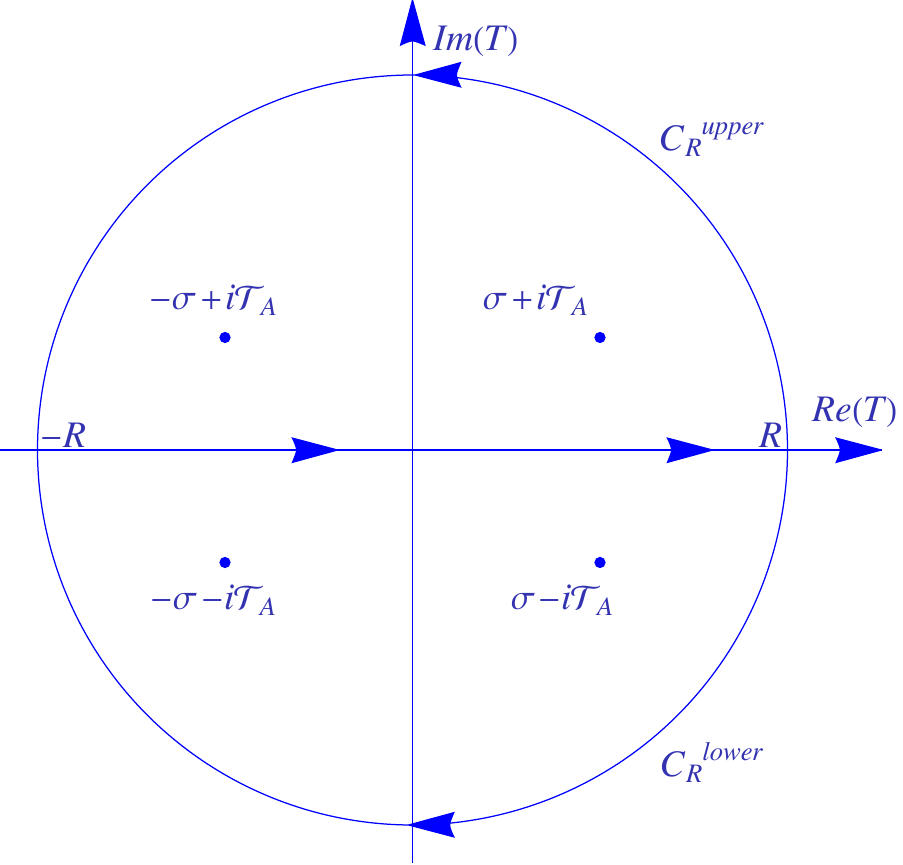}} 
    \caption{Plots showing contours of integral $T$. For an integrand like  $f(T)=e^{\pm ikT}/[(T^{2}-T_{1}^{2})(T^{2}-T_{2}^{2})]$, we choose upper and lower contours for positive and negative sign of the exponential, respectively. For $k=0$, we are free to choose any contour. Both will give same results.}
\label{fig:Contour-U-D}
\end{figure}

The expression of $\mathcal{P}_{j}~(j=A,B)$ is given by (\ref{Pi-Def}). The integrations need to be evaluated for qualitative analysis of efficiency of the cycle. As the interactions are turned on at time $-\mathcal{T}_{j_{k}}/2$ and runs upto time $\mathcal{T}_{j_{k}}/2$, the integration limits should be $-\mathcal{T}_{j_{k}}/2$ to $\mathcal{T}_{j_{k}}/2$. These integrations can not be done analytically due to their finite integration limits. However, we can extend the integration limits to $\pm \infty$ by choosing a suitable switching function. We choose the switching function given by (\ref{switch}), which is non-vanishing for $-\mathcal{T}_{j_{k}}/2<\tau_{j}<\mathcal{T}_{j_{k}}/2$ and approximately vanishing outside this domain. $\mathcal{P}_{j}(\omega)$ can be expressed by (taking $c_{A}=c_{B}=1$)
\begin{eqnarray}\label{PA-1}
\mathcal{P}_{j}( \omega)&=&\int_{-\infty}^{\infty}\int_{-\infty}^{\infty}d\tau_{j}d\tau_{j}' \chi_{j}(\tau_{j}) \chi_{j}(\tau_{j}')e^{i \omega(\tau_{j}-\tau_{j}')}G^{+}(\tau_{j}',\tau_{j}),
\end{eqnarray}
To evaluate this quantity, we choose a coordinate transform 
\begin{eqnarray}\label{CoordTrans}
T=\tau_{A}+\tau_{A}';~~~~ \sigma=\tau_{A}-\tau_{A}'.
\end{eqnarray} 
The Jacobian of this transformation is $1/2$. In the case of evaluating $\mathcal{P}_{B}$, we use (\ref{tAtB}) while detector $B$ is accelerating in RRW and (\ref{tAtB-}) while in LRW to express the integrations in terms of $\tau_{A}$ and $\tau_{A}'$. Use of (\ref{tAtB}) and (\ref{tAtB-}) implies that 
\begin{eqnarray}\label{chiAchiB}
\chi_{B}(\tau_{B}) \chi_{B}(\tau_{B}')&=&\frac{(\mathcal{T}_{B}/2)^2}{\tau_{B}^2+(\mathcal{T}_{B}/2)^2}\frac{(\mathcal{T}_{B}/2)^2}{\tau_{B}^{\prime 2}+(\mathcal{T}_{B}/2)^2}=\frac{(\pm\alpha_{a}\mathcal{T}_{A}/2)^2}{(\pm\alpha_{a}\tau_{A})^2+(\pm\alpha_{a}\mathcal{T}_{A}/2)^2}\frac{(\pm\alpha_{a}\mathcal{T}_{A}/2)^2}{(\pm\alpha_{a}\tau_{A}^{\p})^{2}+(\pm\alpha_{a}\mathcal{T}_{A}/2)^2}\no\\&=&\chi_{A}(\tau_{A}) \chi_{A}(\tau_{A}')
\end{eqnarray}
With use of (\ref{CoordTrans}), we can express the combination of the switching functions as
\begin{eqnarray}\label{chiT}
 \chi_{j}(\tau_{j}) \chi_{j}(\tau_{j}')&=&\frac{(\mathcal{T}_{A}/2)^2}{\tau_{A}^2+(\mathcal{T}_{A}/2)^2}\frac{(\mathcal{T}_{A}/2)^2}{\tau_{A}'^2+(\mathcal{T}_{A}/2)^2}\no\\&=&\frac{\mathcal{T}_{A}^{4}}{(T^{2}-T_{1}^{2})(T^{2}-T_{2}^{2})},
\end{eqnarray}
where, \( T_{1}=\sigma+i \mathcal{T}_{A} \) and \( T_{2}=-\sigma+i \mathcal{T}_{A} \). The Wightman function in (\ref{PA-1}) for detector $A$ can be written as a function of $\sigma$, as $G^{+}(\tau_{A}^{\prime},\tau_{A})=[G^{+}(\tau_{A},\tau_{A}^{\prime})]^{\star}=[G^{+}(\sigma)]^{\star}$. Then we can write (\ref{PA-1}) for $j=A$, as
\begin{eqnarray}\label{PA-11}
\mathcal{P}_{A}( \pm\omega)&=&\frac{1}{2}\int_{-\infty}^{\infty}d\sigma e^{\pm i\omega \sigma}[G^{+}(\sigma)]^{\star}\int_{-\infty}^{\infty}dT \frac{\mathcal{T}_{A}^{4}}{(T^{2}-T_{1}^{2})(T^{2}-T_{2}^{2})},
\end{eqnarray}
To evaluate $T-$integral, we used the upper contour given in (fig:\ref{fig:Contour-U-D}) (see, \cite{Arias:2018wc, Gray:2018uw}). As an integrand like  $f(T)=1/[(T^{2}-T_{1}^{2})(T^{2}-T_{2}^{2})]$ vanishes for both $T\to\pm\infty ~(\pm{i}\infty)$, we are free to choose upper or lower contour. In upper contour, the poles are at $T=T_{1}$ and $T=T_{2}$. The integration results, 

\begin{eqnarray}\label{T-PA}
&&\int_{-\infty}^{\infty} \frac{dT}{(T^{2}-T_{1}^{2})(T^{2}-T_{2}^{2})}=2\pi i\left(Res(f(T))\Big|_{T=T_{1}}+Res(f(T))\Big|_{T=T_{2}}\right)\no\\&&~~~~~~~~~~~~~~~~~~~~~~~~~~~~~~~~~~~=2\pi i\left(\frac{1}{2T_{1}(T_{1}^{2}-T_{2}^{2})}+\frac{1}{2T_{2}(T_{2}^{2}-T_{1}^{2})}\right)\no\\&&~~~~~~~~~~~~~~~~~~~~~~~~~~~~~~~~~~~=\frac{\pi}{2\mathcal{T}_{A}(\sigma^{2}+\mathcal{T}_{A}^{2})} ~.
\end{eqnarray}
Therefore, (\ref{PA-11}) becomes
\begin{eqnarray}\label{PA-2}
\mathcal{P}_{A}(\pm \omega)=\frac{\pi\mathcal{T}_{A}^{3}}{4}\int_{-\infty}^{\infty} \frac{d\sigma~e^{\pm i\omega \sigma}~[G^{+}(\sigma)]^{\star}}{\sigma^{2}+\mathcal{T}_{A}^{2}}~,
\end{eqnarray}•
 The positive frequency Wightman in Minkowski spacetime is given by \cite{book:Birrell}
 \begin{eqnarray}
 G^{+}(\bar{x},\bar{x}')=-\frac{1}{4\pi^{2}[(t-t'-i\epsilon)^{2}-|x-x'|^{2}]}~.
 \end{eqnarray}
 The trajectory of a uniformly accelerating observer moving in RRW and LRW is given in (\ref{B2}) and (\ref{B3}). One finds
 \begin{eqnarray}
 (t_{j}-t_{j}'-i\epsilon)^{2}-|x_{j}-x_{j}'|^{2}&=&\Big(\frac{1}{a_{j}}\sinh(a_{j}\tau_{j})-\frac{1}{a_{j}}\sinh(a_{j}\tau_{j}')-i\epsilon\Big)^{2}-\Big|\pm\frac{1}{a_{j}}\cosh(a_{j}\tau_{j})\mp\frac{1}{a_{j}}\cosh(a_{j}\tau_{j}')\Big|^{2}\no\\&=&-\frac{2}{a_{j}^{2}}+\frac{2}{a_{j}^{2}}\Big(\cosh(a_{j}\tau_{j})\cosh(a_{j}\tau_{j}')-\sinh(a_{j}\tau_{j})\sinh(a_{j}\tau_{j}')\Big)-i\epsilon\no\\&=&-\frac{2}{a_{j}^{2}}\big(1-[\cosh(a_{j}(\tau_{j}-\tau_{j}')-i\epsilon]\big)\no\\&=&-\frac{2}{a_{j}^{2}}\Big(1-\cosh\big(a_{j}(\tau_{j}-\tau_{j}')-i\epsilon\big)\Big)\no\\&=&\frac{4}{a_{j}^{2}}\sinh^{2}\Big(\frac{a_{j}(\tau_{j}-\tau_{j}')}{2}-i\epsilon\Big)~.
 \end{eqnarray}
 Therefore the positive frequency Wightman function of a uniformly accelerating  detector is given by (see also in \cite{book:Birrell})
\begin{eqnarray}\label{G+}
G^{+}(\tau_{j},\tau_{j}^{\prime})=-\frac{a_{j}^{2}}{16\pi^{2}\sinh^{2}(a_{j} (\tau_{j}-\tau_{j}^{\prime})/2-i\epsilon)}=-\frac{1}{4\pi^{2}}\sum_{n=-\infty}^{\infty}\frac{1}{((\tau_{j}-\tau_{j}^{\prime})-i\epsilon-2\pi i n/a_{j})^{2}}~.
\end{eqnarray}
 where we have used $\frac{(a/2)^{2}}{\sinh^{2}(ax/2)}=\sum_{n=-\infty}^{\infty}\frac{1}{(x-i2\pi n/a)^{2}}$. Now using (\ref{G+})  for $j=A$ with  (\ref{CoordTrans}), we can write (\ref{PA-2}) as
\begin{eqnarray}\label{PA-3}
\mathcal{P}_{A}(\pm\omega)&=&-\frac{\mathcal{T}_{A}^{3}}{16\pi}\int_{-\infty}^{\infty}\frac{d\sigma~e^{\pm i\omega \sigma}}{(\sigma^{2}+\mathcal{T}_{A}^{2})}    \sum_{n=-\infty}^{\infty}\frac{1}{(\sigma+i\epsilon-2\pi i n/a_{A})^{2}}~.
\end{eqnarray}
 The summation in this expression can be splited into three parts as
\begin{eqnarray}
\mathcal{P}_{A}( \pm\omega)&=&-\frac{\mathcal{T}_{A}^3}{16 \pi} \int_{-\infty}^{\infty} \frac{d \sigma~e^{\pm i\sigma  \omega }}{ \sigma ^2+\mathcal{T}_{A}^2}\Bigg(\frac{1}{(\sigma +i \epsilon )^2}+ \sum_{n=1}^{\infty}\frac{1}{\left(\sigma -\frac{i 2 \pi  n}{a_{A}}\right)^2}
+ \sum_{n=1}^{\infty}\frac{1}{\left(\sigma +\frac{i 2 \pi  n}{a_{A}}\right)^2}\Bigg)\no\\&=&\mathcal{P}_{A,0}(\pm\omega)+\mathcal{P}_{A,n}(\pm\omega)+\mathcal{P}_{A,-n}(\pm\omega)~.
\end{eqnarray}
Poles contributing to $\mathcal{P}_{A}(\omega)$ are at $i\mathcal{T}_{A},~\frac{2\pi i n}{a_{A}}$ and poles contributing to $\mathcal{P}_{A}(-\omega)$ are at $-i\epsilon,~-i\mathcal{T}_{A},~-\frac{2\pi i n}{a_{A}}$. The corresponding residues with factor $\pm2\pi i$ are ($\epsilon\to0$)
\begin{eqnarray}\label{PAres1}
(2\pi i)\text{Res}(\mathcal{P}_{A}^{(I)}( \omega))\Big|_{\sigma=i\mathcal{T}_{A}}&=&\frac{\mathcal{T}_{A}^{2}e^{-\mathcal{T}_{A} \omega }}{16}\Bigg(\frac{1}{\mathcal{T}_{A}^{2}}+\sum_{n=1}^{\infty}\frac{1}{( \mathcal{T}_{A}-2 \pi  n/a_{A})^2}+\sum_{n=1}^{\infty}\frac{1}{(\mathcal{T}_{A}+2 \pi  n/a_{A})^2}\Bigg)\no\\&=&\frac{(a_{A}\mathcal{T}_{A}/2)^2e^{-\mathcal{T}_{A} \omega }}{16\sin^{2}(a_{A}\mathcal{T}_{A}/2)},~~~~~~~~[\text{using, } \frac{(a/2)^{2}}{\sin^{2}(ax/2)}=\sum_{n=-\infty}^{\infty}\frac{1}{(x-2\pi n/a)^{2}}]~~~~~~\no\\&=&(-2\pi i)\text{Res}(\mathcal{P}_{A}^{(I)}( -\omega))\Big|_{\sigma=-i\mathcal{T}_{A}}~,
\end{eqnarray} 
\begin{eqnarray}\label{PAres2}
&&(2\pi i)\sum_{n=1}^{\infty}\text{Res}(\mathcal{P}_{A,n}^{(I)}(\omega))\Big|_{\sigma=\frac{2\pi i n}{a_{A}}}=\sum_{n=1}^{\infty}\frac{ \mathcal{T}_{A}^3 e^{-\frac{2 \pi  n \omega }{a_{A}}} \left([ \mathcal{T}_{A}^2-\frac{4 \pi ^2 n^2}{a_{A}^{2}}] \omega -\frac{4 \pi n}{  a_{A}} \right)}{8 \left(\mathcal{T}_{A}^2-\frac{4 \pi ^2 n^2}{a_{A}^{2}} \right)^2}\no\\&&~~~~~~~~~~~~~~~~~~~~~~~~~~~~~~~~~~~=(-2\pi i)\sum_{n=1}^{\infty}\text{Res}(\mathcal{P}_{A,-n}^{(I)}(-\omega))\Big|_{\sigma=-\frac{2\pi i n}{a_{A}}}~,
\end{eqnarray}
where
\begin{equation}\label{Sum}
\begin{aligned}
\sum_{n=1}^{\infty}\scalebox{0.9995}{$\frac{ \mathcal{T}_{A}^3 e^{-\frac{2 \pi  n \omega }{a_{A}}} \left([ \mathcal{T}_{A}^2-\frac{4 \pi ^2 n^2}{a_{A}^{2}}] \omega -\frac{4 \pi n}{  a_{A}} \right)}{8 \left(\mathcal{T}_{A}^2-\frac{4 \pi ^2 n^2}{a_{A}^{2}} \right)^2}$}&=\scalebox{0.9995}{$\frac{a_{A}^2 \mathcal{T}_{A}^2 e^{-\frac{2 \pi  \omega }{a_{A}}} }{64 \pi ^2}$}
\left(\Phi \left(e^{-\frac{2 \pi  \omega }{a_{A}}},2,1+\frac{a_{A} \mathcal{T}_{A} }{2 \pi }\right)-\Phi \left(e^{-\frac{2 \pi  \omega }{a_{A}}},2,1-\frac{a_{A} \mathcal{T}_{A}}{2 \pi }\right)\right)\\&+\scalebox{0.9995}{$
\frac{a_{A} \mathcal{T}_{A}^2 \omega  e^{-\frac{2 \pi  \omega }{a_{A}}}}{32 \pi }$}
\left(\Phi \left(e^{-\frac{2 \pi  \omega }{a_{A}}},1,1+\frac{a_{A} \mathcal{T}_{A}}{2 \pi }\right)-\Phi \left(e^{-\frac{2 \pi  \omega }{a_{A}}},1,1-\frac{a_{A} \mathcal{T}_{A}}{2 \pi }\right)\right)~
\end{aligned}
\end{equation}
and
\begin{eqnarray}\label{PAres3}
(-2\pi i)\text{Res}(\mathcal{P}_{A,0}^{(I)}(-\omega))\Big|_{\omega=-i\epsilon}=\frac{\mathcal{T}_{A} \omega }{8}~.
\end{eqnarray}

Here the notation, $\mathcal{P}_{A}^{(I)}$ means the integrand of the the quantity $\mathcal{P}_{A}$. Same notation also will be used in the following calculations. Using (\ref{Sum}) and adding (\ref{PAres1}) and (\ref{PAres2}), we obtain expression of $\mathcal{P}_{A}(\omega)$, given in (\ref{PAexpression}). Similarly, adding  (\ref{PAres1}), (\ref{PAres2})  and (\ref{PAres3}), we obtain the expression of $\mathcal{P}_{A}(-\omega)$, given in (\ref{PAnexpression}).\\

\underline{$\mathcal{P}_{B}$ in RRW:}~Using (\ref{tAtB}), (\ref{chiAchiB}) and (\ref{PA-1}), we can write $\mathcal{P}_{B}$ as
\begin{eqnarray}\label{PB-1}
\mathcal{P}_{B}(\pm\omega)=\alpha_{a}^{2}\int_{-\infty}^{\infty}\int_{-\infty}^{\infty}d\tau_{A}d\tau_{A}^{\p} \chi_{A}(\tau_{A}) \chi_{A}(\tau_{A}^{\p})e^{\pm i\omega\alpha_{a}(\tau_{A}-\tau_{A}')}G^{+}_{B}(\alpha_{a}\tau_{A}^{\p},\alpha_{a}\tau_{A})~.
\end{eqnarray}
The Wightman function for detector $B$ can be written as (from (\ref{G+}))
\begin{eqnarray}\label{G+B}
G^{+}_{B}(\alpha_{a}\tau_{A},\alpha_{a}\tau_{A}^{\p})&=&G^{+}(\tau_{B},\tau_{B}^{\p})=
-\frac{a_{B}^{2}}{16\pi^{2}\sinh^{2}(a_{B} (\tau_{B}-\tau_{B}^{\prime})/2-i\epsilon)}\no\\&=&-\frac{(a_{A}/\alpha_{a})^{2}}{16\pi^{2}\sinh^{2}(a_{A} (\tau_{A}-\tau_{A}^{\prime})/2-i\epsilon)}\no\\&=&-\frac{1}{4\pi^{2}\alpha_{a}^{2}}\sum_{n=-\infty}^{\infty}\frac{1}{(\sigma-i\epsilon-2\pi i n/a_{A})^{2}}~.
\end{eqnarray}
Using the coordinate transformation given in (\ref{CoordTrans}) with (\ref{chiT}), (\ref{T-PA}) and (\ref{G+B}), we can write (\ref{PB-1}) as
\begin{eqnarray}\label{PB-2}
\mathcal{P}_{B}(\pm\omega)&=&-\frac{\mathcal{T}_{A}^{4}}{8\pi^{2}}\int_{-\infty}^{\infty}\int_{-\infty}^{\infty} \frac{d\sigma~dT~e^{\pm i\omega\alpha_{a}\sigma}}{(T^{2}-T_{1}^{2})(T^{2}-T_{2}^{2})}\sum_{n=-\infty}^{\infty}\frac{1}{(\sigma+i \epsilon-\frac{2\pi i n}{a_{A}})^{2}}\no\\&=&-\frac{\mathcal{T}_{A}^{3}}{16\pi}\int_{-\infty}^{\infty} \frac{d\sigma~e^{\pm i\omega\alpha_{a}\sigma}}{\sigma^{2}+\mathcal{T}_{A}^{2}}\sum_{n=-\infty}^{\infty}\frac{1}{(\sigma+i \epsilon-\frac{2\pi i n}{a_{A}})^{2}}\no\\&=&\mathcal{P}_{A}(\pm\omega\alpha_{a})~.
\end{eqnarray}
The last equality comes form comparing expression of $\mathcal{P}_{B}(\pm\omega)$ after the second equality with expression of $\mathcal{P}_{A}(\pm\omega)$, given in (\ref{PA-3}). The only difference is presence of $\alpha_{a}$ with $\omega$ in (\ref{PB-2}). Therefore adding (\ref{PAres1}) and (\ref{PAres2}) with replacing $\omega$ with $\omega\alpha_{a}$ and using $\alpha_{a}=a_{A}/a_{B}$, gives expression of $\mathcal{P}_{B}(\omega)$ in (\ref{PBexpression}). Similary, adding (\ref{PAres1}), (\ref{PAres2}) and (\ref{PAres3}) with replacing $\omega$ with $\omega\alpha_{a}$ and using $\alpha_{a}=a_{A}/a_{B}$, gives expression of $\mathcal{P}_{B}(-\omega)$ in (\ref{PBnexpression}).\\

\underline{$\mathcal{P}_{B}$ in LRW:}~Using (\ref{tAtB-}), (\ref{chiAchiB}) and (\ref{PA-1}), we can write $\mathcal{P}_{B}$ as
\begin{eqnarray}\label{PB-LRW}
\mathcal{P}_{B}(\pm\omega)&=&\int_{-\infty}^{\infty}\int_{-\infty}^{\infty}d\tau_{B}d\tau_{B}^{\p} \chi_{B}(\tau_{B}) \chi_{B}(\tau_{B}^{\p})e^{\pm i\omega(\tau_{B}-\tau_{B}')}G^{+}(\tau_{B}^{\p},\tau_{B})\no\\&=&\alpha_{a}^{2}\int_{-\infty}^{\infty}\int_{-\infty}^{\infty}d\tau_{A}d\tau_{A}^{\p} \chi_{A}(\tau_{A}) \chi_{A}(\tau_{A}^{\p})e^{\mp i\omega\alpha_{a}(\tau_{A}-\tau_{A}')}G^{+}(-\alpha_{a}\tau_{A}^{\p},-\alpha_{a}\tau_{A})\no\\&=&\alpha_{a}^{2}\int_{-\infty}^{\infty}\int_{-\infty}^{\infty}d\tau_{A}d\tau_{A}^{\p} \chi_{A}(\tau_{A}) \chi_{A}(\tau_{A}^{\p})e^{\pm i\omega\alpha_{a}(\tau_{A}-\tau_{A}')}G^{+}(\alpha_{a}\tau_{A}^{\p},\alpha_{a}\tau_{A})~,
\end{eqnarray}
where in the last equality, we changed $\tau\to-\tau$ and we know that $G^{+}(\alpha_{a}\tau_{A}^{\p},\alpha_{a}\tau_{A})=[G^{+}(\alpha_{a}\tau_{A},\alpha_{a}\tau_{A}^{\p})]^{\star}$. This expression of $\mathcal{P}_{B}$ is same as given in (\ref{PB-1}). Thus
\begin{equation}
\mathcal{P}_{B}^{(LRW)}(\pm\omega)=\mathcal{P}_{B}^{(RRW)}(\pm\omega)=\mathcal{P}_{B}(\pm\omega)~.
\end{equation}.

\section{Calculation of $\mathcal{P}_{AB}(\omega,-\omega)-\mathcal{P}_{AB}(-\omega,\omega)$}\label{E}
Using (\ref{tAtB}), (\ref{tAtB-}) and (\ref{chiT}), we obtain the combination of the switching functions for both detectors, as
\begin{eqnarray}\label{switching}
 \chi_{A}(\tau_A)  \chi_{B}(\tau_{B}^{\p})= \chi_{A}(\tau_A)  \chi_{A}(\tau_A^{\p})=\frac{\mathcal{T}_{A}^4}{\left(T^2-T_{1}^2\right)\left(T^2-T_{2}^2\right)}.
\end{eqnarray}
\subsection{Acceleration in Same Direction}\label{E.1}
Using the coordinate transformation (\ref{CoordTrans}) with (\ref{tAtB}), we can write
\begin{eqnarray}\label{expPara}
\omega(\tau_{A}-\tau_{B}^{\p})=\frac{\omega}{2}(\sigma+T-\alpha_{a} T+\alpha_{a}\sigma)=\frac{\omega~T}{2}(1-\alpha_{a})+\frac{\omega\sigma}{2}(1+\alpha_{a}).
\end{eqnarray}
Using the trajectories given in (\ref{accelerationParallel}), denominator of the green function can be evaluated as
\begin{eqnarray}\label{greenDeno}
&&(t_{A}-t_{B}^{\p}-i\epsilon)^2-|x_{A}-x_{B}|^2
\nonumber
\\
&=&\Big(\frac{1}{a_{A}}\sinh(a_{A}\tau_A)-\frac{1}{a_{B}}\sinh(a_{B}\tau_B^{\p})-i\epsilon\Big)^2-\Big(\frac{1}{a_{A}}\cosh(a_{A}\tau_A)-\frac{1}{a_{B}}\cosh(a_{B}\tau_B^{\p})\Big)^2\nonumber\\&=&-\frac{1}{a_{A}^2}-\frac{1}{a_{B}^2}+\frac{2}{a_{A}a_{B}}(\cosh(a_{A}\tau_A)\cosh(a_{B}\tau_B^{\p})-\sinh(a_{A}\tau_A)\sinh(a_{B}\tau_B^{\p}))-i\epsilon\nonumber\\&=&-\frac{2}{a_{A}a_{B}}\frac{1}{2}\Big(\frac{a_{A}}{a_{B}}+\frac{a_{B}}{a_{A}}\Big)+\frac{2}{a_{A}a_{B}}\cosh(a_{A}\sigma-i\epsilon)
\no\\&=&\frac{2}{a_{A}a_{B}}[\cosh(a_{A}\sigma-i\epsilon)-\cosh(a_{A}\kappa)]\no\\&=&\frac{4}{a_{A}a_{B}}\sinh\Big(\frac{a_{A}(\sigma-i\epsilon-\kappa)}{2}\Big)\sinh\Big(\frac{a_{A}(\sigma-i\epsilon+\kappa)}{2}\Big)~,
\end{eqnarray}
here in the third equality, we used (\ref{tAtB}) and (\ref{CoordTrans}), and in the fourth equality we  defined
 \begin{eqnarray}\label{kappa}
\cosh(a_{A}\kappa)=\frac{1}{2}\left(\frac{a_{A}}{a_{B}}+\frac{a_{B}}{a_{A}}\right)~.
\end{eqnarray} The positive frequency Wightman function becomes
\begin{eqnarray}\label{greenPara}
G^{+}(\tau_{A},\tau_{B}^{\p})=G^{+}(\sigma)=-\frac{a_{A}a_{B}}{16\pi^2[\sinh(\frac{a_{A}(\sigma-i\epsilon-\kappa)}{2})\sinh(\frac{a_{A}(\sigma-i\epsilon+\kappa)}{2})]}=[G^{+}(\tau_{B}^{\p},\tau_{A})]^{\star}~, 
\end{eqnarray}
Putting (\ref{expPara}) and (\ref{greenPara}) in (\ref{PAB-Def}), we obtain
\begin{eqnarray}\label{PABpn}
\mathcal{P}_{AB}(\pm\omega,\mp\omega)=-\frac{a_{A}a_{B}~\alpha_{a}\mathcal{T}_{A}^{4}}{32\pi^2}\int_{-\infty}^{\infty}\frac{d\sigma~e^{\pm\frac{i\omega}{2}(1+\alpha_{a})\sigma}}{\sinh(\frac{a_{A}(\sigma+i\epsilon-\kappa)}{2})\sinh(\frac{a_{A}(\sigma+i\epsilon+\kappa)}{2})}
\int_{-\infty}^{\infty} \frac{dT~e^{\pm\frac{i\omega}{2}(1-\alpha_{a})T}}{(T^2-T_1^2)(T^2-T_2^2)}~.
\end{eqnarray}
We are interested to evaluate the quantities $\mathcal{P}_{AB}(\pm\omega)$ when $a_{A}\neq{a_{B}}$. For both the cases $\alpha_{a}>1$ and $\alpha_{a}<1$, we need to evaluate the $T-$integrals separately due to their pole structures in the complex $T-$plane. \\

\begin{figure}[h]
    \centering
    \subfigure[]{\includegraphics[width=0.48\textwidth]{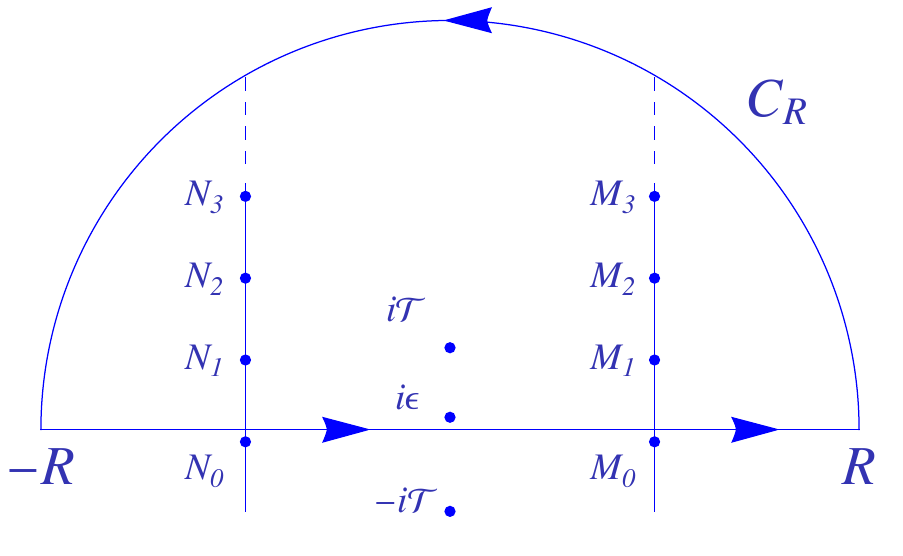}} 
    \subfigure[]{\includegraphics[width=0.49\textwidth]{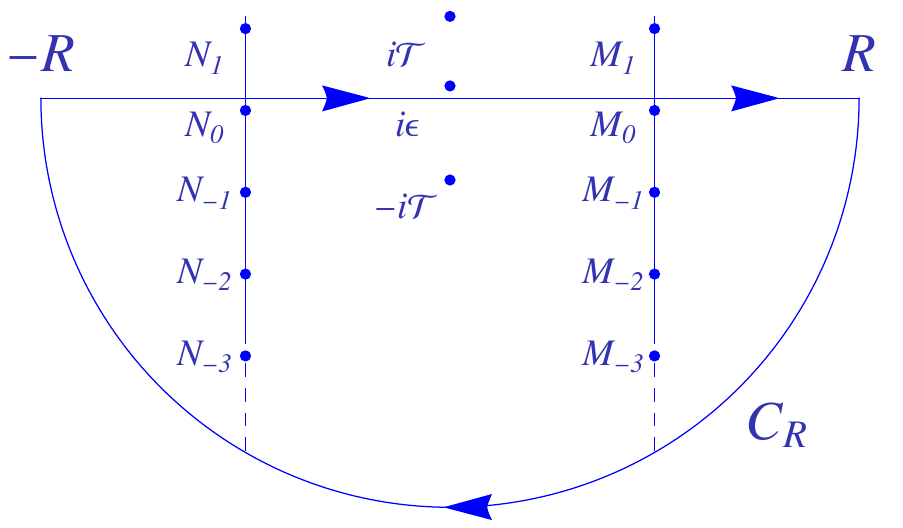}}
  \caption{Plots showing contours of integral (a) \(\mathcal{P}_{AB}(\omega,-\omega)\), (b) \(\mathcal{P}_{AB}(-\omega,\omega)\), where contributing poles are in upper and lower half of the complex-$\sigma$ plane , respectively. The poles \(M_{n}\) and \(N_{n}\) are defined as \(\kappa-i\epsilon+\frac{i2\pi n}{a_{A}}\) and  \(-\kappa-i\epsilon+\frac{i2\pi n}{a_{A}}\), respectively.}
\label{fig:Contour-S}
\end{figure}

\textbf{For $\alpha_{a}<1$:}\\
\underline{$\mathcal{P}_{AB}(\omega,-\omega)$:} The complex $T-$integration for $\mathcal{P}_{AB}(\omega,-\omega)$ has the contributing poles at $T_{1}=\sigma+i\mathcal{T}_{A}$ and $T_{2}=-\sigma+i\mathcal{T}_{A}$ in the upper half plane (see fig:(\ref{fig:Contour-U-D})). The integration yields 
\begin{eqnarray}
\frac{\pi e^{- \omega\mathcal{T}_{A}(1-\alpha_{a})/2 }}{4 ~\sigma ~ \mathcal{T}_{A}}
 \Big(\frac{ e^{-i (1-\alpha_{a})\omega\sigma/2}}{-i\mathcal{T}_{A}+\sigma }+\frac{e^{i (1-\alpha_{a} ) \omega\sigma/2}}{i\mathcal{T}_{A}+\sigma}\Big)~.
\end{eqnarray}
Putting this into (\ref{PABpn}), we will get,
\begin{eqnarray}
&&\mathcal{P}_{AB}(\omega,-\omega)=-\frac{a_{A}a_{B}~\alpha_{a}\mathcal{T}_{A}^{3}e^{-\omega\mathcal{T}_{A}(1-\alpha_{a})/2}}{128\pi}\int_{-\infty}^{\infty}\frac{d\sigma~~}{(\sigma-i\epsilon)[\sinh(\frac{a_{A}(\sigma+i\epsilon-\kappa)}{2})\sinh(\frac{a_{A}(\sigma+i\epsilon+\kappa)}{2})]}\nonumber\\&&~~~~~~~~~~~~~~~~~~~~~~~~~~~~~~~~~~~~~~~~~~~~~\times \Bigg(\frac{e^{i\omega\sigma}}{\sigma+i\mathcal{T}_{A}}+\frac{e^{i\omega\sigma\alpha_{a}}}{\sigma-i\mathcal{T}_{A}}\Bigg)\no\\&&~~~~~~~~~~~~=-\frac{a_{A}a_{B}~\alpha_{a}\mathcal{T}_{A}^{3}e^{-\omega\mathcal{T}_{A}(1-\alpha_{a})/2}}{128\pi}[\mathcal{P}_{AB,1}(\omega,-\omega)+\mathcal{P}_{AB,2}(\omega,-\omega)]~.
\end{eqnarray}
The pole at $\sigma=0$ on the real axis, will be shifted to upper half of the complex $\sigma-$plane. The contributing poles are $+i\epsilon,~+i\mathcal{T}_{A},~\pm\kappa-i\epsilon+ \frac{2\pi i n}{a_{A}}~(n=1,2,\cdots,\infty)$ in the upper half plane (see subfigure (a) in figure (\ref{fig:Contour-S})). The corresponding residues are 
\begin{eqnarray}
(2\pi i)\text{Res}(\mathcal{P}_{AB}^{(I)}(\omega,-\omega))\Big|_{\sigma=i\epsilon}=0~,
\end{eqnarray}
\begin{eqnarray}\label{pole+iTauofPABpn}
(2\pi i)\text{Res}(\mathcal{P}_{AB,2}^{(I)}(\omega,-\omega))\Big|_{\sigma=i\mathcal{T}_{A}}=\frac{2\pi~ e^{-\omega\alpha_{a}  \mathcal{T}_{A}}}{\mathcal{T}_{A}~\sinh\left(\frac{ a_{A} (-\kappa+i\mathcal{T}_{A})}{2}\right) \sinh\left(\frac{a_{A} (\kappa +i \mathcal{T}_{A})}{2} \right)}~,
\end{eqnarray}
\begin{eqnarray}\label{poleP+ofPABpn}
(2\pi i)\text{Res}(\mathcal{P}_{AB}^{(I)}(\omega,-\omega))\Big|_{\sigma=\kappa-i\epsilon+\frac{i2\pi n}{a_{A}}}=\frac{2\pi i}{(\kappa+\frac{2 \pi  ni}{ a_{A}})\frac{a_{A}}{2}\sinh(a_{A}\kappa)}\Bigg[\frac{e^{i \omega [\kappa+\frac{2 \pi  n i}{a_{A}}]}}{\kappa +\frac{2 \pi i n}{a_{A}}+i\mathcal{T}_{A}}
+\frac{e^{i\alpha_{a} \omega [\kappa+\frac{2 \pi  n i}{a_{A}}]}}{\kappa +\frac{2 \pi i n}{a_{A}}-i\mathcal{T}_{A}}\Bigg]~~~
\end{eqnarray}
and
\begin{equation}\label{poleQ+ofPABnp}
(2\pi i)\text{Res}(\mathcal{P}_{AB}^{(I)}(\omega,-\omega))\Big|_{\sigma=-\kappa-i\epsilon+\frac{i2\pi n}{a_{A}}}=\frac{-2\pi i}{(-\kappa+\frac{2 \pi  ni}{ a_{A}})\frac{a_{A}}{2}\sinh(a_{A}\kappa)}\Bigg[\frac{e^{i \omega [-\kappa+\frac{2 \pi  n i}{a_{A}}]}}{-\kappa +\frac{2 \pi i n}{a_{A}}+i\mathcal{T}_{A}}
+\frac{e^{i\alpha_{a} \omega [-\kappa+\frac{2 \pi  n i}{a_{A}}]}}{-\kappa+\frac{2 \pi i n}{a_{A}} -i\mathcal{T}_{A}}\Bigg]~,
\end{equation}
where we have used, $\cosh(in\pi)=(-1)^{n}$ and $\sinh(\pm{a_{A}}\kappa+in\pi)=\pm\sinh(a_{A}\kappa)(-1)^{n}$.

\underline{$\mathcal{P}_{AB}(-\omega,\omega)$:}  From (\ref{PABpn}), we have
\begin{eqnarray}\label{PABnp}
\mathcal{P}_{AB}(-\omega,\omega)=-\frac{a_{A}a_{B}~\alpha_{a}\mathcal{T}_{A}^{4}}{32\pi^2}\int_{-\infty}^{\infty}\frac{d\sigma~e^{-i\frac{\omega\sigma}{2}(1+\alpha_{a})}}{\sinh(\frac{a_{A}(\sigma+i\epsilon-\kappa)}{2})\sinh(\frac{a_{A}(\sigma+i\epsilon+\kappa)}{2})}
\int_{-\infty}^{\infty} \frac{dT~e^{-i\frac{\omega{T}}{2}(1-\alpha_{a})}}{(T^2-T_1^2)(T^2-T_2^2)}.
\end{eqnarray}
The complex $T-$integration has the contributing poles at $T=-T_{1}=-\sigma-i\mathcal{T}_{A}$ and $T=-T_{2}=\sigma-i\mathcal{T}_{A}$ in the lower half plane (see fig:(\ref{fig:Contour-U-D})). The integration yields
\begin{eqnarray}
\frac{\pi  e^{- \omega \mathcal{T}_{A}(1-\alpha_{a} )/2}}{4 \sigma  \mathcal{T}_{A}} \left(\frac{e^{-\frac{i\omega\sigma}{2}  (1-\alpha_{a} )}}{\sigma -i \mathcal{T}_{A}}+\frac{e^{\frac{i\omega \sigma}{2}(1-\alpha_{a}) }}{\sigma +i \mathcal{T}_{A}}\right)~.
\end{eqnarray}
Putting this into (\ref{PABnp}), we obtain
\begin{eqnarray}
&&\mathcal{P}_{AB}(-\omega,\omega)=-\frac{a_{A}a_{B}~\alpha_{a}\mathcal{T}_{A}^{3}e^{-\omega\mathcal{T}_{A}(1-\alpha_{a})/2}}{128\pi}\int_{-\infty}^{\infty}\frac{d\sigma~~}{(\sigma-i\epsilon)[\sinh(\frac{a_{A}(\sigma+i\epsilon-\kappa)}{2})\sinh(\frac{a_{A}(\sigma+i\epsilon+\kappa)}{2})]}\nonumber\\&&~~~~~~~~~~~~~~~~~~~~~~~~~~~~~~~~~~~~~~~~~~~~~\times \Bigg(\frac{e^{-i\omega\sigma}}{\sigma-i\mathcal{T}_{A}}+\frac{e^{-i\omega\sigma\alpha_{a}}}{\sigma+i\mathcal{T}_{A}}\Bigg)\no\\&&~~~~~~~=-\frac{a_{A}a_{B}~\alpha_{a}\mathcal{T}_{A}^{3}e^{-\omega\mathcal{T}_{A}(1-\alpha_{a})/2}}{128\pi}[\mathcal{P}_{AB,1}(-\omega,\omega)+\mathcal{P}_{AB,2}(-\omega,\omega)]~.
\end{eqnarray}

The contributing poles are $-i\mathcal{T}_{A},~\pm\kappa+i\epsilon- \frac{2\pi i n}{a_{A}}~(n=0,1,\cdots,\infty)$ in the lower half plane (see subfigure (b) in figure (\ref{fig:Contour-S})). The corresponding residues are
\begin{eqnarray}\label{pole-iTau-ofPABnp}
(-2\pi i)\text{Res}(\mathcal{P}_{AB,2}^{(I)}(-\omega,\omega))\Big|_{\sigma=-i\mathcal{T}_{A}}=\frac{2\pi~ e^{-\omega\alpha_{a}  \mathcal{T}_{A}}}{\mathcal{T}_{A}~\sinh\left(\frac{ a_{A} (\kappa-i\mathcal{T}_{A})}{2}\right) \sinh\left(\frac{a_{A} (-\kappa -i \mathcal{T}_{A})}{2} \right)}~,
\end{eqnarray}
\begin{equation}\label{poleP-ofPABnp}
(-2\pi i)\text{Res}(\mathcal{P}_{AB}^{(I)}(-\omega,\omega))\Big|_{\sigma=\kappa-i\epsilon-\frac{i2\pi n}{a_{A}}}=\frac{-2\pi i}{(k-\frac{2\pi i n}{a_{A}})\frac{a_{A}}{2}\sinh(a_{A} \kappa )}\Bigg[
\frac{e^{-i\omega[k-\frac{2\pi i n}{a_{A}}]}}{k-\frac{2\pi i n}{a_{A}}-i\mathcal{T}_{A}}+\frac{e^{-i\omega\alpha_{a}[k-\frac{2\pi i n}{a_{A}}]}}{k-\frac{2\pi i n}{a_{A}}+i\mathcal{T}_{A}}\Bigg]~,
\end{equation}
and
\begin{equation}\label{poleQ-ofPABnp}
(-2\pi i)\text{Res}(\mathcal{P}_{AB}^{(I)}(-\omega,\omega))\Big|_{\sigma=-\kappa-i\epsilon-\frac{i2\pi n}{a_{A}}}=\frac{-2\pi i}{(k+\frac{2\pi i n}{a_{A}})\frac{a_{A}}{2}\sinh(a_{A} \kappa )}\Bigg[
\frac{e^{i\omega[k+\frac{2\pi i n}{a_{A}}]}}{-k-\frac{2\pi i n}{a_{A}}-i\mathcal{T}_{A}}+\frac{e^{i\omega\alpha_{a}[k+\frac{2\pi i n}{a_{A}}]}}{-k-\frac{2\pi i n}{a_{A}}+i\mathcal{T}_{A}}\Bigg]~,
\end{equation}
where we have used, $\cosh(-in\pi)=(-1)^{n}$ and $\sinh(\pm{a_{A}}\kappa-in\pi)=\pm\sinh(a_{A}\kappa)(-1)^{n}$.\\

In \( \mathcal{P}_{AB}(\omega,-\omega) -  \mathcal{P}_{AB}(-\omega,\omega) \), residue given in (\ref{pole+iTauofPABpn}) will cancel out with residue (\ref{pole-iTau-ofPABnp}), and for all \( n>0 \) residues given in (\ref{poleP+ofPABpn}) and (\ref{poleQ+ofPABnp}) cancel out with residues in (\ref{poleQ-ofPABnp}) and (\ref{poleP-ofPABnp}), respectively. The difference, \( \mathcal{P}_{AB}(\omega,-\omega) -  \mathcal{P}_{AB}(-\omega,\omega) \) can be obtained by adding (\ref{poleP-ofPABnp}) and(\ref{poleQ-ofPABnp}) for $n=0$ (multiplied with the factor $\frac{-a_{A}a_{B}~\alpha_{a}\mathcal{T}_{A}^{3}e^{-\omega\mathcal{T}_{A}(1-\alpha_{a})/2}}{128\pi}$), given in (\ref{deltaPABa}).\\

\textbf{For $\alpha_{a}>1$: }\\
\underline{$\mathcal{P}_{AB}(\omega,-\omega)$:}  From (\ref{PABpn}), we have
\begin{eqnarray}\label{PABpnII}
\mathcal{P}_{AB}(\omega,-\omega)=-\frac{a_{A}a_{B}~\alpha_{a}\mathcal{T}_{A}^{4}}{32\pi^2}\int_{-\infty}^{\infty}\frac{d\sigma~e^{\frac{i\omega\sigma}{2}(1+\alpha_{a})}}{\sinh(\frac{a_{A}(\sigma+i\epsilon-\kappa)}{2})\sinh(\frac{a_{A}(\sigma+i\epsilon+\kappa)}{2})}
\int_{-\infty}^{\infty} \frac{dT~e^{-\frac{i\omega{T}}{2}(\alpha_{a}-1)}}{(T^2-T_1^2)(T^2-T_2^2)}~.
\end{eqnarray}
The complex $T-$integration has the contributing poles at $T=-T_{1}=-\sigma-i\mathcal{T}_{A}$ and $T=-T_{2}=\sigma-i\mathcal{T}_{A}$ in the lower half plane (see fig:(\ref{fig:Contour-U-D})). The integration yields

\begin{eqnarray}
\frac{\pi  e^{-\omega \mathcal{T}_{A}(\alpha_{a} -1)/2}}{4 \sigma  \mathcal{T}_{A}} \left(\frac{e^{-\frac{i \omega \sigma }{2} (\alpha_{a} -1)}}{\sigma -i \mathcal{T}_{A}}+\frac{e^{\frac{i \omega\sigma}{2} (\alpha_{a} -1)  }}{\sigma +i \mathcal{T}_{A}}\right)~.
\end{eqnarray}
Putting this into (\ref{PABpnII}), we will get,
\begin{eqnarray}
&&\mathcal{P}_{AB}(\omega,-\omega)=-\frac{a_{A}a_{B}~\alpha_{a}\mathcal{T}_{A}^{3}e^{-\omega\mathcal{T}_{A}(\alpha_{a}-1)/2}}{128\pi}\int_{-\infty}^{\infty}\frac{d\sigma~~}{(\sigma-i\epsilon)[\sinh(\frac{a_{A}(\sigma+i\epsilon-\kappa)}{2})\sinh(\frac{a_{A}(\sigma+i\epsilon+\kappa)}{2})]}\nonumber\\&&~~~~~~~~~~~~~~~~~~~~~~~~~~~~~~~~~~~~~~~~~~~~~\times \Bigg(\frac{e^{i\omega\sigma}}{\sigma-i\mathcal{T}_{A}}+\frac{e^{i\omega\sigma\alpha_{a}}}{\sigma+i\mathcal{T}_{A}}\Bigg)\no\\&&~~~~~~~~~~~~~=-\frac{a_{A}a_{B}~\alpha_{a}\mathcal{T}_{A}^{3}e^{-\omega\mathcal{T}_{A}(\alpha_{a}-1)/2}}{128\pi}[\mathcal{P}_{AB,1}(\omega,-\omega)+\mathcal{P}_{AB,2}(\omega,-\omega)]~.
\end{eqnarray}
The contributing poles are $i\epsilon,~ i\mathcal{T}_{A},~\pm\kappa-i\epsilon+ \frac{2\pi i n}{a_{A}}~(n=1,2,\cdots,\infty)$ in the upper half plane (see subfigure (a) in figure (\ref{fig:Contour-S})). The corresponding residues are
\begin{eqnarray}
(2\pi i)\text{Res}(\mathcal{P}_{AB}^{(I)}(\omega,-\omega))\Big|_{\sigma=i\epsilon}=0~,
\end{eqnarray}
\begin{eqnarray}\label{pole+iTauofPABpna}
(2\pi i)\text{Res}(\mathcal{P}_{AB,1}^{(I)}(\omega,-\omega))\Big|_{\sigma=i\mathcal{T}_{A}}=\frac{2\pi~ e^{-\omega  \mathcal{T}_{A}}}{\mathcal{T}_{A}~\sinh\left(\frac{ a_{A} (-\kappa+i\mathcal{T}_{A})}{2}\right) \sinh\left(\frac{a_{A} (\kappa +i \mathcal{T}_{A})}{2} \right)}~,
\end{eqnarray}

\begin{equation}\label{poleP+ofPABpna}
(2\pi i)\text{Res}(\mathcal{P}_{AB}^{(I)}(\omega,-\omega))\Big|_{\sigma=\kappa-i\epsilon+\frac{i2\pi n}{a_{A}}}=\frac{2\pi i}{(\kappa+\frac{2 \pi  ni}{ a_{A}})\frac{a_{A}}{2}\sinh(a_{A}\kappa)}\Bigg[\frac{e^{i \omega [\kappa+\frac{2 \pi  n i}{a_{A}}]}}{\kappa +\frac{2 \pi i n}{a_{A}}-i\mathcal{T}_{A}}
+\frac{e^{i\alpha_{a} \omega [\kappa+\frac{2 \pi  n i}{a_{A}}]}}{\kappa+\frac{2 \pi i n}{a_{A}} +i\mathcal{T}_{A}}\Bigg]~,
\end{equation}
and
\begin{equation}\label{poleQ+ofPABnpa}
(2\pi i)\text{Res}(\mathcal{P}_{AB}^{(I)}(\omega,-\omega))\Big|_{\sigma=-\kappa-i\epsilon+\frac{i2\pi n}{a_{A}}}=\frac{-2\pi i}{(-\kappa+\frac{2 \pi  ni}{ a_{A}})\frac{a_{A}}{2}\sinh(a_{A}\kappa)}\Bigg[\frac{e^{i \omega [-\kappa+\frac{2 \pi  n i}{a_{A}}]}}{-\kappa +\frac{2 \pi i n}{a_{A}}-i\mathcal{T}_{A}}
+\frac{e^{i\alpha_{a} \omega [-\kappa+\frac{2 \pi  n i}{a_{A}}]}}{-\kappa +\frac{2 \pi i n}{a_{A}}+i\mathcal{T}_{A}}\Bigg]~,
\end{equation}
where we have used, $\cosh(in\pi)=(-1)^{n}$ and $\sinh(\pm{a_{A}}\kappa+in\pi)=\pm\sinh(a_{A}\kappa)(-1)^{n}$.\\

\vskip 10mm

\underline{$\mathcal{P}_{AB}(-\omega,\omega)$:}  From (\ref{PABpn}), we have
\begin{eqnarray}\label{PABnpII}
\mathcal{P}_{AB}(-\omega,\omega)=-\frac{a_{A}a_{B}~\alpha_{a}\mathcal{T}_{A}^{4}}{32\pi^2}\int_{-\infty}^{\infty}\frac{d\sigma~e^{-\frac{i\omega\sigma}{2}(1+\alpha_{a})}}{\sinh(\frac{a_{A}(\sigma+i\epsilon-\kappa)}{2})\sinh(\frac{a_{A}(\sigma+i\epsilon+\kappa)}{2})}
\int_{-\infty}^{\infty} \frac{dT~e^{\frac{i\omega{T}}{2}(\alpha_{a}-1)}}{(T^2-T_1^2)(T^2-T_2^2)}~.
\end{eqnarray}
The complex $T-$integration has the contributing poles at $T=T_{1}=\sigma+i\mathcal{T}_{A}$ and $T=T_{2}=-\sigma+i\mathcal{T}_{A}$ in the upper half plane (see fig:(\ref{fig:Contour-U-D})). The integration yields
\begin{eqnarray}
\frac{\pi  e^{- \omega \mathcal{T}_{A}(\alpha_{a} -1)/2}}{4 \sigma  \mathcal{T}_{A}} \left(\frac{e^{-i (\alpha_{a} -1) \omega \sigma /2 }}{\sigma -i \mathcal{T}_{A}}+\frac{e^{ i (\alpha_{a} -1)\omega\sigma /2}}{\sigma +i \mathcal{T}_{A}}\right)~.
\end{eqnarray}
Putting this into (\ref{PABnpII}), we obtain
\begin{eqnarray}
&&\mathcal{P}_{AB}(-\omega,\omega)=-\frac{a_{A}a_{B}~\alpha_{a}\mathcal{T}_{A}^{3}e^{-\omega\mathcal{T}_{A}(\alpha_{a}-1)/2}}{128\pi}\int_{-\infty}^{\infty}\frac{d\sigma~~}{(\sigma-i\epsilon)[\sinh(\frac{a_{A}(\sigma+i\epsilon-\kappa)}{2})\sinh(\frac{a_{A}(\sigma+i\epsilon+\kappa)}{2})]}\nonumber\\&&~~~~~~~~~~~~~~~~~~~~~~~~~~~~~~~~~~~~~~~~~~~~~\times \Bigg(\frac{e^{-i\omega\sigma}}{\sigma+i\mathcal{T}_{A}}+\frac{e^{-i\omega\sigma\alpha_{a}}}{\sigma-i\mathcal{T}_{A}}\Bigg)\no\\&&~~~~~~~~~~~~~~~~=-\frac{a_{A}a_{B}~\alpha_{a}\mathcal{T}_{A}^{3}e^{-\omega\mathcal{T}_{A}(\alpha_{a}-1)/2}}{128\pi}[\mathcal{P}_{AB,1}(-\omega,\omega)+\mathcal{P}_{AB,2}(-\omega,\omega)]~.
\end{eqnarray}
The contributing poles are $-i\mathcal{T}_{A},~\pm\kappa-i\epsilon- \frac{2\pi i n}{a_{A}}~(n=0,1,\cdots,\infty)$ in the lower half plane (see subfigure (b) in figure (\ref{fig:Contour-S})). The corresponding residues are
\begin{eqnarray}\label{pole-iTau-ofPABnpa}
(-2\pi i)\text{Res}(\mathcal{P}_{AB,1}^{(I)}(-\omega,\omega))\Big|_{\sigma=-i\mathcal{T}_{A}}=\frac{2\pi~ e^{-\omega  \mathcal{T}_{A}}}{\mathcal{T}_{A}~\sinh\left(\frac{ a_{A} (\kappa-i\mathcal{T}_{A})}{2}\right) \sinh\left(\frac{a_{A} (-\kappa -i \mathcal{T}_{A})}{2} \right)}~,
\end{eqnarray}
\begin{equation}\label{poleP-ofPABnpa}
(-2\pi i)\text{Res}(\mathcal{P}_{AB}^{(I)}(-\omega,\omega))\Big|_{\sigma=\kappa-i\epsilon-\frac{i2\pi n}{a_{A}}}=\frac{-2\pi i}{(k-\frac{2\pi i n}{a_{A}})\frac{a_{A}}{2}\sinh(a_{A} \kappa )}\Bigg[
\frac{e^{-i\omega[k-\frac{2\pi i n}{a_{A}}]}}{k-\frac{2\pi i n}{a_{A}}+i\mathcal{T}_{A}}+\frac{e^{-i\omega\alpha_{a}[k-\frac{2\pi i n}{a_{A}}]}}{k-\frac{2\pi i n}{a_{A}}-i\mathcal{T}_{A}}\Bigg]~,
\end{equation}
and
\begin{equation}\label{poleQ-ofPABnpa}
(-2\pi i)\text{Res}(\mathcal{P}_{AB}^{(I)}(-\omega,\omega))\Big|_{\sigma=-\kappa-i\epsilon-\frac{i2\pi n}{a_{A}}}=\frac{2\pi i}{(-k-\frac{2\pi i n}{a_{A}})\frac{a_{A}}{2}\sinh(a_{A} \kappa )}\Bigg[
\frac{e^{i\omega[k+\frac{2\pi i n}{a_{A}}]}}{-k-\frac{2\pi i n}{a_{A}}+i\mathcal{T}_{A}}+\frac{e^{i\omega\alpha_{a}[k+\frac{2\pi i n}{a_{A}}]}}{-k-\frac{2\pi i n}{a_{A}}-i\mathcal{T}_{A}}\Bigg]~,
\end{equation}
where we have used, $\cosh(-in\pi)=(-1)^{n}$ and $\sinh(\pm{a_{A}}\kappa-in\pi)=\pm\sinh(a_{A}\kappa)(-1)^{n}$.\\

In \( \mathcal{P}_{AB}(\omega,-\omega) -  \mathcal{P}_{AB}(-\omega,\omega) \), residue given in (\ref{pole+iTauofPABpna})  cancel out with residue in (\ref{pole-iTau-ofPABnpa}) and for all \( n>0 \) residues given in (\ref{poleP+ofPABpna}) and (\ref{poleQ+ofPABnpa}) cancel out with residues in (\ref{poleQ-ofPABnpa}) and (\ref{poleP-ofPABnpa}), respectively. The difference, \( \mathcal{P}_{AB}(\omega,-\omega) -  \mathcal{P}_{AB}(-\omega,\omega) \) can be obtained by adding (\ref{poleQ-ofPABnpa}) and (\ref{poleP-ofPABnpa}) for $n=0$ (multiplied with the factor $\frac{-a_{A}a_{B}~\alpha_{a}\mathcal{T}_{A}^{3}e^{-\omega\mathcal{T}_{A}(1-\alpha_{a})/2}}{128\pi}$), given in (\ref{deltaPAB}).\\

\subsection{Acceleration in Opposite Direction}\label{E.2}
If the detector A and  B are accelerating anti-parallely, their trajectories are given by (\ref{accelerationAntiParallel}).
Therefore, denominator of the green function can be calculated as
\begin{eqnarray}
&&(t_{A}-t_{B}^{\p}-i\epsilon)^2-|\bar{x}_{A}-\bar{x}_{B}^{\p}|^2\no\\
&=&\Big(\frac{1}{a_{A}}\sinh(a_{A}\tau_A)-\frac{1}{a_{B}}\sinh(a_{B}\tau_B^{\p})-i\epsilon\Big)^2-\Big(\frac{1}{a_{A}}\cosh(a_{A}\tau_A)+\frac{1}{a_{B}}\cosh(a_{B}\tau_B^{\p})\Big)^2\nonumber\\&=&-\frac{1}{a_{A}^2}-\frac{1}{a_{B}^2}-\frac{2}{a_{A}a_{B}}(\cosh(a_{A}\tau_A+a_{B}\tau_B^{\p})+i\epsilon)\nonumber\\&=&-\frac{2}{a_{A}a_{B}}(\cosh(a_{A}\sigma+i\epsilon)+\cosh(a_{A}\kappa))\nonumber\\&=&-\frac{4}{a_{A}a_{B}}\cosh\Big(\frac{a_{A}(\sigma+i\epsilon+\kappa)}{2}\Big)\cosh\Big(\frac{a_{A}(\sigma+i\epsilon-\kappa)}{2}\Big)~,
\end{eqnarray}
where in the third equality, we have used (\ref{tAtB-}), (\ref{CoordTrans}) and (\ref{kappa}). The corresponding Wightman function is 
\begin{eqnarray}\label{G+op}
G^{+}(\tau_{A},\tau_{B}^{\p})=G^{+}(\sigma)=+\frac{a_{A}a_{B}}{16\pi^2[\cosh(\frac{a_{A}(\sigma+i\epsilon-\kappa)}{2})\cosh(\frac{a_{A}(\sigma+i\epsilon+\kappa)}{2})]}=[G^{+}(\tau_{B}^{\p},\tau_{A})]^{\star}~.
\end{eqnarray}
 The Wightman function has poles at 
\begin{eqnarray}
\sigma =\pm\kappa-i \epsilon \pm {i}\frac{\pi(2n+1)}{a_{A}}; ~~~n=0,1,2,\cdots\infty
\end{eqnarray}
As there is no poles on the real axis, the absence or presence of the $`i\epsilon$' in the green function, will not effect the complex integral of $\sigma$. We will drop the $`i\epsilon$' from the Wightman function. We also can check that 
\begin{eqnarray}\label{expop}
\omega(\tau_{A}- \tau_{B})= \omega( \tau_{A}+ \alpha_{a} \tau_{A}^{\p})= \frac{ \omega \sigma}{2}(1- \alpha_{a})+\frac{ \omega{T}}{2}(1+ \alpha_{a})~.
\end{eqnarray}

 \begin{figure}[h]
    \centering
    \subfigure[]{\includegraphics[width=0.48\textwidth]{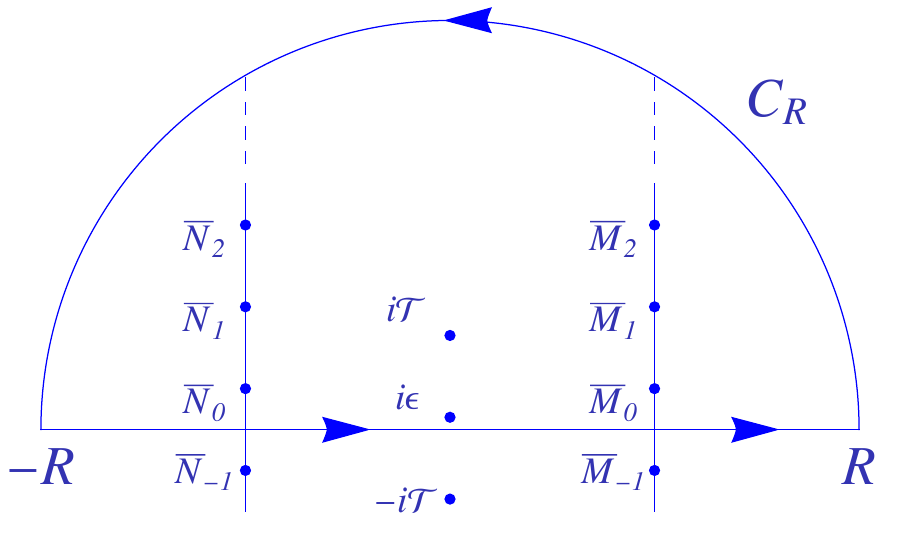}} 
    \subfigure[]{\includegraphics[width=0.49\textwidth]{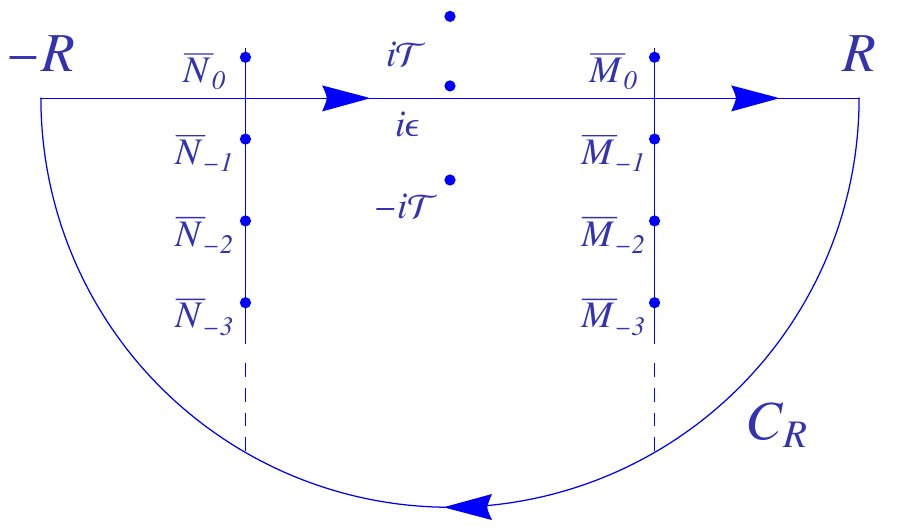}}
  \caption{Plots showing contours of integral (a) \(\mathcal{P}_{AB}(\omega,-\omega)\), (b) \(\mathcal{P}_{AB}(-\omega,\omega)\), where contributing poles are in upper and lower half of the complex-$\sigma$ plane, respectively. The poles \(\bar{M}_{n}\) and \(\bar{N}_{n}\) are defined as \(\kappa+\frac{i\pi(2 n+1)}{a_{A}}\) and  \(-\kappa+\frac{i\pi(2 n+1)}{a_{A}}\), respectively.}
\label{fig:Contour-O}
\end{figure} 

Using (\ref{G+op}), (\ref{expop}) and (\ref{switching}) in (\ref{PAB-Def}), we obtain expression of $\mathcal{P}_{AB}$, as
\begin{eqnarray}\label{PABpnOp}
\mathcal{P}_{AB}( \pm\omega,\mp \omega)= -\frac{a_{A}a_{B}~\alpha_{a}\mathcal{T}_{A}^{4}}{32\pi^2}\int_{-\infty}^{\infty}\frac{d\sigma~e^{\pm\frac{ i\omega}{2}(1-\alpha_{a})\sigma}}{\cosh(\frac{a_{A}(\sigma-\kappa)}{2})\cosh(\frac{a_{A}(\sigma+\kappa)}{2})}\int_{-\infty}^{\infty} \frac{dT~e^{\pm\frac{ i\omega}{2}(1+\alpha_{a})T}}{(T^2-T_1^2)(T^2-T_2^2)}~.
\end{eqnarray}

\underline{$\mathcal{P}_{AB}(\omega,-\omega)$:}  From (\ref{PABpnOp}), the complex $T-$integration has the contributing poles at $T=T_{1}=\sigma+i\mathcal{T}_{A}$ and $T=T_{2}=-\sigma+i\mathcal{T}_{A}$ in the upper half plane (see fig:(\ref{fig:Contour-U-D})). The integration yields
\begin{eqnarray}
\frac{\pi  e^{-\frac{1}{2} \omega\mathcal{T}_{A} (\alpha_{a} +1) }}{4 \sigma  \mathcal{T}_{A}}
 \left(\frac{e^{\frac{i}{2} (\alpha_{a} +1) \omega\sigma }}{\sigma +i \mathcal{T}_{A}}+\frac{e^{-\frac{i}{2} (\alpha +1)  \omega\sigma }}{\sigma -i \mathcal{T}_{A}}\right)~.
 \end{eqnarray}
Putting this into (\ref{PABpnOp})
\begin{eqnarray}
\mathcal{P}_{AB}(\omega,- \omega)&=&-\frac{a_{A}a_{B}~\alpha_{a}\mathcal{T}_{A}^{3}~e^{-(1+\alpha_{a})\mathcal{T}_{A} \omega/2}}{128\pi} 
\no\\
&&\times\int_{-\infty}^{\infty}\frac{d\sigma}{(\sigma-i\epsilon)[\cosh(\frac{a_{A}(\sigma-\kappa)}{2})\cosh(\frac{a_{A}(\sigma+\kappa)}{2})]}\Big(\frac{e^{i\omega\sigma}}{\sigma+i\mathcal{T}_{A}}+\frac{e^{-i\alpha_{a} \omega\sigma}}{\sigma -i \mathcal{T}_{A}}\Big)\no\\&=&-\frac{a_{A}a_{B}~\alpha_{a}\mathcal{T}_{A}^{3}~e^{-(1+\alpha_{a})\mathcal{T}_{A} \omega/2}}{128\pi} [\mathcal{P}_{AB,1}(\omega,- \omega)+\mathcal{P}_{AB,2}(\omega,- \omega)]~.
\end{eqnarray}
 The contributing poles are $+i\epsilon,~\pm\kappa\pm\frac{i \pi(2n+1)}{a_{A}}~(n=0,1,\cdots,\infty)$ for the upper and lower half planes.
The residues of the first term in the upper half plane (see subfigure (a) of figure (\ref{fig:Contour-O})) are obtained as
\begin{eqnarray}\label{Pole+iepPABop1}
\text{Res}(\mathcal{P}_{AB,1}^{(I)}(\omega,- \omega))\Big|_{\sigma=i\epsilon}=\frac{2\pi}{\mathcal{T}_{A}\cosh^2\big(\frac{a_{A} \kappa }{2}\big)}
\end{eqnarray}and
\begin{equation}\label{PoleUpperPABop1}
\begin{aligned}
&(2\pi i)\text{Res}(\mathcal{P}_{AB,1}^{(I)}(\omega,- \omega))\Big|_{\sigma=\kappa+\frac{\pi i(2 n+1)}{a_{A}}}+(2\pi i)\text{Res}(\mathcal{P}_{AB,1}^{(I)}(\omega,- \omega))\Big|_{\sigma=-\kappa+\frac{\pi i (2n+1)}{a_{A}}}\\&=
\frac{4\pi i~\text{csch}(a_{A}\kappa)}{a_{A}\sinh^{2}(\frac{(2n+1)i\pi}{2})}\Bigg(\frac{e^{i\omega\big[k+\frac{(2n+1)i\pi}{a_{A}}\big]}}{(k+\frac{(2n+1)i\pi}{a_{A}})(k+\frac{(2n+1)i\pi}{a_{A}}+i\mathcal{T}_{A})}- \frac{e^{i\omega\big[-k+\frac{(2n+1)i\pi}{a_{A}}\big]}}{(k-\frac{(2n+1)i\pi}{a_{A}})(k-\frac{(2n+1)i\pi}{a_{A}}-i\mathcal{T}_{A})}\Bigg)~,
\end{aligned}
\end{equation}
where we have used $\cosh(\pm{ak}\pm(2n+1)\frac{i\pi}{2})=\sinh(\pm{ak})\sinh(\pm(2n+1)\frac{i\pi}{2})$ and for the second term in the lower half plane  (see subfigure (b) of figure (\ref{fig:Contour-O})), we obtain
\begin{equation}\label{PoleLowerPABop1}
\begin{aligned}
&(-2\pi i)\text{Res}(\mathcal{P}_{AB,2}^{(I)}(\omega,- \omega))\Big|_{\sigma=\kappa-\frac{\pi i(2 n+1)}{a_{A}}}+(-2\pi i)\text{Res}(\mathcal{P}_{AB,2}^{(I)}(\omega,- \omega))\Big|_{\sigma=-\kappa-\frac{\pi i (2n+1)}{a_{A}}}\\&=
\frac{-4\pi i~\text{csch}(a_{A}\kappa)}{a_{A}\sinh^{2}(-\frac{(2n+1)i\pi}{2})}\Bigg(\frac{e^{-i\omega\alpha_{a}\big[k-\frac{(2n+1)i\pi}{a_{A}}\big]}}{(k-\frac{(2n+1)i\pi}{a_{A}})(k-\frac{(2n+1)i\pi}{a_{A}}-i\mathcal{T}_{A})}- \frac{e^{-i\omega\alpha_{a}\big[-k-\frac{(2n+1)i\pi}{a_{A}}\big]}}{(k+\frac{(2n+1)i\pi}{a_{A}})(k+\frac{(2n+1)i\pi}{a_{A}}+i\mathcal{T}_{A})}\Bigg)~.
\end{aligned}
\end{equation}

\underline{$\mathcal{P}_{AB}(-\omega,\omega)$:}  From (\ref{PABpnOp}), the complex $T-$integration has the contributing poles at $T=-T_{1}=-\sigma-i\mathcal{T}_{A}$ and $T=-T_{2}=\sigma-i\mathcal{T}_{A}$ in the lower half plane (see fig:(\ref{fig:Contour-U-D})). The integration yields
\begin{eqnarray}
\frac{\pi  e^{-\frac{1}{2} (\alpha_{a} +1) \mathcal{T}_{A} \omega }}{4 \sigma  \mathcal{T}_{A}}\Bigg(\frac{e^{\frac{i}{2}(\alpha_{a} +1) \sigma  \omega }}{\sigma +i \mathcal{T}_{A}}+\frac{e^{-\frac{i}{2}  (\alpha_{a} +1) \sigma  \omega }}{\sigma -i \mathcal{T}_{A}}\Bigg)~.
\end{eqnarray}
Putting this into (\ref{PABpnOp}), we obtain
\begin{eqnarray}
\mathcal{P}_{AB}(-\omega,\omega)&=&-\frac{a_{A}a_{B}~\alpha_{a}\mathcal{T}_{A}^{3}~e^{-(1+\alpha_{a})\mathcal{T}_{A} \omega/2}}{128\pi} 
\no\\
&&\times\int_{-\infty}^{\infty}\frac{d\sigma}{(\sigma-i\epsilon)[\cosh(\frac{a_{A}(\sigma-\kappa)}{2})\cosh(\frac{a_{A}(\sigma+\kappa)}{2})]}\Big(\frac{e^{i\alpha_{a}\omega\sigma}}{\sigma+i\mathcal{T}_{A}}+\frac{e^{-i \omega\sigma}}{\sigma -i \mathcal{T}_{A}}\Big)\no\\&=&-\frac{a_{A}a_{B}~\alpha_{a}\mathcal{T}_{A}^{3}~e^{-(1+\alpha_{a})\mathcal{T}_{A} \omega/2}}{128\pi} [\mathcal{P}_{AB,1}(-\omega,\omega)+\mathcal{P}_{AB,2}(-\omega,\omega)]~.
\end{eqnarray}
 The contributing poles are $+i\epsilon,~\pm\kappa\pm\frac{i \pi(2n+1)}{a_{A}}~(n=0,1,\cdots,\infty)$ for the upper half plane and lower half plane. The corresponding residues for the first term in the upper half plane (see subfigure (a) of figure (\ref{fig:Contour-O})),  are
 \begin{eqnarray}\label{Pole+iepPABop2}
\text{Res}(\mathcal{P}_{AB,1}^{(I)}(-\omega, \omega))\Big|_{\sigma=i\epsilon}=\frac{2\pi}{\mathcal{T}_{A}\cosh^2\big(\frac{a_{A} \kappa }{2}\big)}
\end{eqnarray}
and
\begin{equation}\label{PoleUpperPABop2}
\begin{aligned}
&(2\pi i)\text{Res}(\mathcal{P}_{AB,1}^{(I)}(-\omega, \omega))\Big|_{\sigma=\kappa+\frac{\pi i(2 n+1)}{a_{A}}}+(2\pi i)\text{Res}(\mathcal{P}_{AB,1}^{(I)}(-\omega,\omega))\Big|_{\sigma=-\kappa+\frac{\pi i (2n+1)}{a_{A}}}\\&=
\frac{4\pi i~\text{csch}(a_{A}\kappa)}{a_{A}\sinh^{2}(\frac{(2n+1)i\pi}{2})}\Bigg(\frac{e^{i\omega\alpha_{a}\big[k+\frac{(2n+1)i\pi}{a_{A}}\big]}}{(k+\frac{(2n+1)i\pi}{a_{A}})(k+\frac{(2n+1)i\pi}{a_{A}}+i\mathcal{T}_{A})}- \frac{e^{i\omega\alpha_{a}\big[-k+\frac{(2n+1)i\pi}{a_{A}}\big]}}{(k-\frac{(2n+1)i\pi}{a_{A}})(k-\frac{(2n+1)i\pi}{a_{A}}-i\mathcal{T}_{A})}\Bigg)~.
\end{aligned}
\end{equation}
For the second term in the lower half plane (see subfigure (b) of figure (\ref{fig:Contour-O})), we obtain
\begin{equation}\label{PoleLowerPABop2}
\begin{aligned}
&(-2\pi i)\text{Res}(\mathcal{P}_{AB,2}^{(I)}(-\omega, \omega))\Big|_{\sigma=\kappa-\frac{\pi i(2 n+1)}{a_{A}}}+(-2\pi i)\text{Res}(\mathcal{P}_{AB,2}^{(I)}(-\omega,\omega))\Big|_{\sigma=-\kappa-\frac{\pi i (2n+1)}{a_{A}}}\\&=
\frac{-4\pi i~\text{csch}(a_{A}\kappa)}{a_{A}\sinh^{2}(-\frac{(2n+1)i\pi}{2})}\Bigg(\frac{e^{-i\omega\big[k-\frac{(2n+1)i\pi}{a_{A}}\big]}}{(k-\frac{(2n+1)i\pi}{a_{A}})(k-\frac{(2n+1)i\pi}{a_{A}}-i\mathcal{T}_{A})}- \frac{e^{-i\omega\big[-k-\frac{(2n+1)i\pi}{a_{A}}\big]}}{(k+\frac{(2n+1)i\pi}{a_{A}})(k+\frac{(2n+1)i\pi}{a_{A}}+i\mathcal{T}_{A})}\Bigg)~.
\end{aligned}
\end{equation}

 So, in \( \Delta\mathcal{P}_{AB} \), residues in (\ref{Pole+iepPABop1}), (\ref{PoleUpperPABop1}), and (\ref{PoleLowerPABop1}) cancel out with residues in (\ref{Pole+iepPABop2}), (\ref{PoleLowerPABop2}) and  (\ref{PoleUpperPABop2}), respectively. Hence, $\Delta\mathcal{P}_{AB}$ vanishes for anti-parallel acceleration of the detectors.

\end{appendix}

\bibliography{EUOE_Dipankar2}
\bibliographystyle{ieeetr}



\end{document}